\documentclass[%
 reprint,
 superscriptaddress,
 amsmath,amssymb,
 aps,
]{revtex4-2}

\usepackage{graphicx}
\usepackage{dcolumn}
\usepackage{bm}
\usepackage{hyperref}

\hypersetup{
  colorlinks   = true, 
  urlcolor     = blue, 
  linkcolor    = blue, 
  citecolor   = blue 
}

\usepackage[normalem]{ulem}
\usepackage{xcolor}
\usepackage{mathdots}
\usepackage{comment}
\usepackage{verbatim}
\usepackage{bbm}
\usepackage{amsthm}
\usepackage{makecell}

\newcommand{\abs}[1]{\lvert #1 \rvert}
\newcommand{\ket}[1]{\lvert #1 \rangle}
\newcommand{\bra}[1]{\langle #1 \rvert}

\newcommand{\tr}{\operatorname{tr}}

\usepackage{textcomp}

\usepackage[smallerops]{newtxmath}

\begin{document}

\preprint{APS/123-QED}

\title{Solvable Quantum Circuits from Spacetime Lattices}

\author{Michael A. Rampp}
\affiliation{Max Planck Institute for the Physics of Complex Systems, 01187 Dresden, Germany}

\author{Suhail A. Rather}
\affiliation{Max Planck Institute for the Physics of Complex Systems, 01187 Dresden, Germany}
\affiliation{Dahlem Center for Complex Quantum Systems, Freie Universit\"at Berlin, 14195 Berlin, Germany}

\author{Pieter W. Claeys}
\affiliation{Max Planck Institute for the Physics of Complex Systems, 01187 Dresden, Germany}

\date{\today}

\begin{abstract}
In recent years dual-unitary circuits and their multi-unitary generalizations have emerged as exactly solvable yet chaotic models of quantum many-body dynamics.
However, a systematic picture for the solvability of multi-unitary dynamics remains missing. 
We present a framework encompassing a large class of such non-integrable models with exactly solvable dynamics, which we term \emph{completely reducible} circuits. 
In these circuits, the entanglement membrane determining operator growth and entanglement dynamics can be characterized analytically.
Completely reducible circuits extend the notion of space-time symmetry to more general lattice geometries, breaking dual-unitarity globally but not locally, and allow for a rich phenomenology going beyond dual-unitarity.
As example, we introduce circuits that support four and five directions of information flow.
We derive a general expression for the entanglement line tension in terms of the pattern of information flow in spacetime. 
The solvability is shown to be related to the absence of knots of this information flow, connecting entanglement dynamics to the Kauffman bracket as knot invariant. 
Building on these results, we propose that in general non-integrable dynamics the curvature of the entanglement line tension can be interpreted as a density of information transport. 
Our results provide a new and unified framework for exactly solvable models of many-body quantum chaos, encompassing and extending known constructions.
\end{abstract}

\maketitle

\tableofcontents


\section{Introduction}

The quantum dynamics of interacting many-body systems presents a formidable challenge. 
This issue is especially pressing for nonintegrable (``chaotic'') dynamics, where exact results are typically out of reach and numerical simulations require an exponential growth of resources.
In recent years, solvable models of quantum chaos have been identified in the study of dual-unitary circuits, leading to an improved understanding of quantum chaos and thermalization~\cite{Akila2016,Bertini2018,Gopalakrishnan2019,Bertini2019,Bertini2019a,Piroli2020,Claeys2020,Claeys2021,Aravinda2021,Bertini2025}. Dual-unitary circuits enable the exact computation of certain dynamical quantities such as dynamical two-point correlation functions~\cite{Bertini2019}, entanglement dynamics~\cite{Bertini2019a,Piroli2020}, and the spectral form factor~\cite{Bertini2018} (for a recent review, see Ref.~\cite{Bertini2025}). Nevertheless, dual-unitary circuits are not integrable -- in general they possess no conservation laws at all. In fact, away from known fine-tuned points they are provably chaotic, and the dynamics are in accordance with general expectations from random matrix theory~\cite{Bertini2018,Flack2020,Bertini2021}. Dual-unitary circuits hence present a new paradigm of exactly solvable models, avoiding the nonergodicity of `conventional' integrability~\cite{calabrese_introduction_2016}.

These results in turn motivated a wide range of extensions and generalizations of dual-unitarity, such as different lattice geometries or different algebraic conditions leading to solvability~\cite{Jonay2021,Milbradt2023,Sommers2023,Mestyan2024,Sommers2024,Yu2024,Rampp2025,Liu2025,Breach2025}.
The solvability of most of these extensions is typically established on an ad hoc basis, and it is not a priori clear how the solvability constrains the dynamics.
In this paper, we address the problem of finding a common framework for at least part of the zoo of these exactly solvable models. This framework is based on spacetime lattices of dual-unitary interactions, where the combination of the lattice geometry with local dual-unitarity strongly constrains the information flow in spacetime, as directly reflected in the operator and entanglement dynamics. This framework can be applied to systematically understand and identify new solvable models.

More specifically, we introduce a large class of exactly solvable models of non-integrable quantum dynamics in one dimension and analyze their properties in a unified manner. We show that placing dual-unitary gates, or more generally biunitary connections~\cite{Reutter2019,Claeys2024,Rampp2025}, on various lattices in spacetime gives rise to models where entanglement dynamics and correlation functions can be analyzed exactly. 
Usual brickwork dual-unitary circuits are understood to live on the square lattice, and the space-time duality resulting in their solvability directly reflects the symmetry of the square lattice. 
Different lattice geometries then allow for entanglement and correlation dynamics going beyond the strong constraints imposed by dual-unitarity.
As one example, dual-unitarity fixes every velocity to be maximal, resulting in their denomination of maximum velocity circuits~\cite{Claeys2020} and the vanishing of correlations away from the maximum velocity light rays~\cite{Bertini2019}. These more general lattice geometries allow for different velocities and can support correlations on multiple light rays with different velocities, approaching the general physics of models with a continuum of (quasi-)particle velocities. Our discussion also provides a novel perspective on dual-unitary circuits, showing how their space-time duality is composed of two distinct parts: the local dual-unitarity condition and the spacetime geometry. For other geometries where this duality is broken, dual-unitarity can still lead to solvability. 

Crucially, not every possible lattice leads to solvability. We identify a subclass of lattices that leads to dynamics we term {completely reducible}, which we identify as the universal feature underlying their solvability. 
Completely reducible circuits share a number of interconnected features that set them apart from generic circuits: (i) information is constrained to flow along a finite number of directions in spacetime, (ii) the operator entanglement spectrum is flat, (iii) correlation functions vanish away from isolated light rays, (iv) they exhibit a vanishing Thouless time -- a form of maximal quantum chaos, and (v) the dynamics does not exhibit any `knots' in spacetime. 
For the latter, we find that completely reducible circuits are related to the mathematical theory of knots. We show that complete reducibility implies that certain knots in spacetime that can be formed from the circuit are not knotted (more precisely these are links: knots made up of multiple components). Furthermore, the entanglement line tension is determined by the Kauffman bracket of the link, a polynomial invariant. This reveals a surprising connection between the contractions of certain tensor networks, the dynamics of non-integrable many-body systems, and knot theory.

In the remainder of this section, we first present a high-level overview of our main results (Sec.~\ref{subsec:overview}), followed by an introduction to dual-unitarity (Sec.~\ref{subsec:du_circuits}) and entanglement membrane theory as the framework through which we understand solvability (Sec.~\ref{subsec:EMT}).

\subsection{Overview of the main results}
\label{subsec:overview}

\begin{figure*}[t]
    \centering
    \includegraphics[width = 0.9\textwidth]{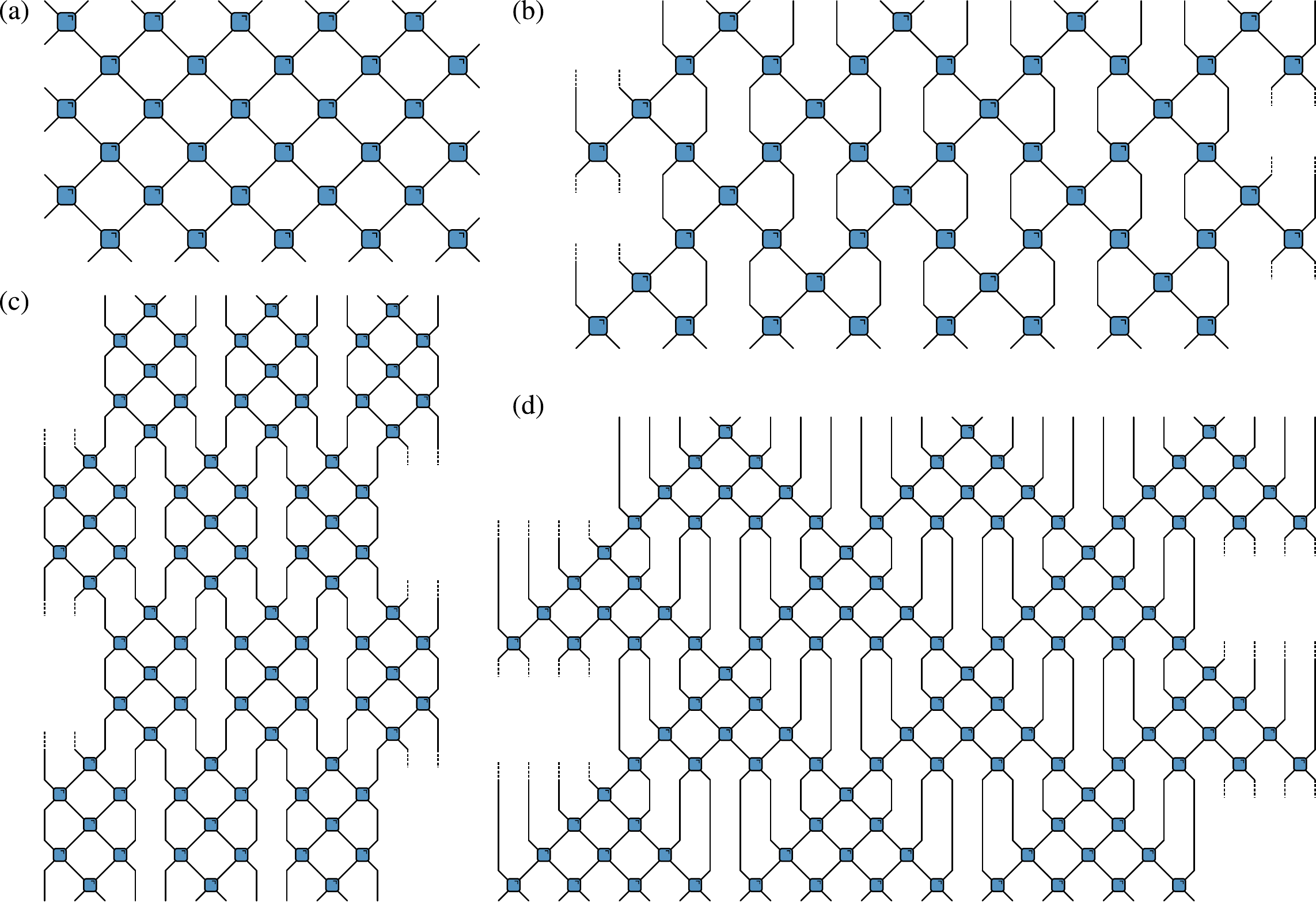}
    \caption{Examples of spacetime lattice circuits of dual-unitary gates (as represented by blue squres). The horizontal and vertical directions correspond to space and time respectively. (a) A square lattice returns a dual-unitary brickwork circuit, (b) A Kagome lattice of dual-unitary interactions returns triunitary dynamics, (c) a generic lattice which does not enable any simplifications of tensor network diagrams representing observables, (d) a lattice which gives rise to completely reducible dynamics.}
    \label{fig:lattices}
    \vspace{-\baselineskip}
\end{figure*}

In this paper, we investigate the properties of quantum circuits composed of dual-unitary interactions placed on general lattices in spacetime, thus breaking the space-time duality of usual brickwork dual-unitary circuits. In these lattices the horizontal (vertical) direction corresponds to space (time).
In Fig.~\ref{fig:lattices} we illustrate four possible choices of lattices. The first two correspond to known solvable models: Fig.~\ref{fig:lattices}(a) to a standard dual-unitary brickwork circuit, Fig.~\ref{fig:lattices}(b) to dual-unitary interactions on the Kagome lattice, returning triunitary circuit dynamics~\cite{Jonay2021,Rampp2025}. The latter two illustrate more general lattices which, while superficially similar, behave qualitatively differently. The circuit of Fig.~\ref{fig:lattices}(c) does not enable any simplifications in the calculation of the dynamics, in spite of the dual-unitarity of the local interactions, whereas Fig.~\ref{fig:lattices}(d) remains solvable and e.g. the entanglement line tension (ELT) underlying entanglement membrane theory can be exactly evaluated. This example however does not satisfy any known solvability constraints. 

These solvable lattice share a property which we term \emph{complete reducibility}. Complete reducibility is a purely geometric property of the lattice, allowing tensor networks that represent physical quantities such as the ELT to be fully contracted using only the unitarity and dual unitarity of the interactions (as formalized in Sec.~\ref{sec:framework}). 
The lattice geometries in Fig.~\ref{fig:lattices}(a), (b) and (d) are completely reducible, while (c) is not.
The resulting circuits in turn expand the phenomenology of dual-unitarity, and we introduce various solvable circuits throughout this paper giving rise to hitherto unobserved behavior.
For example, in the circuit of Fig.~\ref{fig:lattices}(d), correlations and information spread along four discrete rays in spacetime.
This behavior is reflected in the entanglement dynamics, with the entanglement line tension being a piecewise linear function of the ray velocity $v$ with kinks at the directions of information flow. 
This is unexpected from the point of view of symmetry: previous constructions of exactly solvable circuits have exploited rotational symmetries of spacetime lattices to generate additional unitary directions, prohibiting dynamics with four or more directions of information flow in $(1+1)$ dimensions. The presented constructions do not rely on symmetry, highlighting that the mechanism generating solvability is more subtle.

Complete reducibility implies that a large class of physical quantities, but not all, can be evaluated to an expression that does not depend on the choice of gates in the circuit.
Specifically, we focus on physical quantities that can be represented as tensor networks with boundary conditions expressed in terms of permutation states.
As illustrative example we focus on the ELT, but this includes more general quantifiers of (operator) entanglement, as well as operator-averaged OTOCs and their higher-order generalizations, the frame potential, etc.~\cite{HunterJones2019}.
Such objects are ubiquitous in the study of entanglement, chaos, and scrambling, such that this is not a strong restriction.
The independence of the choice of gates expresses a particular insensitivity to the microscopic details of the dynamics, in that the result of the contraction only depends on the boundary conditions and the geometry of the circuit. 
It can be understood as a robustness to integrability-breaking interactions, as long as the interactions remain dual unitary: completely reducible quantities evaluate to the same result for free evolution made of SWAP gates (the simplest dual-unitary gates) as for interacting evolution in the same geometry. 
While the former is trivially solvable, nonscrambling, and highly nonergodic, the latter is generically chaotic, scrambling, and ergodic, such that complete reducibility highlights an unexpected universality in their corresponding dynamics.
While correlation functions remain analytically tractable in these circuits, they cannot be expressed in this completely reducible way, as also implied by the observation that correlation functions behave qualitatively differently in circuits with and without interactions.

The robustness to interactions can in turn be used to identify which spacetime lattices lead to completely reducible dynamics. 
Starting from a spacetime lattice where all dual-unitary gates are tuned to the non-interacting SWAP point, in Sec.~\ref{sec:defects} we re-introduce interactions for some of the gates and check if this is sufficient to spoil reducibility of the circuit. 
By collecting elementary non-reducible diagrams we find an infinite number of conditions excluding complete reducibility. 
We use these results to give a physical criterion for the absence of complete reducibility. 
If information is transported along a ray with velocity $v=0$ by worldlines that cross the bipartition between $x>0$ and $x<0$, i.e. if information is locally flowing ``back and forth'', then the circuit cannot have a completely reducible operator entanglement for sufficiently large times.

In Sec.~\ref{sec:consequences} we derive several consequences of complete reducibility that follow from the robustness to interactions.
Information always flows along a discrete set of rays in spacetime, determined by the worldlines of the associated non-interacting circuit, leading to an entanglement line tension that is piecewise linear and correlation functions that are supported only along the rays of information flow. From the knowledge of the entanglement line tension, we can infer several other properties such as the butterfly velocity and the decay rate of out-of-time-order correlators. The operator entanglement spectrum is flat, despite the circuits being generically non-Clifford. Together with the constraints on the information flow this implies that the entanglement velocity is quantized and set by the Schmidt rank $\mathcal{R}$ of the unit cell of the lattice (which can itself be thought of as a unitary gate) as
\begin{align}
    v_E = \frac{\log\mathcal{R}}{\log q^2},
\end{align}
with $q^2$ being the Hilbert space dimension of the unit cell. This generalizes results on DU and DU2 circuits.
Next to these analyical results, we establish using numerical evidence that the models we construct are indeed non-integrable or rather quantum chaotic. We numerically compute the spectral form factor for a variety of spacetime lattice geometries hosting completely reducible dynamics and demonstrate good agreement with the predictions of random matrix theory, a common indicator of many-body quantum chaos. Furthermore, we find that the Thouless time, the time scale associated with early-time deviations from the random matrix theory prediction, vanishes, positioning completely reducible circuits as ``maximally chaotic''.

For completely reducible circuits the entanglement line tension is determined purely by the associated non-interacting circuit, such that it can be expressed in terms of the number and direction of the worldlines, showing how the entanglement dynamics is determined by the information flow. This can be thought of as a ``quasiparticle picture'' for non-integrable circuits. We propose that this picture can be extended to general non-integrable circuits by replacing the discrete directions of information flow by a continuous density. We show that the curvature of the line tension is then given by the density of information flow, providing a physical reason for the convexity of the line tension. 

We also present a surprising connection between completely reducible circuits and the mathematical theory of knots (Sec.~\ref{sec:knots}). We construct a mapping associating a tensor network diagram representing the operator entanglement to a link (a knot consisting of multiple components). We show that if the tensor network diagram is completely reducible then the link is equivalent to a disconnected collection of unknots, i.e. not knotted or linked. This connection gives an intuitive perspective on the solvability of completely reducible circuits. 
Moreover, we show that for completely reducible circuits the operator entanglement is given by the Kauffman bracket of the associated link, explicitly connecting a polynomial invariant of links to the entanglement dynamics of the circuit.

The connection to knot theory warrants comparing our results on generically non-integrable many-body dynamics to the theory of integrable models. The theory of integrability has many long-standing and deep connections to the theory of knots, the most intuitive of which is the analogy between the Yang-Baxter equation and the Reidemeister III move, a particular deformation of knot diagrams. Similarly, in our theory, unitarity and dual-unitarity of the interactions are analogous to the Reidemeister II move.

\subsection{Dual-unitary circuits}
\label{subsec:du_circuits}

In this section, we briefly introduce dual-unitary interactions and fix some notation.
We consider a one-dimensional chain of $q$-state quantum systems (\emph{qudits}), with local Hilbert space spanned by $\{\ket{a},a=1\dots q\}$. Interactions are encoded in two-site unitary gates $U$, corresponding to $q^2 \times q^2$ unitary matrices. In tensor network notation~\cite{Orus2014}, the matrix elements of these unitary gates and their Hermitian conjugate are represented as
\begin{align}
    \bra{ab}U\ket{cd} = \vcenter{\hbox{\includegraphics[width=0.07\textwidth]{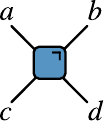}}}, \quad
    \bra{ab}U^{\dagger}\ket{cd} = \vcenter{\hbox{\includegraphics[width=0.07\textwidth]{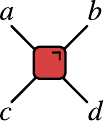}}}\,.
\end{align}
Tensor contractions are graphically represented by connecting two open legs. Unitarity fixes $U U^{\dagger} = U^{\dagger}U = \mathbbm{1}$, which is graphically represented as
\begin{align}
    \vcenter{\hbox{\includegraphics[width=0.053\textwidth]{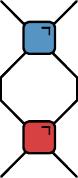}}}\,\,=\,\,\vcenter{\hbox{\includegraphics[width=0.053\textwidth]{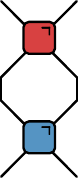}}}=\,\,\vcenter{\hbox{\includegraphics[width=0.035\textwidth]{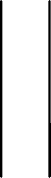}}}\,\,,
\end{align}
with a simple line representing the identity matrix on a qudit.

It is often convenient to work in the so-called \emph{folded representation}, where instead of the unitary gates the superoperator $U\otimes U^\ast$ is considered. Graphically this superoperator is represented as
\begin{align}
    \vcenter{\hbox{\includegraphics[width=0.061\textwidth]{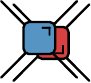}}} \,\,\equiv\,\,\vcenter{\hbox{\includegraphics[width=0.053\textwidth]{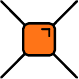}}}\,\,.
\end{align}
In this folded representation, the (normalized) $q \times q$ identity matrix is represented as 
\begin{align}
    \vcenter{\hbox{\includegraphics[width=0.014\textwidth]{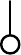}}} \,\equiv\, \frac{1}{\sqrt{q}}\,\, \vcenter{\hbox{\includegraphics[width=0.0166\textwidth]{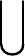}}}\,\,.
\end{align}
Unitarity is then represented as an eigenvalue equation,
\begin{align}
    \vcenter{\hbox{\includegraphics[width=0.063\textwidth]{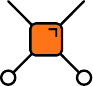}}}\,\, =\,\,  \vcenter{\hbox{\includegraphics[width=0.053\textwidth]{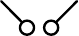}}}\,\,,\qquad
    \vcenter{\hbox{\includegraphics[width=0.063\textwidth]{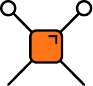}}}\,\, =\,\, \vcenter{\hbox{\includegraphics[width=0.053\textwidth]{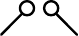}}}\,\,.
    \label{eq:unitarity_folded}
\end{align}

Throughout this work, we will focus on \emph{dual-unitary} interactions.
Dual-unitary (DU) gates are a subclass of unitary gates that additionally satisfy the following unitarity conditions in the spatial directions~\cite{Bertini2025}:
\begin{align}
    \vcenter{\hbox{\includegraphics[width=0.067\textwidth]{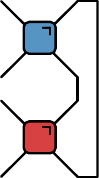}}}\,\,=\,\,\vcenter{\hbox{\includegraphics[width=0.022\textwidth]{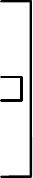}}}\,\,,\qquad\vcenter{\hbox{\includegraphics[width=0.067\textwidth]{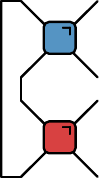}}}\,\,=\,\,\vcenter{\hbox{\includegraphics[width=0.022\textwidth]{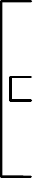}}}\,\,.
\end{align}
Working directly in the folded picture, these correspond to eigenvalue equations along the horizonal (spatial) direction:
\begin{align}
    \vcenter{\hbox{\includegraphics[width=0.058\textwidth]{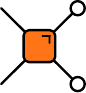}}}\,\, =\,\,  \vcenter{\hbox{\includegraphics[width=0.024\textwidth]{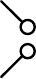}}}\,\,,\qquad
    \vcenter{\hbox{\includegraphics[width=0.058\textwidth]{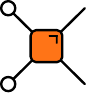}}}\,\, =\,\, \vcenter{\hbox{\includegraphics[width=0.024\textwidth]{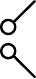}}}\,\,.
    \label{eq:dual_unitarity_folded}
\end{align}
The simplest example of a dual-unitary gate is given by the SWAP gate, which is defined for any dimension of the local Hilbert space and acts as $\textrm{SWAP}\ket{ab} = \ket{ba}$. The SWAP gate is graphically represented as
\begin{align}\label{eq:SWAP_crossing}
    \textrm{SWAP} = \,\,\vcenter{\hbox{\includegraphics[width=0.053\textwidth]{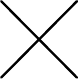}}}\,\,.
\end{align}
This gate gives rise to non-interacting dynamics. More general dual-unitary gates can be systematically realized and generically lead to interacting dynamics~\cite{Bertini2019,Claeys2021,prosen2021many,rather_construction_2022,Borsi2022,Bertini2025}.

In order to realize (discrete-time) dynamics for the full lattice of qudits, unitary gates can be applied to pairs of qubits in a regular or a random pattern, with the number of gates fixing the number of discrete time steps. 
We refer to the full unitary operator acting on the chain of qudits as the \emph{unitary circuit} $\mathcal{U}(t)$. By combining the properties of unitarity and locality, unitary circuits constitute minimal models for many questions of interest in many-body quantum dynamics~\cite{Fisher2023}. Typically, \emph{brickwork circuits} are studied in which the gates are applied in a regular brickwork geometry on pairs of neighboring qubits, as illustrated in Fig.~\ref{fig:lattices}(a). Such a geometry arises naturally when approximating the continuous-time dynamics of a two-local Hamiltonian with a Suzuki-Trotter expansion~\cite{Trotter1959,Suzuki1976}.

Remarkably, when arranging DU gates in a brickwork geometry, the resulting circuit admits the exact analytical calculation of many dynamical quantities of interest, despite being generically non-integrable~\cite{Bertini2025}. 
Particularly striking are the maximal entanglement growth~\cite{Bertini2019a} and the maximal butterfly velocity~\cite{Claeys2020}, both indicative of a maximal spreading of quantum information.
It was additionally proven that brickwork DU circuits conform to the predictions of random matrix theory for the spectral form factor, a common indicator of quantum chaos~\cite{Bertini2018}. This exact solvability is often traced back to the space-time duality that the DU conditions endow the circuit with. However, other types of exactly solvable non-integrable models have recently been found that explicitly break this space-time duality~\cite{Jonay2021,Yu2024,Sommers2024,Rampp2025,Liu2025,Breach2025}. It is hence desirable to have a different framework incorporating a wider class of exactly solvable models. Furthermore, in contrast to the theory of integrable models, which can be systematically constructed starting from solutions to the Yang-Baxter equation~\cite{Baxter2016,Eckle2019}, there is a lack of an algebraic framework for exactly solvable non-integrable systems.

\subsection{Entanglement membrane theory}
\label{subsec:EMT}

Throughout this work, we will focus on exact solvability through the lens of entanglement membrane theory.
When quenching from states with low entanglement, many-body quantum systems typically generate entanglement between distant regions. Investigations in recent years have shown that in non-integrable systems the leading contributions to entanglement can be understood as a hydrodynamic quantity, with the associated conservation law being the conservation of information provided by unitarity~\cite{Nahum2017,Nahum2018,Keyserlingk2018}. The hydrodynamic theory describing its dynamics is known as \emph{entanglement membrane theory}~\cite{Nahum2017,Jonay2018,Mezei2018,Zhou2019,Zhou2020,Mezei2020,Agon2021,Rampp2024,Sommers2024}. Its use is not restricted to the entanglement of states, but it also determines the macroscopic dynamics of local operator entanglement~\cite{Jonay2018} and out-of-time-order correlation functions~\cite{Zhou2020}.

Entanglement membrane theory is the generalization of the idea of a minimal cut through a tensor network~\cite{Hayden2016} that is applicable to Poissonian random circuits~\cite{Nahum2017}. In this theory, the bipartite entanglement entropy of a region $A$ is given by the free energy of the minimal membrane pinned to the boundaries of $A$. The line tension per unit time $\mathcal{E}(v)$, known as the \emph{entanglement line tension} (ELT), determines the free energy cost of a membrane along the cut in spacetime $x/t=v$. It contains all the microscopic information about the underlying model relevant for the macroscopic entanglement dynamics. The entanglement entropy of $A$ can be expressed as
\begin{equation}
    S_{A}(t) = s_{\textrm{eq}} \min_{\{v(t')\}} \left[ \int_0^t \mathrm{d}t' \mathcal{E}(v(t'))\right],
\end{equation}
where $s_{\textrm{eq}}$ is the equilibrium entropy density. For an infinite half chain and a translationally invariant initial state and circuit, this is always minimized by a vertical membrane, leading to
\begin{equation}
    S_{A}(t) = s_{\textrm{eq}}\,\mathcal{E}(0)t.
\end{equation}
The quantity
\begin{equation}
    v_E \equiv \mathcal{E}(0),
\end{equation}
is known as the entanglement velocity and sets the rate at which subsystem entanglement grows in time~\cite{Nahum2017}.
The ELT can be interpreted as the amount of information flowing across the membrane~\cite{Sommers2024}. It has to be a convex function of $v$ in order for the description to be consistent~\cite{Jonay2018}. However, a physical interpretation of this convexity is not known.

Computing the ELT in a given model is generally exponentially hard. However, brickwork DU circuits and certain generalizations~\cite{Zhou2020,Rampp2024,Sommers2024,Foligno2024} enable the exact computation of this quantity. The ELT can be extracted from the operator entanglement of the time-evolution operator. We consider a partition of the input and output legs of the time-evolution operator of a chain of $L$ sites in such a way that the entanglement cuts are connected by a ray of velocity $v$, i.e.
\begin{align}
    \mathcal{U}(t) = \,\vcenter{\hbox{\includegraphics[height = .48\columnwidth]{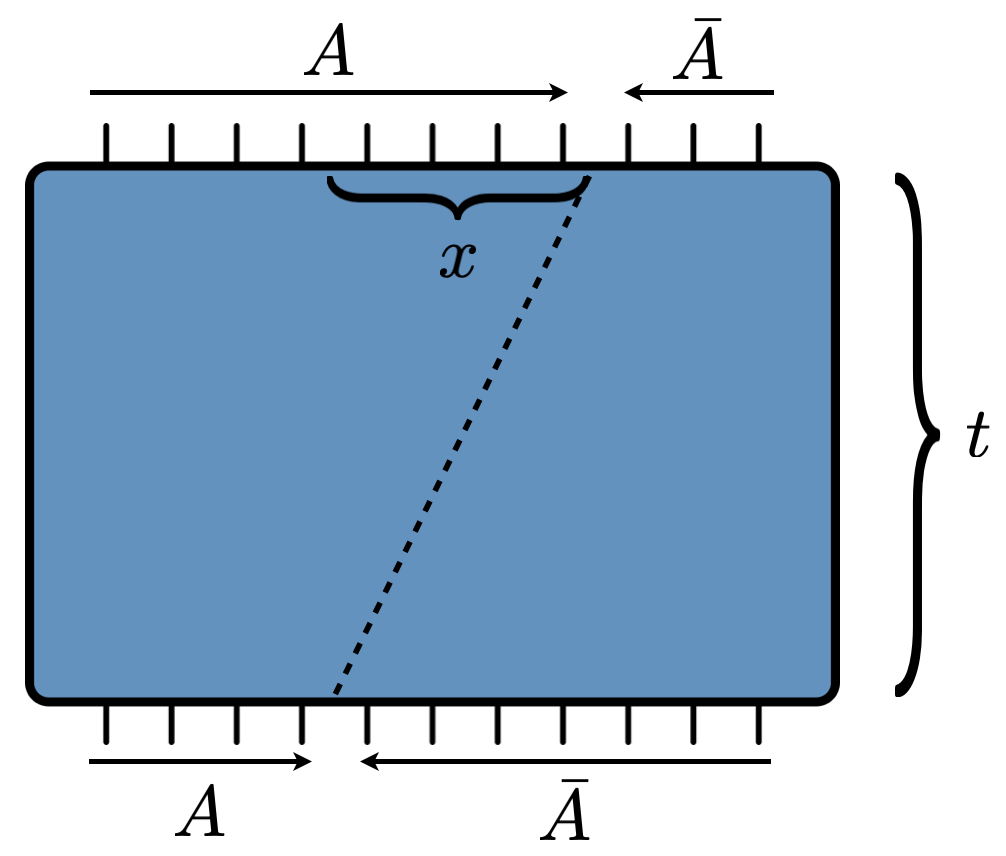}}}.
\end{align}
where we have made abstraction of the precise choice of circuit.
Performing an operator-to-state mapping of the unitary evolution operator
\begin{align}
    \mathcal{U}(t) &= \sum_{\substack{a_1,\dots,a_L\\ b_1,\dots,b_L}} U(t)^{a_1,\dots,a_L}_{b_1,\dots,b_L} \ket{a_1,\dots,a_L}\bra{b_1,\dots,b_L}
\end{align}
to
\begin{align}  
    \ket{\mathcal{U}(t)} \equiv {q^{-\frac{L}{2}}}\!\!\sum_{\substack{a_1,\dots,a_L\\ b_1,\dots,b_L}} & U(t)^{a_1,\dots,a_L}_{b_1,\dots,b_L} \nonumber\\
    &\times \ket{a_1,\dots,a_L}\otimes \ket{b_1,\dots,b_L},
\end{align}
where we have fixed the computational basis, enables to define the R\'{e}nyi-$\alpha$ operator entanglement as the R\'{e}nyi-$\alpha$ entanglement entropy of the state $\ket{\mathcal{U}(t)}$
\begin{equation}\label{eq:Renyi_a}
    S_{\alpha} (x,t) \equiv \frac{1}{1-\alpha} \log \tr\left[ \left(\operatorname{tr}_A \ket{\mathcal{U}(t)}\bra{\mathcal{U}(t)}\right)^\alpha \right].
\end{equation}
In the scaling limit $x,t\rightarrow\infty,\,x/t=v$, the leading contribution of the entanglement entropy is determined by the ELT as
\begin{equation}
    S_{\alpha} (x,t) \approx s_{\mathrm{eq}}\, \mathcal{E}_{\alpha}(v)t. \label{eq:ELT_def}
\end{equation}
Note that the ELT depends on the R\'{e}nyi index in general.

Returning to unitary circuits expressed in terms of local two-qudit gates, it is convenient to introduce a generalization of the folded gates to higher numbers of replicas, 
\begin{align}
\vcenter{\hbox{\includegraphics[width=0.053\textwidth]{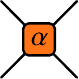}}}\,= (U \otimes U^*)^{\otimes\alpha}, \label{fig:folded_gate}
\end{align}
along with the permutation states
\begin{align}
\vcenter{\hbox{\includegraphics[width=0.014\textwidth]{figs/circle.pdf}}} \,= q^{-\frac{\alpha}{2}}\, \,\overbrace{\vcenter{\hbox{\includegraphics[width=0.1\textwidth]{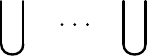}}}}^{2\alpha}\,, \qquad \vcenter{\hbox{\includegraphics[width=0.014\textwidth]{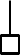}}}\, = q^{-\frac{\alpha}{2}}\, \,\overbrace{\vcenter{\hbox{\includegraphics[width=0.1\textwidth]{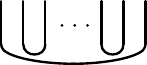}}}}^{2\alpha}\,. \,\,
\end{align}
These are normalized and their overlap can be evaluated as
\begin{align}
    \vcenter{\hbox{\includegraphics[width=0.014\textwidth]{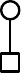}}} = q^{-\alpha}\,\,\vcenter{\hbox{\includegraphics[width=0.1\textwidth]{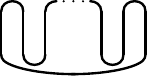}}}\,\, = \left(\frac{1}{q}\right)^{\alpha-1}\,.
\end{align}
To represent the operator entanglement as a tensor network, we first express it as
\begin{equation}
     S_{\alpha} (x,t) \equiv \frac{1}{1-\alpha} \log  Z_\alpha(m,n),
\end{equation} 
with $Z_\alpha$ defined as
\begin{align}
    Z_\alpha(m,n) \equiv \tr\left[ \left(\operatorname{tr}_A \ket{\mathcal{U}(t)}\bra{\mathcal{U}(t)}\right)^\alpha \right]. \label{eq:z_alpha_def}
\end{align}
Graphically, $Z_\alpha$ is represented in terms of the global unitary $\mathcal{U}(t)$ as
\begin{align}
    Z_\alpha = \,\vcenter{\hbox{\includegraphics[height = .48\columnwidth]{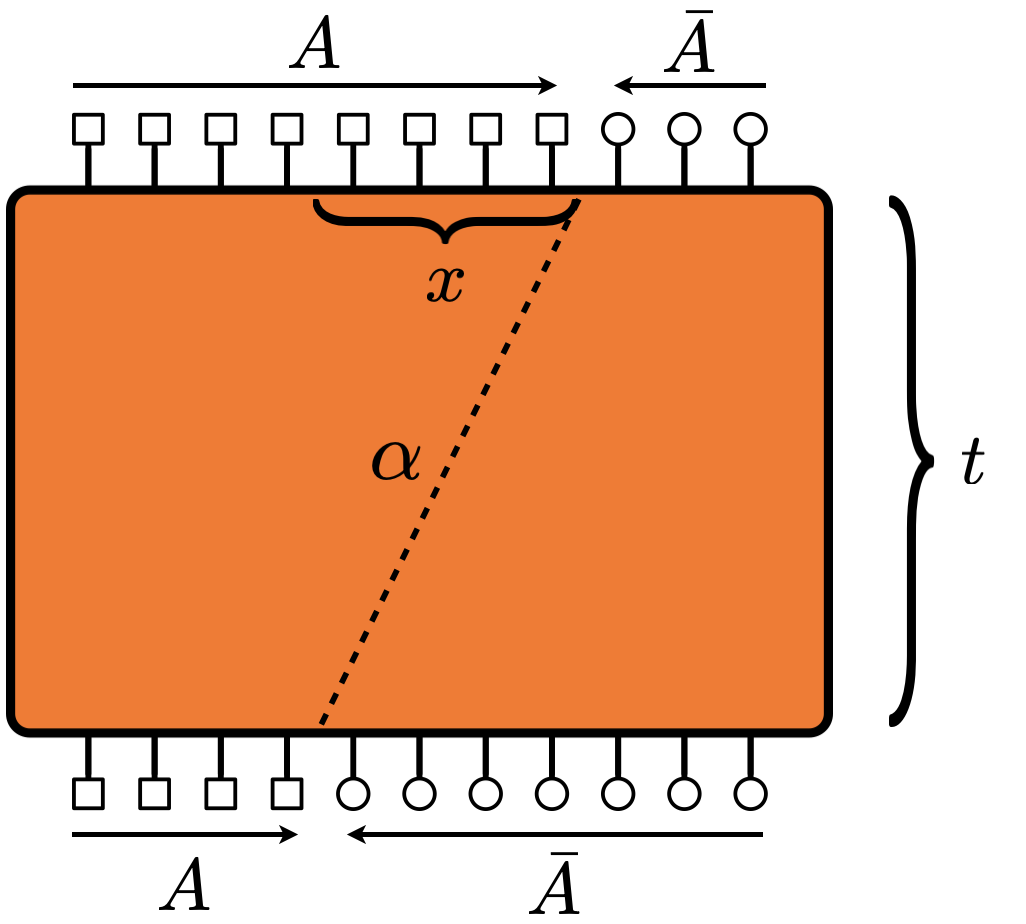}}}. \label{eq:z_alpha}
\end{align}
Using the brickwork structure of the dynamics, unitarity enables us to remove gates outside the intersection of light cones emanating from the endpoints of the entanglement cut and simplify Eq.~\eqref{eq:z_alpha} to
\begin{align}
    Z_\alpha = \,\vcenter{\hbox{\includegraphics[height = .34\textwidth]{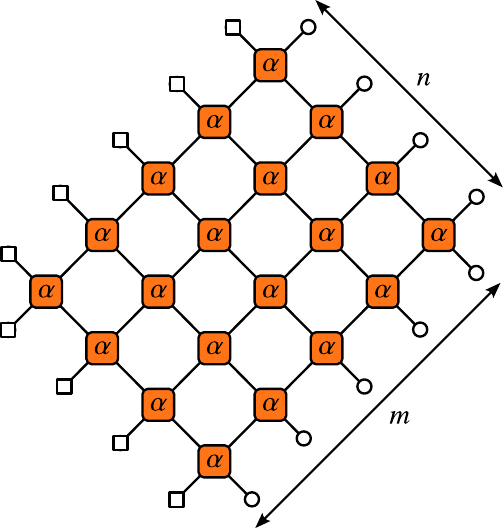}}}\,. \label{eq:z_alpha_reduced}
\end{align}
Here, the size of the tensor network is set by the coordinates of the entanglement cut as
\begin{equation}
    n = \frac{t-x -(x \,\mathrm{mod}\, 2)}{2}, \quad m = \frac{t+x -(x\, \mathrm{mod}\, 2)}{2}.
\end{equation}
The different boundary conditions encode the different ordering of taking the power $\alpha$ and taking the trace in $A$ and $\bar{A}$. 

In dual-unitary circuits, dual-unitarity can be used to fully contract this tensor network~\cite{Zhou2020}. The folded dual-unitary gates~\eqref{fig:folded_gate} again satisfy eigenvalue equations along the horizontal and vertical directions,
\begin{align}
    \label{eq:unitarity_folded_circ}
    \vcenter{\hbox{\includegraphics[width=0.063\textwidth]{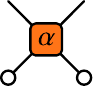}}}\,\, &=&\,\,  \vcenter{\hbox{\includegraphics[width=0.053\textwidth]{figs/folded_eig_2}}}\,\,,\qquad
    \vcenter{\hbox{\includegraphics[width=0.063\textwidth]{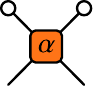}}}\,\, &=&\,\, \vcenter{\hbox{\includegraphics[width=0.053\textwidth]{figs/folded_eig_4}}}\,\,, \\
    \vcenter{\hbox{\includegraphics[width=0.058\textwidth]{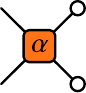}}}\,\, &=&\,\,  \vcenter{\hbox{\includegraphics[width=0.024\textwidth]{figs/folded_eig_6}}}\,\,,\qquad
    \vcenter{\hbox{\includegraphics[width=0.058\textwidth]{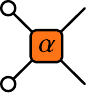}}}\,\, &=&\,\, \vcenter{\hbox{\includegraphics[width=0.024\textwidth]{figs/folded_eig_8}}}\,\,,
    \label{eq:dualunitarity_folded_circ}
\end{align}
as well as the equivalent conditions
\begin{align}
    \label{eq:unitarity_folded_sq}
    \vcenter{\hbox{\includegraphics[width=0.063\textwidth]{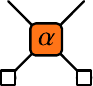}}}\,\, &=&\,\,  \vcenter{\hbox{\includegraphics[width=0.053\textwidth]{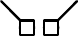}}}\,\,,\qquad
    \vcenter{\hbox{\includegraphics[width=0.063\textwidth]{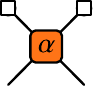}}}\,\, &=&\,\, \vcenter{\hbox{\includegraphics[width=0.053\textwidth]{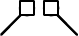}}}\,\,, \\
    \vcenter{\hbox{\includegraphics[width=0.058\textwidth]{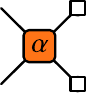}}}\,\, &=&\,\,  \vcenter{\hbox{\includegraphics[width=0.024\textwidth]{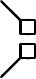}}}\,\,,\qquad
    \vcenter{\hbox{\includegraphics[width=0.058\textwidth]{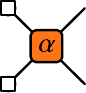}}}\,\, &=&\,\, \vcenter{\hbox{\includegraphics[width=0.024\textwidth]{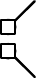}}}\,\,.
    \label{eq:dualunitarity_folded_sq}
\end{align}
Through the repeated application of Eq.~\eqref{eq:dualunitarity_folded_circ} from the right and Eq.~\eqref{eq:dualunitarity_folded_sq} from the left, the tensor network~\eqref{eq:z_alpha} can be fully contracted to return
\begin{equation}
    Z_\alpha(m,n) = \left(\,\vcenter{\hbox{\includegraphics[width=0.038\textwidth]{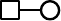}}}\,\right)^{m+n}  = \left(\frac{1}{q}\right)^{(\alpha-1)(m+n)} \!\!\!= q^{(1-\alpha)t}. \label{eq:Z_du}
\end{equation}
From this result it directly follows that the ELT is flat, $ \mathcal{E}_{\alpha}(v) = 1$, for all $v$ and $\alpha$, which in turn directly implies the maximal entanglement growth and maximal butterfly velocity of dual-unitary dynamics~\cite{Zhou2020}.
Remarkably, similar manipulations can be performed to evaluate the ELT of circuits of dual-unitary gates with a more involved lattice geometry.


\section{Framework: completely reducible quantum circuits}
\label{sec:framework}

In this section, we introduce the class of models that we are investigating in this paper: spacetime lattices of dual unitary gates.
We discuss different notions of solvability and term the strongest notion as \emph{complete reducibility}, highlighting the role of a structural robustness to dual-unitary perturbations as a crucial feature. 
Various examples of completely reducible circuits are presented throughout and used to illustrate the general phenomenology of such circuits.

\begin{figure*}[t]
    \centering
    \includegraphics[width = 0.55\textwidth]{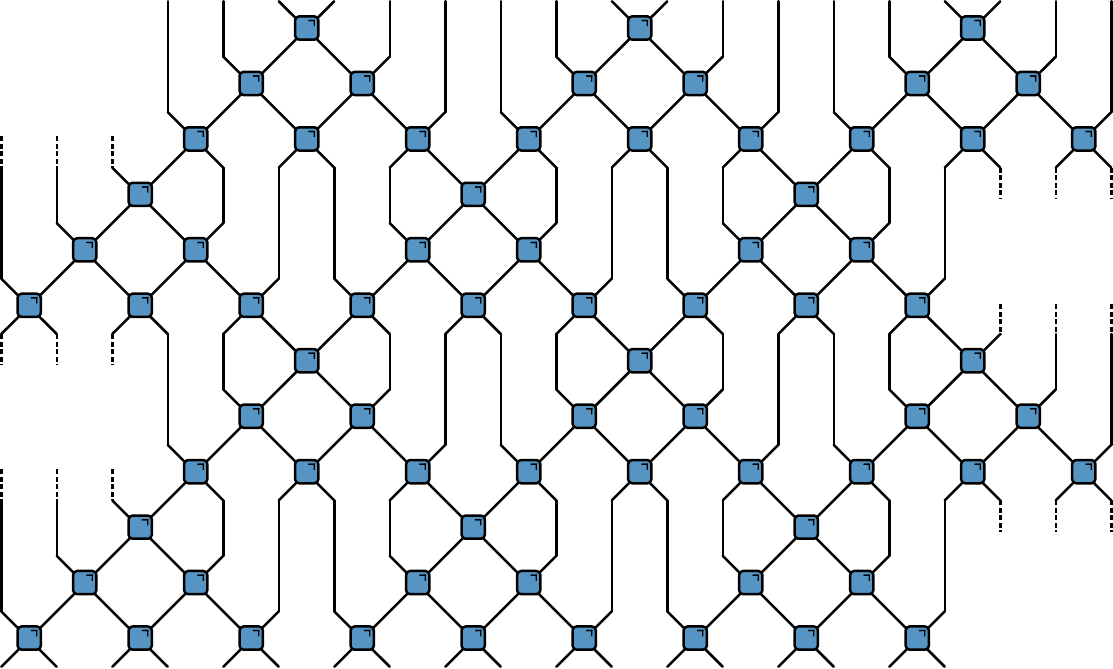}
    \caption{Illustration of the spacetime lattice generated from a base gate~\eqref{eq:basegate}. The horizontal and vertical directions correspond to space and time respectively. While the local interactions are dual-unitary, the full circuit breaks space-time duality.}
    \label{fig:basegate}
\end{figure*}

\subsection{Spacetime lattices of dual-unitary gates}

To investigate non-equilibrium phenomena with unitary circuits, typically a brickwork geometry is considered. 
However, for these inherently driven systems the particular choice of gate sequence -- and thus also the spacetime geometry -- can be seen as part of the driving protocol.
We fix the spatial lattice to be a chain and restrict the two-site unitary gates to be dual unitary. The sequence of bonds on which the gates are applied defines the spacetime lattice of the circuit. The much-investigated brickwork dual-unitary circuits are recovered by alternately applying DU gates to even bonds and odd bonds. For this choice the DU gates are placed on the vertices of a square lattice. However, other regular lattices are possible as well, see Fig.~\ref{fig:lattices} for examples.
In this work we consider regular lattices constructed in the following way, making sure that the evolution is unitary: we consider a circuit of DU gates acting on $2N$ qudits of dimension $d$ and take this to define a two-site gate for composite qudits of dimension $q=d^N$. By arranging the so-constructed gate in a brickwork fashion we generate the spacetime lattice circuit on the original qudits. We refer to these as the \emph{$d$-qudits} and to the composite qudits as \emph{$q$-qudits}. We call the gate acting on the $q$-qudits the \emph{base gate}, acting as the unit cell of the lattice. This construction is illustrated in Fig.~\ref{fig:basegate} for the following choice of base gate:
\begin{align}\label{eq:basegate}
\vcenter{\hbox{\includegraphics[width=0.24\textwidth]{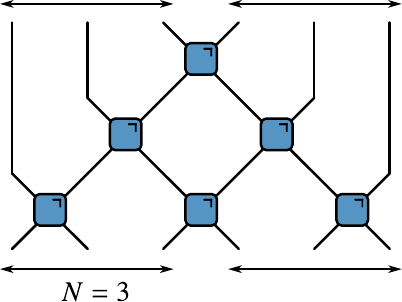}}} 
\end{align}
We will return to this lattice in the next section.
Different unit cells can lead to the same bulk spacetime lattice, and we discuss the ways in which different base gates can lead to equivalent spacetime lattices in App.~\ref{app:more_examples}.

\subsection{Complete reducibility}

What do we mean by exactly solvable many-body quantum dynamics? Exact solvability is always understood with respect to a given physical quantity, such as a two-point dynamical correlation function or the bipartite entanglement entropy after a quench from a particular state. If this quantity can be evaluated in the scaling limit with a cost at most polynomial in $t$, then we say that it can be solved exactly in this circuit. 
Dual-unitary circuits initially gained attention since their dynamical correlation functions could be solved exactly in this manner.
However, other properties can be considered, which provide a stricter notion of solvability.
In this work, we focus on the entanglement line tension (ELT) as a measure of solvability. Crucially, the ELT is independent of any observable, and acts as a probe of intrinsic properties of the dynamics. 
On the level of tensor network diagrams, the boundary contractions are given only by permutations of replicas. When computing the ELT in DU brickwork circuits, the diagram can be reduced to overlaps of permutation states only, as apparent from Eq.~\eqref{eq:Z_du}. The ELT can hence be expressed as a power of the local Hilbert space dimension, completely independent of the particular choice of dual-unitary gates or their degree of ergodicity. \emph{All} brickwork DU circuits yield the same result.

This surprising robustness is not restricted to brickwork DU circuits. It turns out to depend crucially on the spacetime lattice defining the circuit. We say that a quantity is \emph{completely reducible} in a circuit, if the tensor network can be contracted until it consists of overlaps of permutation states only.
In other words, any dependence on the constituting gates is lost and the ELT equals a power of the local Hilbert space dimension $q$, with the exponent depending only on the coordinates $m,n$ and the replica number $\alpha$. 
Complete reducibility is thus a stronger notion than exact solvability, as highlighted by its insensitivity to microscopic details. This work is primarily concerned with the study of completely reducible circuits.

\subsection{Examples}

\begin{figure*}[t]
    \centering
    \includegraphics[width = .9\textwidth]{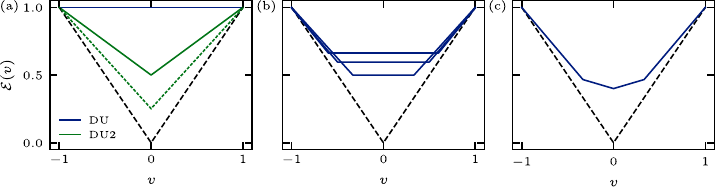}
    \caption{Entanglement line tension (ELT) $\mathcal{E}(v)$ of classes of completely reducible circuits. (a) DU and DU2 circuits [Eqs.~\eqref{eq:du_elt}, \eqref{eq:du2_elt}, and~\eqref{eq:du2_elt_2}]. 
    (b) 4-pyramid lattice [Eq.~\eqref{eq:4pyr_elt}] and larger unit cell generalizations [Eq.~\eqref{eq:family_1}].
    (c) Lattice with five directions of information flow [Eq.~\eqref{eq:5ray_elt}].}
    \label{fig:elt_sketches}
\end{figure*}

Let us give some examples of spacetime lattices giving rise to an ELT that is completely reducible, exactly solvable, and not solvable, respectively. 
These are organized by the number of rays in spacetime along which information flows, as discussed further in the text.
These examples include known and novel dual-unitary and hierarchical dual-unitary constructions, as well as previously unidentified classes of circuits giving rise to a phenomenology different from either.
The ELT $\mathcal{E}(v)$ is illustrated in Fig.~\ref{fig:elt_sketches} for some representative examples.
In all cases, the calculation of the ELT is straightforward and can be performed essentially algorithmically through the repeated applications of Eqs.~\eqref{eq:dualunitarity_folded_circ} and \eqref{eq:dualunitarity_folded_sq}.

\textbf{Two rays.}
As mentioned above, brickwork DU circuits are the prototypical example of models with a completely reducible ELT~\cite{Zhou2020}, and have the characteristic property that all information propagates along the two lightrays $x = \pm t$. 
The base gate corresponds to a single dual-unitary gate, i.e. $N=1$,
\begin{align}\label{eq:du_gate}
    \vcenter{\hbox{\includegraphics[width=0.06\textwidth]{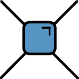}}}\,.
\end{align}

The ELT can be derived from Eq.~\eqref{eq:Z_du}. Note that the appearance of $\alpha-1$ in the exponent implies that the spectrum of ELTs is independent of the R\'{e}nyi index. This turns out to be a general feature of completely reducible circuits, such that we drop the index $\alpha$ in the following. The ELT follows as [see also Fig.~\ref{fig:elt_sketches}(a)]
\begin{align}
    \mathcal{E}(v) = \begin{cases}
        1, & \quad \abs{v}\leq1,\\
        \abs{v}, & \quad \abs{v}>1.
    \end{cases} \label{eq:du_elt}
\end{align}
In the interval $-1<v<1$, the ELT attains the maximal possible value of $\mathcal{E}(v)=1$, expressing the maximal capacity of generating entanglement characterizing dual-unitary gates and their maximal entanglement velocity $v_E=1$ (see also Ref.~\cite{zhou_maximal_2022}). This is particular to brickwork DU circuits. Completely reducible circuits defined on different spacetime lattices generate less entanglement with an entanglement velocity $v_E<1$.

\textbf{Three rays.}
An example for this submaximal entanglement velocity is given by the Kagome lattice circuit~\cite{Sommers2023,Rampp2025}, as illustrated in Fig.~\ref{fig:lattices}(b). 
This circuit is generated by the $N=2$ base gate
\begin{align}
    \vcenter{\hbox{\includegraphics[width=0.14\textwidth]{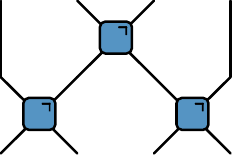}}}\,\, \label{eq:gate_n2_du2}.
\end{align}
Entanglement dynamics in a Kagome lattice of dual-unitary gates has been previously studied in Ref.~\cite{Rampp2025}, and the entanglement line tension in general DU2 circuits was characterized in Refs.~\cite{Foligno2024,Rampp2024}.
In this model the entanglement velocity is reduced to $v_E=1/2$, and the ELT follows as [see Fig.~\ref{fig:elt_sketches}(a)]
\begin{align}
    \mathcal{E}(v) = \begin{cases}
        ({1+\abs{v}})/{2}, & \quad \abs{v}\leq1,\\
        \abs{v}, & \quad \abs{v}>1.
    \end{cases} \label{eq:du2_elt}
\end{align}
The ELT is piecewise linear, with kinks at $v=0$ and $|v|=1$. 
These results were previously understood by observing that the gate~\eqref{eq:gate_n2_du2} satisfies the DU2 condition, a solvability condition involving two copies of a folded gate that provides one particular way of generalizing dual unitarity~\cite{Yu2024}.
The form of the ELT reflects that DU2 circuits can transport information along three lines in spacetime: the light rays $x = \pm t$ and the static worldine $x=0$.

The nested Kagome lattice provides another example of a completely reducible lattice, as anticipated in Ref.~\cite{Rampp2025}. This lattice is generated by the $N=4$ base gate
\begin{align}
    \vcenter{\hbox{\includegraphics[width=0.31\textwidth]{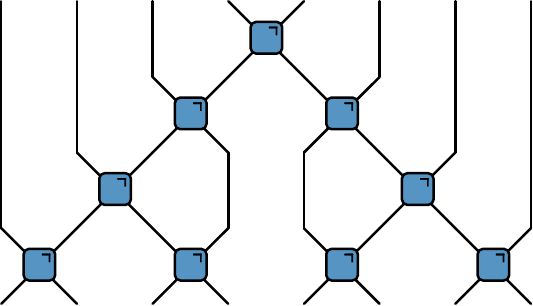}}}\,\,. \label{eq:gate_n4_nested}
\end{align}
This gate satisfies the DU2 condition with $v_E=1/4$, again leading to information dynamics along three lines in spacetime, with the corresponding ELT given by
\begin{align}
    \mathcal{E}(v) = \begin{cases}
        ({1+3\abs{v}})/{4}, & \quad \abs{v}\leq1,\\
        \abs{v}, & \quad \abs{v}>1.
    \end{cases} \label{eq:du2_elt_2}
\end{align}
Alternative constructions for $N=3$ are presented in App.~\ref{app:more_examples}.

\textbf{Four rays.}
Let us now discuss novel circuits constructed from spacetime lattices. We introduce the following circuit acting on eight qudits of dimension $d$ as a $N=4$ base gate
\begin{align}
    \vcenter{\hbox{\includegraphics[width=0.31\textwidth]{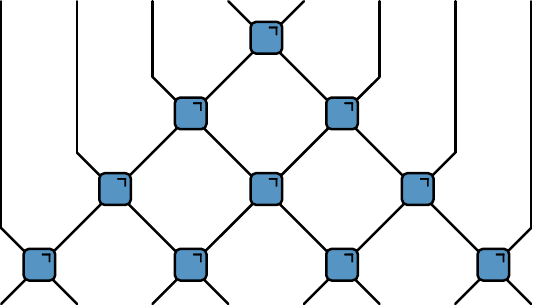}}}\,\, \label{eq:4pyr}.
\end{align}
Arranging this in a brickwork geometry yields the circuit depicted in Fig.~\ref{fig:lattices}(d).
We call this the \emph{4-pyramid lattice} because of the shape and size of the building block. Remarkably, this lattice gives rise to completely reducible dynamics even though it goes beyond previously known solvability conditions. 

To illustrate the complete reducibility, we observe that the diagram for the ELT along $v=0$ can be explicitly contracted, yielding
\begin{widetext}
\begin{equation}
    Z_{\alpha}(n,n) = \,\vcenter{\hbox{\includegraphics[width=0.6\textwidth]{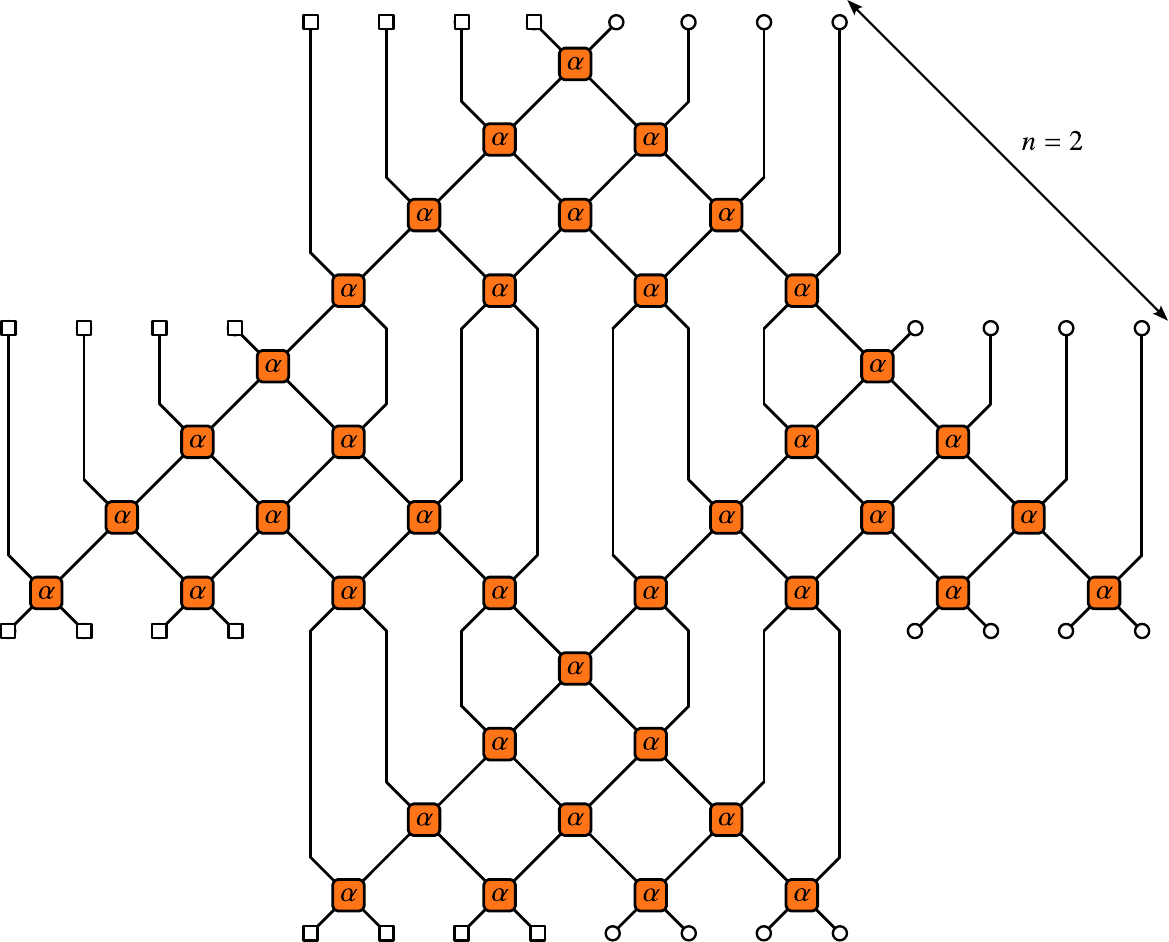}}}\, = \left(\frac{1}{d^{\alpha-1}}\right)^{4n} = \left(\frac{1}{q^{\alpha-1}}\right)^{n},
\end{equation}
\end{widetext}
here illustrated for $n=2$. 
The derivation consists of the repeated application of Eqs.~\eqref{eq:unitarity_folded_circ}, \eqref{eq:dualunitarity_folded_circ}, \eqref{eq:unitarity_folded_sq} and \eqref{eq:dualunitarity_folded_sq} where possible, which reduces this expression to a product of overlaps between permutation states.
For convenience, we choose to use the coordinates of the brickwork circuit defined on the $q$-qudits. The entanglement velocity follows from the asymptotic scaling of the above result, using that $n\sim t/2$,
\begin{equation}
    v_E^{(\alpha)} = \mathcal{E}_\alpha(v=0) = -\lim_{n\rightarrow\infty}\frac{\log Z_\alpha(n,n)}{2n(\alpha-1)\log q} = \frac{1}{2}.
\end{equation}
The entanglement velocity equals $v_E=1/2$, again independent of the R\'{e}nyi index. The full ELT reads [see also Fig.~\ref{fig:elt_sketches}(b)]
\begin{align}
    \mathcal{E}(v) = \begin{cases}
        {1}/{2}, & \quad \abs{v}\leq\frac{1}{3},\\
        ({1+3\abs{v}})/{4}, & \quad\frac{1}{3}<\abs{v}\leq 1,\\
        \abs{v}, & \quad\abs{v}>1.
    \end{cases} \label{eq:4pyr_elt}
\end{align}
This is the first explicit example of a circuit having a piecewise linear ELT with four kinks. The four kinks suggest that the information flow is along four directions in spacetime. This picture is supported by the observation that the correlation functions are non-vanishing only along four rays in spacetime, $\abs{v}={1}/{3}$ and $1$, coincident with the kinks in the ELT (see also App.~\ref{app:correlations}). 
This presence of four directions of information flow defies the intuition that directions of information flow should be related to symmetries of the spacetime lattice -- clearly a 2D lattice cannot have an eightfold rotational symmetry. Indeed, the 4-pyramid lattice does not have any apparent symmetries beyond a mirror symmetry. Rather, the information flow can be traced back to the directions in which particles propagate in the non-interacting limit. 
We will return to this point in Sec.~\ref{sec:information_flow}.
The fact that this feature survives the inclusion of integrability-breaking interactions is highly nontrivial, and goes beyond previous notions of solvable non-integrable dynamics.

Similar dynamics is exhibited by the following $N=4$ base gate:
\begin{align}
    \vcenter{\hbox{\includegraphics[width=0.3\textwidth]{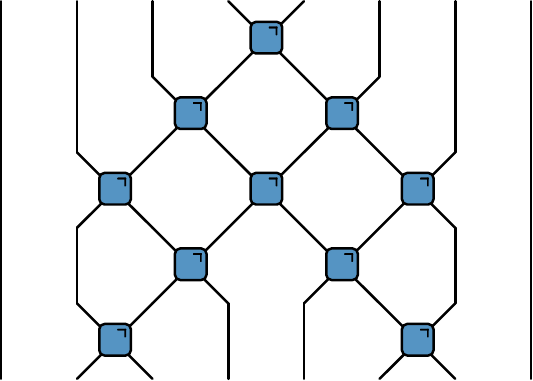}}}\,\,. \label{eq:gate_n4_rocket}
\end{align}
We call it the \emph{4-rocket} gate. It leads to completely reducible dynamics similar to the 4-pyramid gate, where information flows along the four directions $\abs{v}=1/3,1$. The resulting ELT is given by
\begin{align}
    \mathcal{E}(v) = \begin{cases}
        {1}/{2}, & \quad \abs{v}\leq\frac{1}{3},\\
        ({1+3\abs{v}})/{4}, & \quad\frac{1}{3}<\abs{v}\leq 1,\\
        \abs{v}, & \quad\abs{v}>1,
    \end{cases} \label{eq:4rock_elt}
\end{align}
identical to Eq.~\eqref{eq:4pyr_elt}. In App.~\ref{app:more_examples} we show that both gates are members of an infinite family of completely reducible circuits.

\textbf{Five rays.}
We now consider an example of a completely reducible gate with information flow along five distinct rays in spacetime. We note that $N=5$ is the smallest size such that this is possible, as follows from counting the multiplicities of worldlines in the swap circuit.
The $N=5$ base gate reads
\begin{align}
    \vcenter{\hbox{\includegraphics[width=0.34\textwidth]{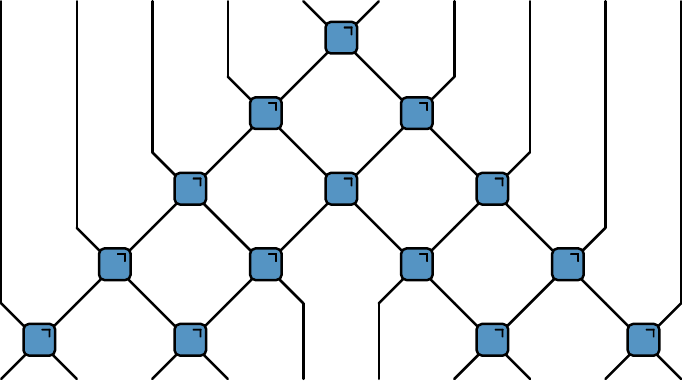}}}\,\, \label{eq:gate_n5_5ray},
\end{align}
and the ELT can be evaluated as [see also Fig.~\ref{fig:elt_sketches}(d)]
\begin{align}
    \mathcal{E}(v) = \begin{cases}
        ({2+\abs{v}})/{5}, & \quad \abs{v}\leq\frac{1}{3},\\
        ({1+4\abs{v}})/{5}, & \quad \frac{1}{3}<\abs{v}\leq1,\\
        \abs{v}, & \quad \abs{v}>1.
    \end{cases} \label{eq:5ray_elt}
\end{align}
The ELT shows that information flows along the five discrete directions $\abs{v}=0,{1}/{3},1$. The corresponding circuit presents the first example of a solvable model of quantum many-body dynamics with five rays of information flow, which again goes beyond the phenomenology predicted by any known solvability condition.

\textbf{Exact solvability vs. complete reducibility.}
We conclude this list of examples with some explicit constructions highlighting that exact solvability differs from being completely reducible.
We consider a circuit with $N=3$ base gate,
\begin{align}
    \vcenter{\hbox{\includegraphics[width=0.2\textwidth]{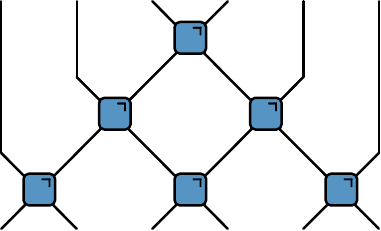}}}\,\, \label{eq:3pyr}.
\end{align}
We call the resulting circuit the \emph{3-pyramid} circuit. In this circuit, the ELT along $v=0$ cannot be reduced to overlaps of permutations. Writing down the tensor network~\eqref{eq:z_alpha} and applying unitarity and dual-unitarity where possible, this diagram can be simplified to (here illustrated for $n=4$):
\begin{widetext}
\begin{align}
    \vcenter{\hbox{\includegraphics[height=0.715\textwidth]{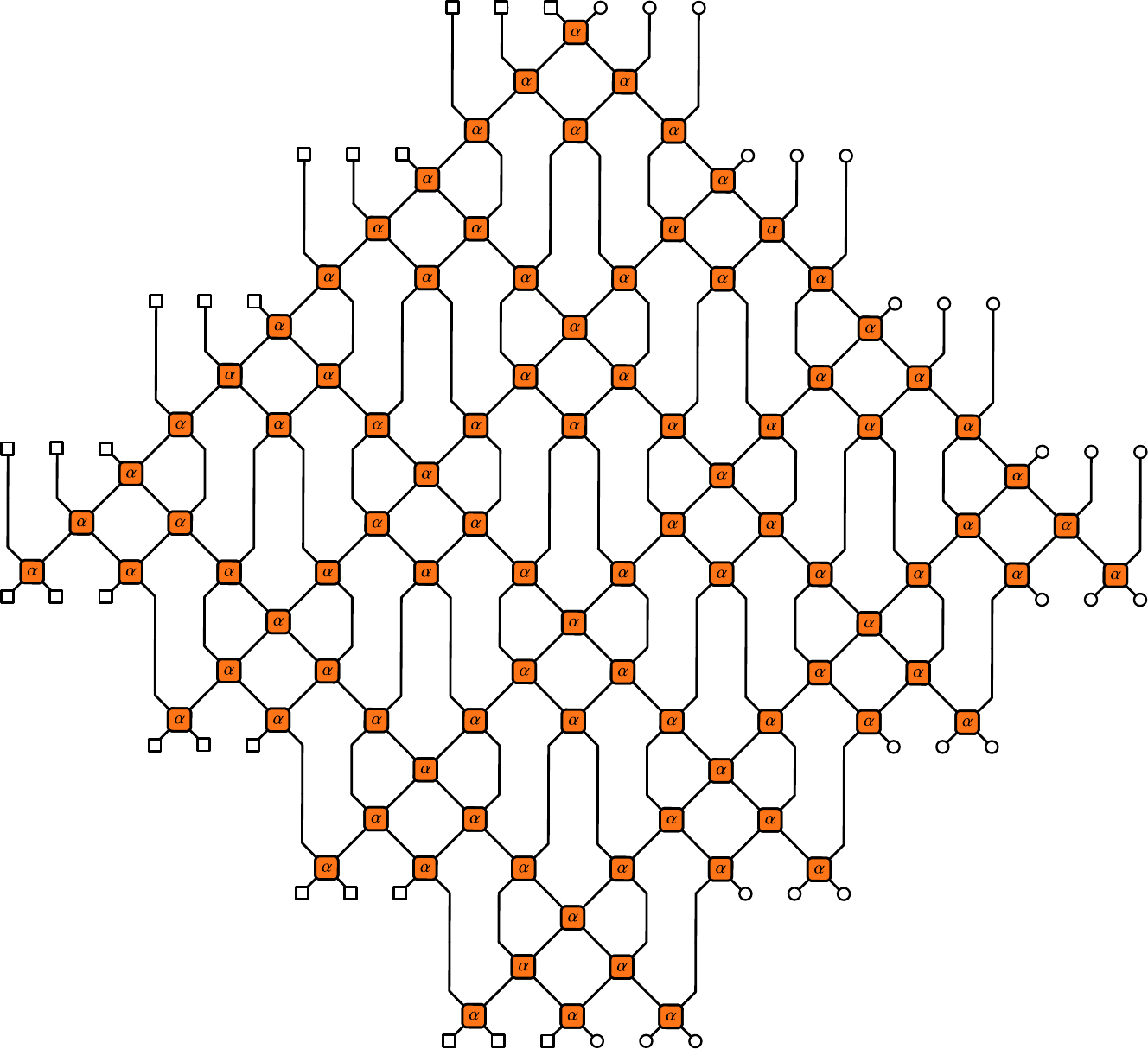}}} \,\,=\,\, \vcenter{\hbox{\includegraphics[height=0.65\textwidth]{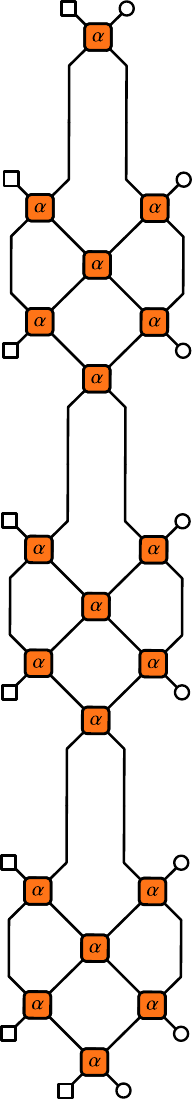}}}\,\,\nonumber.
\end{align}
\end{widetext}
While this diagram does not simplify further, it can be efficiently evaluated numerically through the repeated application of a low-dimensional quantum channel (the transfer matrix along the vertical direction), making it equivalent to a one-dimensional tensor network contraction. Therefore, the ELT along $v=0$ can be numerically computed exactly, even though it is not completely reducible. 

We note that the circuit is equivalent to a composite triunitary circuit~\cite{Jonay2021}. An alternative unit cell can be identified, corresponding to a three-site gate acting on three Hilbert spaces with different dimensions, by grouping together the two central $d$-qudits:
\begin{align}
 \bra{abc}U_{\rm tri}\ket{def} = \vcenter{\hbox{\includegraphics[width=0.13\textwidth]{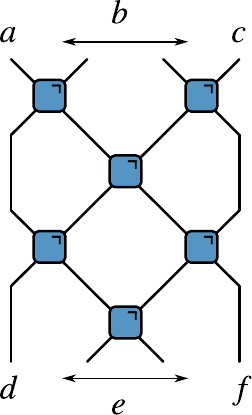}}}\,\, \label{eq:3pyr_threesite}.
\end{align}
It is direct to check that the gate $U_{\rm tri}$ satisfies the triunitary condition from Ref.~\cite{Jonay2021}.

Finally, let us present an example of a spacetime lattice of DU gates which is not solvable. We define a two-site gate acting on $q=d^2$-dimensional qudits through
\begin{align}
    \vcenter{\hbox{\includegraphics[width=0.12\textwidth]{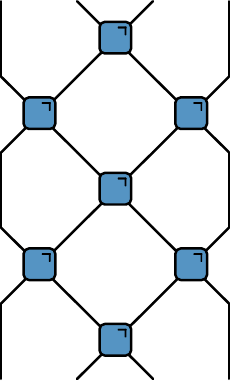}}}\,\, \label{eq:2loc},
\end{align}
which yields the circuit in Fig.~\ref{fig:lattices}(c).
The diagram corresponding to $Z_\alpha(n,n)$ follows as (here illustrated for $n=3$):
\begin{align}
    \vcenter{\hbox{\includegraphics[height=0.8\textwidth]{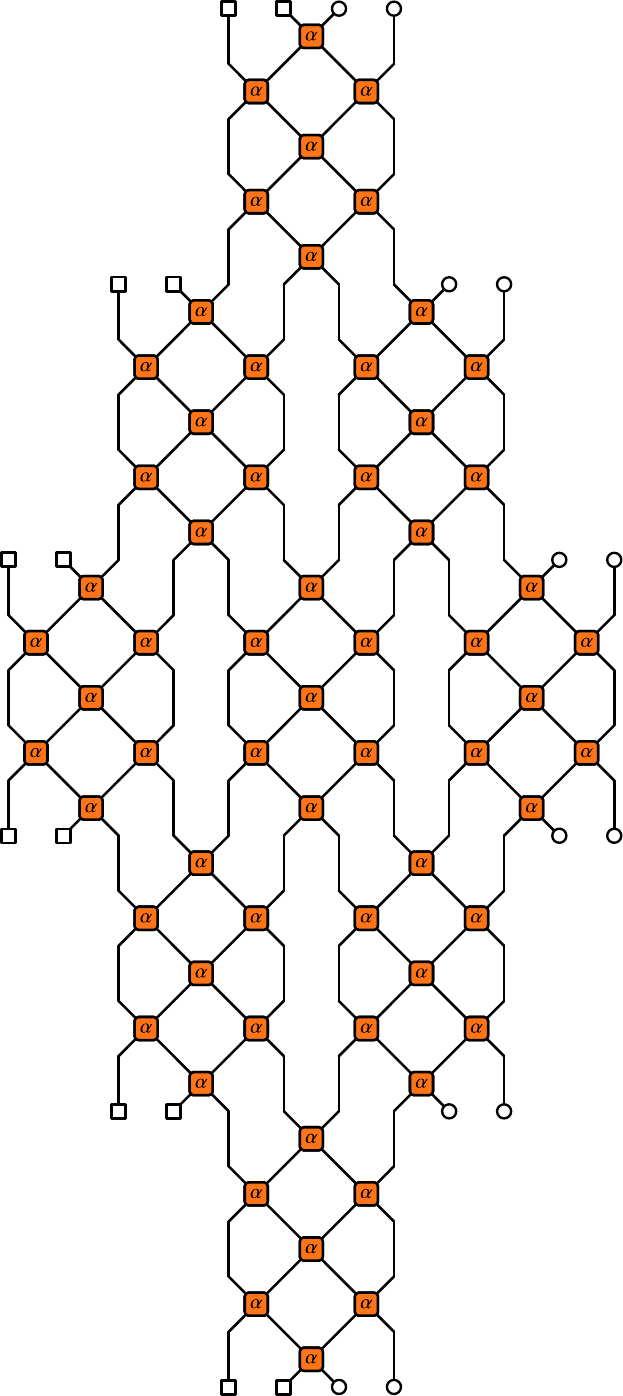}}}
\end{align}
This diagram cannot be reduced at all beyond the causal light-cone: there is no point on the boundary where (dual) unitarity can be used to remove a gate. The evaluation of this diagram and determination of the corresponding ELT is hence exponentially hard in $t$ for generic choices of DU gates.


\section{Contractability defects}
\label{sec:defects}

In this section, we ask the question: what makes a spacetime lattice support completely reducible dynamics? Complete reducibility, as we have defined it, implies an independence of the microscopic details of the gate. The results are identical for noninteracting SWAP gates and for interacting dual-unitary gates.
It is thus a geometric property of the spacetime lattice. 
In this section, we derive necessary conditions for a circuit to be completely reducible, which in turn lead to physical constraints on the information dynamics.

In order to derive simple conditions on the circuit, we can look at the invariance in the following manner. Replacing every DU gate on a given spacetime lattice with the SWAP gate, we obtain a circuit describing non-interacting particles. When tuning some of the gates away from the free point, the solvability is generically lost. However, if the diagram is completely reducible, locally turning on dual-unitary interactions does not matter and the solvability is preserved.
We can thus systematically identify conditions \emph{excluding} complete reducibility in the following manner: we consider a spacetime lattice with all gates being SWAPs, then we reintroduce a small number $N$ of generic DU gates (``defects''). If the diagram can no longer be contracted to an overlap of permutations, it cannot be completely reducible. By collecting all not completely reducible diagrams containing $N$ defects, we systematically obtain a set of sufficient conditions for the absence of completely reducibility.  
This approach has the advantage that the contractions in the SWAP circuit can be simply read off, for arbitrarily complicated geometries, as will be apparent in the following, and contractions involving a small number $N$ of generic DU gates can similarly be straightforwardly evaluated.
We then go on to use this concept to give a simple criterion excluding complete reducibility: the presence of crossing $v=0$ worldlines.

\subsection{Irreducible diagrams}

Let us begin by illustrating this procedure with an example. We consider the (not solvable) gate defined in Eq.~\eqref{eq:2loc} and consider $Z_\alpha(1,1)$. We here illustrate the case where either all gates equal the SWAP gate, indicating how this expression can be trivially evaluated, as well as the case where all gates except one equal the SWAP gate:
\begin{align}
  \vcenter{\hbox{\includegraphics[width=0.1\textwidth]{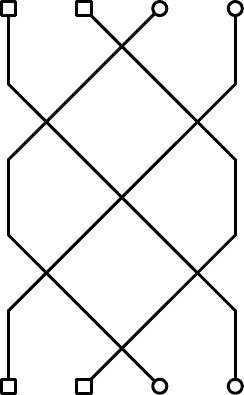}}}\, \, \to \,\,  \vcenter{\hbox{\includegraphics[width=0.1\textwidth]{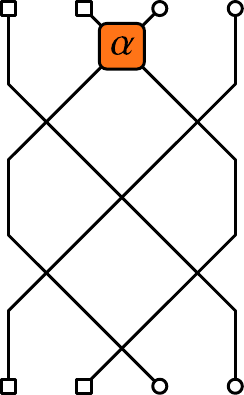}}}\,\,=\frac{1}{d^{2(\alpha-1)}}\,\,\vcenter{\hbox{\includegraphics[width=0.063\textwidth]{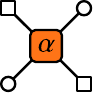}}}\,\,. \label{eq:one_defect_example}
\end{align}
The expression on the right-hand side cannot be further simplified, such that this circuit is not completely reducible. This observation immediately implies that the full lattice, with generic DU gates, is also not completely reducible.

There are in fact two positions in the diagram where an inserted defect spoils reducibility. The irreducible factor
\begin{align}
    \vcenter{\hbox{\includegraphics[width=0.063\textwidth]{figs/one_defect.pdf}}} \label{eq:one_body_irred}
\end{align}
is the only irreducible diagram that can be formed using a single gate. Note that we could form more irreducible diagrams, if we would allow worldlines to cross themselves or form closed loops in spacetime, e.g., 
\begin{align}
    \vcenter{\hbox{\includegraphics[width=0.09\textwidth]{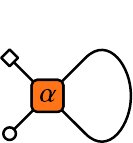}}}\,\, \qquad \vcenter{\hbox{\includegraphics[width=0.099\textwidth]{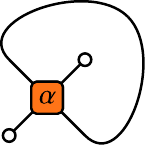}}}\,\,, \label{eq:forbidden_defects}
\end{align}
which is however forbidden by unitarity.

How can we identify the presence of such irreducible diagrams in general spacetime lattices? The above diagram can only appear after insertion of a defect if a worldline connecting a circle to a circle and a worldline connecting a square to a square cross. These crossings have an interpretation in terms of physical processes. A worldline connecting circles can be thought of as an operator starting and ending in $A$, while a worldline connecting squares can be thought of as an operator starting and ending in $\bar{A}$. The crossing of the worldlines can be thought of as an interaction between these operators, which destroys complete reducibility. Note that there are in total four kinds of worldlines. In addition to the previously discussed ones, worldlines can also connect a circle to a square and vice versa. This can be thought of as an operator moving from $A$ to $\bar{A}$ ($\bar{A}$ to $A$).

As a further example, we discuss the 3-pyramid circuit [Eq.~\eqref{eq:3pyr}]. Here, the presence of non-reducible crossings spoils complete reducibility but leaves solvability intact. We mark the positions where inserted defects yield the irreducible diagram \eqref{eq:one_body_irred} in the following:
\begin{align*}
    \vcenter{\hbox{\includegraphics[width=\columnwidth]{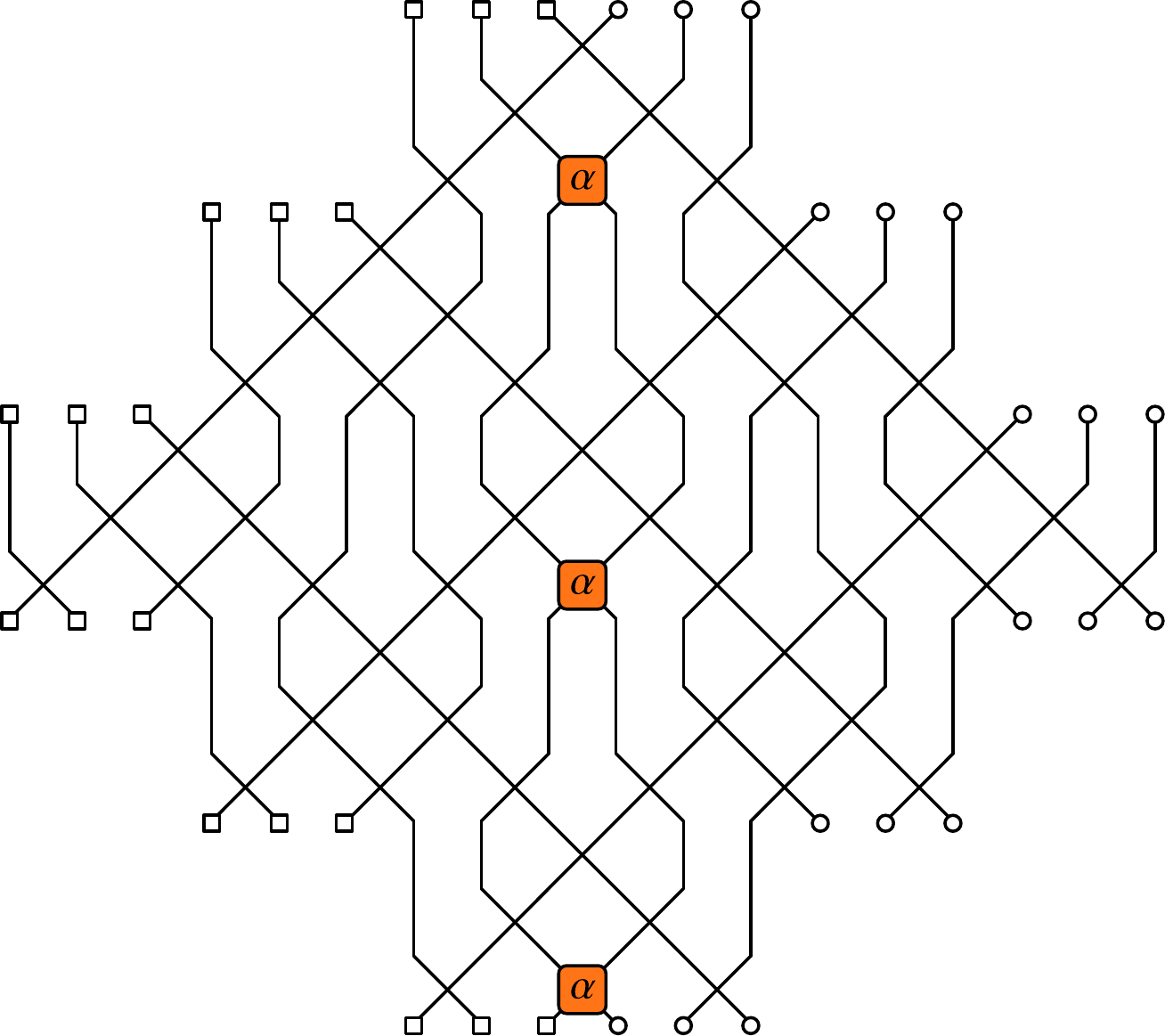}}}
\end{align*}
Choosing a generic DU gate as either of these three gates and fixing the other two as SWAP gates returns an irreducible diagram.
While the number of such possible defect insertions grows with $t$, they remain confined close to $x=0$. Therefore, contraction of a large part of the diagram is still possible, and the resulting simplified diagram can be efficiently evaluated numerically -- consistent with the previously observed solvability.

Even if the insertion of a single defect preserves reducibility of the diagram, this can change when multiple defects are inserted. We can now catalogue all irreducible diagrams involving two DU gates. Before proceeding, let us introduce a definition. We call a diagram $N$-body irreducible, if it is not reducible and if it becomes reducible upon replacing any of the gates by a SWAP. E.g., the one-gate diagram introduced above is $1$-body irreducible. The purpose of this definition is to exclude diagrams that would have been irreducible already when inserting fewer defects and thus to avoid generating redundant conditions.

For two defects, we have the following $2$-body irreducible diagrams
\begin{align}
    (\textrm{i})\quad\vcenter{\hbox{\includegraphics[width=0.064\textwidth]{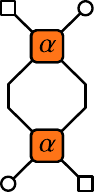}}}\,\,,\qquad (\textrm{ii})\quad\vcenter{\hbox{\includegraphics[width=0.115\textwidth]{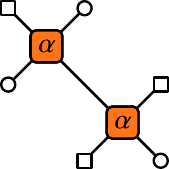}}}\,\,.  \label{eq:two_body_irred}
\end{align}
Note that the orientation of the gates is arbitrary, as is the orientation of the diagrams.

An example of a spacetime lattice that does not have the $1$-body irreducible diagram but has $2$-body irreducible ones is given by the following base gate
\begin{align}
    \vcenter{\hbox{\includegraphics[width=0.227\textwidth]{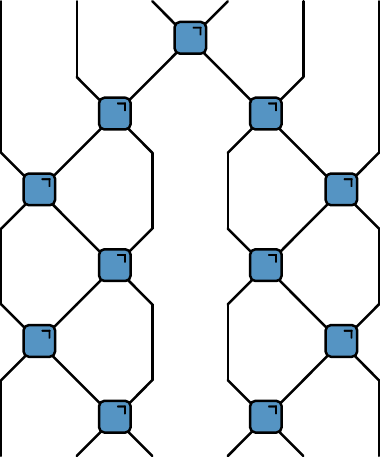}}}\,. \label{eq:gate_3unsolv}
\end{align}
Considering $Z_\alpha(2,2)$, we can choose all gates to be SWAP gates except for two generic gates to observe the presence of the type (i) diagram:
\begin{align}
    \vcenter{\hbox{\includegraphics[width=0.65\columnwidth]{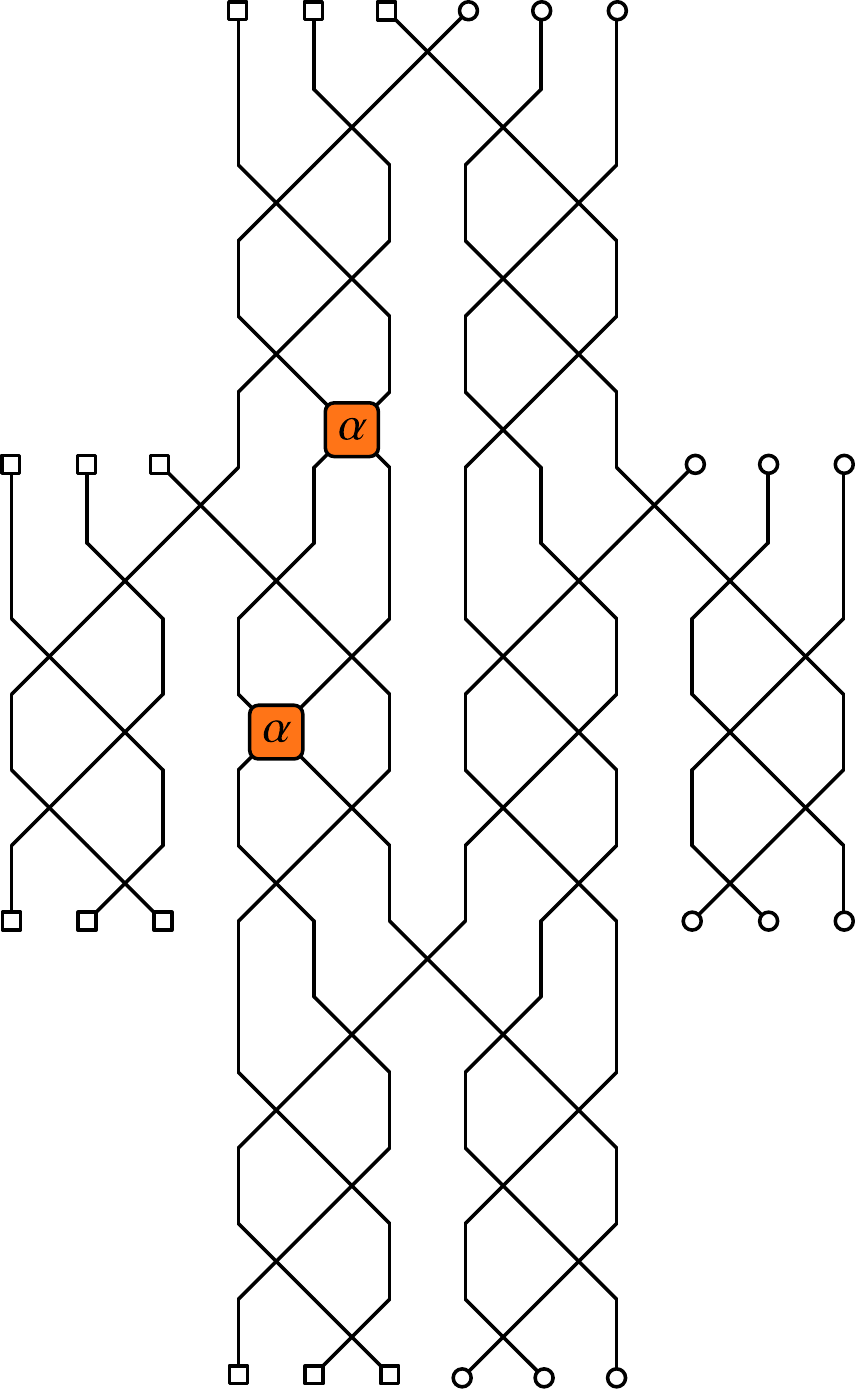}}}
\end{align}

This approach maps the problem of finding all conditions excluding complete reducibility to the mathematical problem of identifying all $N$-body irreducible diagrams for any $N$. We also have the additional constraints that the diagrams should be consistent with unitarity and the boundary conditions imposed by $Z_\alpha$. Checking the latter constraint is not straightforward. While a full enumeration of all diagrams satisfying these conditions is beyond the scope of this work, we can explicitly construct a family of $N$-body irreducible diagrams for any $N$, thereby showing the existence of infinitely many inequivalent sufficient conditions for the absence of complete reducibility. Indeed, for any number of gates $N$, the following diagram is $N$-body irreducible
\begin{align}
    \vcenter{\hbox{\includegraphics[width=0.32\textwidth]{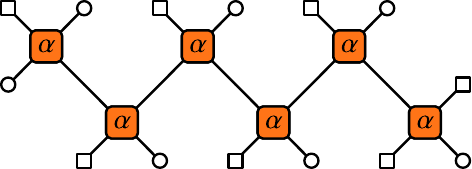}}}\,\,, \label{eq:N_body_irred}
\end{align}
which again holds for any orientation of the constituting gates, e.g., the following diagram is similarly $N$-body irreducible:
\begin{align}
    \vcenter{\hbox{\includegraphics[width=0.187\textwidth]{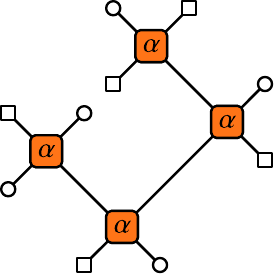}}}\,\,.
\end{align}
It is however a non-trivial problem to determine if these diagrams are consistent with the boundary conditions imposed by $Z_\alpha$.
There are also diagrams for $N\geq3$ falling outside the form of the families constructed above, further complicating a systematic classification of completely reducible circuits. As one example, for $N=4$ the following example is irreducible and falls outside the class of diagrams presented above:
\begin{align}
    \vcenter{\hbox{\includegraphics[width=0.19\textwidth]{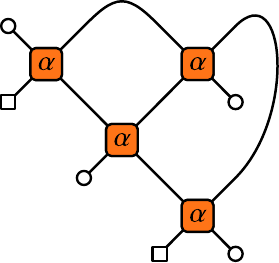}}}\,\,. \label{eq:4_body_irred_2}
\end{align}

\subsection{Crossing $v=0$ worldlines}
\label{sec:crossing}

\begin{figure*}[t]
    \centering
    \includegraphics[width = 0.72\textwidth]{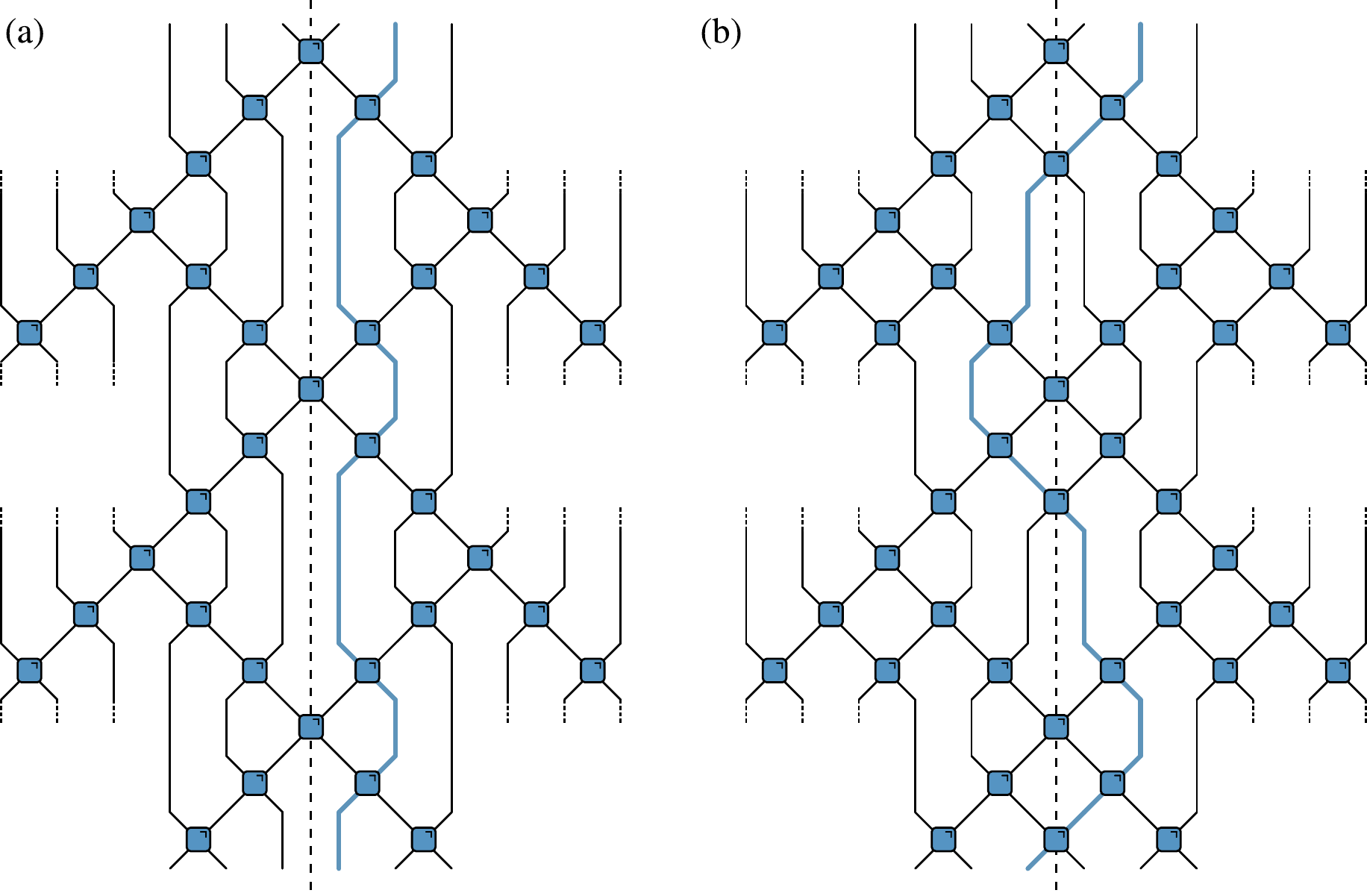}
    \caption{Two spacetime lattice circuits with highlighted (a) non-crossing $v=0$ worldline and (b) crossing $v=0$ worldline. The former leads to completely reducible dynamics, whereas the latter does not.}
    \label{fig:zero_crossings}
\end{figure*}

While a full classification of completely reducible spacetime lattices is currently out of reach, we can use the insights obtained in the previous section to find simple criteria that can aid in the identification and construction of completely reducible lattices. We here show that a large number of spacetime lattices cannot be completely reducible along $v=0$ based on a simple property of their worldlines, namely, the presence of crossing $v=0$ worldlines. We consider $Z_\alpha(m,m)$ for a sufficiently large $m$ and fix a $v=0$ worldline that crosses the bipartition at least twice. By time translation symmetry there is a complementary worldline crossing the first worldline as often as that worldline crosses the bipartition line. We now consider defects placed on some of these crossings. If the number of crossings is even, it is sufficient to place a defect on one of the crossings and we obtain the $1$-body irreducible diagram~\eqref{eq:one_body_irred}. If the number of crossings is odd, placing two defects at subsequent crossings yields the $2$-body irreducible diagram Eq.~\eqref{eq:two_body_irred} (i). 
We have already seen two examples of base gates with such crossing $v=0$-worldlines: the 3-pyramid and Eq.~\eqref{eq:2loc}. 
Examples for non-crossing and crossing $v=0$ worldlines are illustrated in Fig.~\ref{fig:zero_crossings}.

To conclude, we note that these considerations provide a new perspective on the failure of the solvability of DU brickwork circuits of finite size. The presence of boundaries, either periodic or closed, introduces crossings of this form into brickwork lattices at time scales larger than the system size. These crossings are in turn fundamental in establishing the universality of dual-unitary brickwork circuits at late times for finite system sizes, which underlies the computational hardness of obtaining exact results there~\cite{suzuki_computational_2022}.

\section{Consequences of complete reducibility}
\label{sec:consequences}

In this section, we investigate the general properties of completely reducible circuits. First, we show that the operator entanglement spectrum is flat. Then, we discuss information flow and show that it is constrained to happen along a finite number of rays in spacetime. We derive a general formula for the ELT, finding that it is piecewise linear with kinks corresponding to the directions of information flow. We conjecture a generalization to non-solvable circuits and discuss the origin of convexity of the ELT. Then, we prove that the entanglement velocity is determined by the Schmidt rank of the base gate. 
Finally, we present numerical evidence that completely reducible circuits are indeed quantum chaotic.

\subsection{Flatness of operator entanglement spectrum}
\label{sec:flatness}

If the diagram~\eqref{eq:z_alpha} is completely reducible, this means that we can reduce it to a product of disconnected overlaps of permutations. Let $\mathcal{N}$ be the number of overlaps between unequal permutation states. Using normalized permutation states, we obtain
\begin{equation}
    Z_\alpha(m,n) =  \left(\frac{1}{d}\right)^{(\alpha-1)\mathcal{N}}.
\end{equation}
This implies via $\mathcal{E}_\alpha\sim {\log (Z_\alpha)}/({1-\alpha)}$ that the ELT is independent of the R\'{e}nyi index, as we have already seen for brickwork DU circuits.

For quenches from generic states, the entanglement spectrum of the time-evolved wave function is in general not flat even in completely reducible circuits. The flatness of the ELT spectrum rather implies that asymptotically, the entanglement spectrum with respect to macroscopic regions is flat.
This result generalizes previous results on DU and DU2 circuits~\cite{Zhou2020,Foligno2024}. For these cases, it is additionally known that there exist special initial states, so-called solvable
states, which yield exactly flat entanglement spectra after quenches~\cite{Piroli2020,Foligno2024}. The observation of flat entanglement spectra in all known examples of solvable non-integrable many-body systems points to a deeper connection between these seemingly unrelated properties.

\subsection{Information flow along discrete directions}
\label{sec:information_flow}

Completely reducible circuits have the special feature that information can flow only along a finite number of certain discrete directions in spacetime (assuming the lattice is periodic in space and time). This can be seen by invoking that the ELT must be the same as the one of the corresponding swap circuit. In the swap circuit, the evolution is non-interacting and information can flow only along the worldlines imposed by the spacetime lattice. This implies that two-point correlation functions of one-site observables of $d$-qudits are only non-vanishing if the operators lie on the same worldline. It also implies that the ELT possesses kinks at the velocities corresponding to the worldlines~\cite{Sommers2024}. 

The knowledge of the ELT enables us to deduce a variety of further consequences. First, solvability of the ELT also implies that the dynamical two-point correlation functions can be solved exactly~\cite{Rampp2024}. This follows because both quantities can be computed from the influence matrix~\cite{Lerose2021}, which in this case has area law temporal entanglement, as further discussed in App.~\ref{app:correlations}. As for the non-interacting case, two-point correlation functions of one-site $d$-qudit observables are only non-vanishing on the same worldline. However, in contrast to the non-interacting case, correlation functions decay exponentially along the worldline with a rate determined by a low-dimensional quantum channel~(following, e.g., Refs.~\cite{Bertini2019,Claeys2021}).

The presence of a finite number of information flow directions implies that the butterfly velocity $v_B$ corresponds to the fastest such direction. Entanglement membrane theory further predicts the decay of the out-of-time-ordered correlator, a signal of scrambling, to be~\cite{Zhou2020}
\begin{align}
    \textrm{OTOC}(x=vt, t) \sim \exp\left(-\left(\mathcal{E}(v)-v\right)t\right), \quad \abs{v}<v_B.
\end{align}
The absence of curvature of the ELT further implies the absence of a diffusively broadening operator front. Again, this is consistent with the information being transported only along a finite set of directions.

All these features have already been observed in the simplest case of brickwork DU circuits. In this case, the directions of information flow are simply given by the directions of the exact causal light cone, $v=\pm 1$. This leads to an ELT which is flat in the interval $-1<v<1$ with kinks at $v=\pm 1$ [c.f. Fig.~\ref{fig:elt_sketches}(a)]. The influence matrix takes the form of a product state of identities, corresponding to a perfectly dephasing bath~\cite{Lerose2021,Gopalakrishnan2019}. The correlation functions along the worldlines are generated by two quantum channels~\cite{Bertini2019} (see again App.~\ref{app:correlations})
\begin{align}
    \mathcal{M}(v=+1) =\vcenter{\hbox{\includegraphics[width=0.063\textwidth]{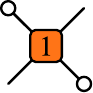}}},\quad \mathcal{M}(v=-1) =\vcenter{\hbox{\includegraphics[width=0.063\textwidth]{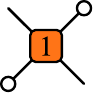}}}
\end{align}
It has also been observed that brickwork DU circuits possess a maximal velocity of information spreading and that the operator front is sharp~\cite{Claeys2020}.

Our findings clarify which of these properties are more general properties of solvable non-integrable models and which are particular to brickwork DU circuits only.

\subsubsection{Entanglement line tension}

\begin{figure*}[t]
    \centering
    \includegraphics[width = 0.7\textwidth]{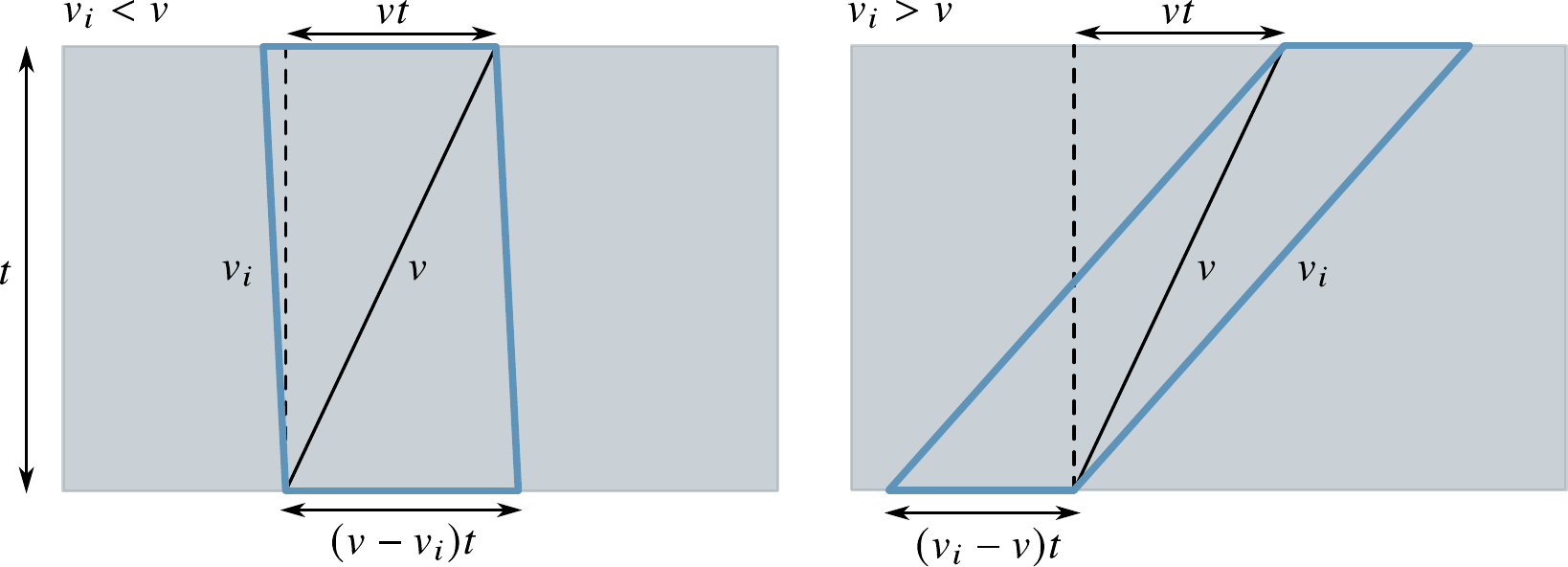}
    \caption{Illustration how in completely reducible circuits worldlines transporting information along a ray of velocity $v_i$ in spacetime contribute to the ELT. Generally, a worldline along $v_i$ entangles the interval $[x_1,x_2]$ at $t=0$ with the interval $[x_1+v_it,x_2+v_it]$ at the final time $t$. (a) For $v_i<v$ this corresponds to the contribution coming from the initial region $[0,(v-v_i)t]$. (b) For $v_i>v$ the constribution comes from $[-(v_i-v)t,0]$.}
    \label{fig:informationflow}
\end{figure*}

We now compute the ELT in the SWAP circuit and obtain an explicit expression in terms of the worldlines, which can in turn be applied to obtain the form of the ELT for generic completely reducible circuits. For a lattice constructed from a base gate of size $N$, we say that a worldline has velocity $v_i$ if it asymptotically goes along $x=v_i t$. We count all the worldlines with velocity $v_i$ emanating from a single base gate and call this number the multiplicity $n_i$. The multiplicities have to sum to $2N$.

Without loss of generality we consider a velocity $v>0$ and unequal to any of the worldline velocities, along which we aim to compute the ELT. Since the permutation states on the boundaries of $Z_\alpha(m,n)$ are normalized, the only worldlines that contribute a nontrivial factor of $1/d$ are those that connect a circle to a square or vice versa. We can understand these contributions in a geometric picture: they correspond to the worldlines crossing the ray $x=vt$.  Asymptotically, these can be counted by only considering the velocities and multiplicities $v_i$ and $n_i$. First, we consider worldlines that cross the ray $x=vt$ from right to left. These have $v_i<v$ and each species contributes
\begin{align}
    d^{-n_i{(v-v_i)t}/{2}}.
\end{align}
The factor of $1/2$ in the exponent arises because a unit cell of the chain consists of two sites. Analogously, worldlines with $v_i>v$ can only cross the ray $x=vt$ from the left. Their contribution is
\begin{align}
    d^{-n_i{(v_i-v)t}/{2}}.
\end{align}
This argument is illustrated geometrically in Fig.~\ref{fig:informationflow}. Overall, we obtain the asymptotic result
\begin{align}
    Z_2(m,n) = d^{-c(m,n)} \prod_{i:\,v_i<v}d^{-n_i(v-v_i)t/2} \prod_{i:\,v_i>v}d^{-n_i(v_i-v)t/2},
\end{align}
where $c(m,n)$ depends on $m,n$ but crucially is $\mathcal{O}(1)$. Using $q=d^N$ we find for the ELT
\begin{equation}
    \mathcal{E}(v) = \sum_{i:\,v_i<v} \frac{n_i}{2N}(v-v_i) + \sum_{i:\,v_i>v} \frac{n_i}{2N}(v_i-v), \quad v>0. \label{eq:ELTqp}
\end{equation}
This formula expresses the ELT in terms of the properties of non-interacting evolution but, remarkably, because of complete reducibility it also holds in chaotic systems. The formula directly relates the pattern of information flow in spacetime to entanglement and operator dynamics, in accordance with the picture put forward in Ref.~\cite{Sommers2024} that the ELT measures the information flow across the membrane. We note the similarity to a formula proposed in Ref.~\cite{Sommers2024}, which was proposed to hold in some but not all multi-unitary models. Eq.~\eqref{eq:ELTqp} describes a piecewise linear function with kinks at the velocities corresponding to directions of information flow. The ELTs of some exemplary completely reducible circuits are illustrated in Fig.~\ref{fig:elt_sketches}.

In mirror symmetric circuits we can simplify Eq.~\eqref{eq:ELTqp} further. Now, for any $v_i>0$ there is a corresponding $v_j=-v_i$ with $n_j=n_i$. We write
\begin{align}
    \mathcal{E}(v) = \frac{n_0}{2N}v + \sum_{i:\,0<v_i<v}\frac{n_i}{N}v + \sum_{i:\,v_i>v}\frac{n_i}{N}v_i, \quad v>0. \label{eq:ELTqpsymm}
\end{align}
The entanglement velocity is given by
\begin{equation}\label{eq:v_E_symme}
    v_E = \mathcal{E}(0) = \sum_{i:\,v_i>0}\frac{n_i}{N}v_i.
\end{equation}
It receives contributions from all wordlines with $v\neq0$, telling us that all information flow in space generates entanglement.
We now apply the above formula to some simple examples. 

\textbf{Brickwork dual-unitary circuits.} This result directly implies the constant ELT of dual-unitary brickwork circuits [Eq.~\eqref{eq:du_elt}].
In these circuits we have one worldline going along $v=1$ and one going along $v=-1$. With $N=1$ and $n(v=\pm 1)=1$ we can apply Eq.~\eqref{eq:ELTqp} for $|v|<1$ to find
\begin{align}
    \mathcal{E}(|v|<1) = \frac{1}{2} (v+1)+\frac{1}{2}(1-v) = 1,
\end{align}
explicitly observating the cancellation of the linear term and returning the flat ELT with entanglement velocity $v_E=1$.

\textbf{DU2 circuits.} For the Kagome lattice DU2 circuit from Eq.~\eqref{eq:gate_n2_du2}, we have $N=2$ with two worldlines going along $v=0$ and single worldlines along $v=\pm 1$. The resulting ELT follows from Eq.~\eqref{eq:ELTqpsymm}, e.g. for $0<v<1$, as
\begin{align}
    \mathcal{E}(0<v<1) &= \frac{1}{2}+\frac{1}{2}v,
\end{align}
reproducing Eq.~\eqref{eq:du2_elt}. Note that the $v=0$ worldlines are responsible for the linear term $v/2$. 
The entanglement velocity similarly follows as $v_E=1/2$, with the only contribution in Eq.~\eqref{eq:v_E_symme} coming from the single worldline along $v=1$.

For the nested Kagome lattice circuit from Eq.~\eqref{eq:gate_n4_nested} with $N=4$, it follows from direct inspection that there are six worldlines along $v=0$ and again single worldlines along $v=\pm 1$, which directly returns the ELT from Eq.~\eqref{eq:du2_elt_2}.

\textbf{4-pyramid circuit.} A slightly more involved example is given by the 4-pyramid circuit from Eq.~\eqref{eq:4pyr} with $N=4$. In the corresponding lattice there is a set of three worldlines along $v=1/3$, a single worldline $v=1$, and corresponding worldlines for negative velocities. For $0<v<1/3$ we obtain from Eq.~\eqref{eq:ELTqpsymm}
\begin{align}
    \mathcal{E}(0<v<1/3) =  \frac{3}{4}\times\frac{1}{3}+\frac{1}{4}\times 1 = \frac{1}{2}
\end{align}
whereas for $1/3<v<1$ we find that
\begin{equation}
    \mathcal{E}(1/3<v<1) = \frac{3}{4}v+\frac{1}{4},
\end{equation}
reproducing Eq.~\eqref{eq:4pyr_elt}.

All presented expressions for the ELT can be obtained in a similar manner.
Altogether, we conclude that completely reducible circuits possess an ELT with kinks at a finite number of velocities corresponding to directions of information flow. Furthermore, periodicity of the lattice implies that these velocities must be rational numbers (in units of the lattice spacing of the $q$-qudits).

\subsubsection{Convexity of the ELT}

What if the relation between information flow and the ELT could be extended to more general systems? For circuits that are not solvable, we expect that information is spread in all directions with a continuous (except possibly at isolated points) weight function $n(v)$ which we normalize as
\begin{align}
    \int_{-1}^1\textrm{d}v\,n(v)=1.
\end{align}
Taking the continuum limit of Eq.~\eqref{eq:ELTqp} yields
\begin{equation}
    \mathcal{E}(v) = \int_{-1}^v\textrm{d}v' n(v')(v-v') + \int_{v}^1\textrm{d}v' n(v')(v'-v), \quad v\geq0. \label{eq:elt_continuous}
\end{equation}
This can be seen as a 'quasiparticle picture' of entanglement dynamics in chaotic systems, where such quasiparticles do not exist. In the general case this will depend on the R\'{e}nyi index $\alpha$ via the density of information flow $n(v)=n_\alpha(v)$. Eq.~\eqref{eq:elt_continuous} generally describes a function with non-vanishing curvature. Interestingly, the curvature is exactly given by (twice) the density of information flow
\begin{equation}
    \frac{\textrm{d}^2}{\textrm{d}v^2}\mathcal{E}(v) = 2n(v). \label{eq:convexity}
\end{equation}
This ensures the convexity of the ELT and provides a physical picture for its necessity. It generalizes the picture of isolated directions of information flow leading to kinks in the ELT to the general setting, where a continuous density of information flow gives rise to the curvature of the ELT.

\subsection{Entanglement velocity and Schmidt rank}
\label{sec:vE_Schmidt}

In this section, we show that in completely reducible circuits there is a simple relation between the entanglement velocity and the Schmidt rank of the base gate $\mathcal{R}$ (which we introduce below)
\begin{equation}
    v_E = \frac{\log\mathcal{R}}{\log q^2}. \label{eq:logR}
\end{equation}
This formula has already been shown to hold in DU and DU2 circuits~\cite{Foligno2024}.
Consider the operator Schmidt decomposition~\cite{Nielsen2003} of a base gate $U$
\begin{equation}
    U = \sum_{i=1}^{q^2}\lambda_iX_i\otimes Y_i, \quad \tr\left[X_i^\dagger X_j\right]=\tr\left[Y_i^\dagger Y_j\right]=\delta_{ij},
\end{equation}
where the $\lambda_i$ form the operator entanglement spectrum of the base gate. The Schmidt rank $\mathcal{R}$ is the number of nonzero $\lambda_i$. As shown in Sec.~\ref{sec:flatness}, this spectrum is flat for completely reducible gates. This enables us to compute $\mathcal{R}$ as
\begin{equation}
    \mathcal{R} = \frac{1}{Z_2(1,1)}\,.
\end{equation}
We can compute $Z_2(1,1)$ from the number of worldlines that cross the bipartition. This number is given by
\begin{equation}
    \sum_{i:\,v_i\neq0}n_i\abs{v_i} + c_0\,,
\end{equation}
where $c_0$ denotes the number of worldlines of velocity $v=0$, that connect $A$ to $\bar{A}$. We already showed in Sec.~\ref{sec:crossing} that the presence of such crossing $v=0$ worldlines implies that the circuit cannot be completely reducible, such that we can express
\begin{equation}
    \mathcal{R} = d^{\sum_{i:\,v_i\neq0}n_i\abs{v_i}}.
\end{equation}
Using the quasiparticle formula for the ELT, Eq.~\eqref{eq:ELTqp}, we again conclude
\begin{equation}
    \frac{\log\mathcal{R}}{\log q^2} = \sum_{i:\,v_i\neq0}\frac{n_i\abs{v_i}}{2N} = v_E\,.
\end{equation}
The relation between the flat entanglement spectrum and the entanglement velocity was also discussed in Refs.~\cite{Foligno2024} and \cite{Rampp2024} in the context of DU2 circuits. The former pointed out that the flat entanglement spectrum implies that the dynamics corresponds to an isometry in the spatial direction, as opposed to the unitarity exhibited by dual-unitary brickwork circuits, and identified this as the key property underpinning the solvability of DU2 circuits. We here observe that the same property applies in completely reducible circuits, and refine the relation between the ELT, the Schmidt rank, and the directions of information propagation.

\subsection{Maximal quantum chaos}
\label{sec:chaos}

\begin{figure*}[t]
    \centering
    \includegraphics[width = 0.7\textwidth]{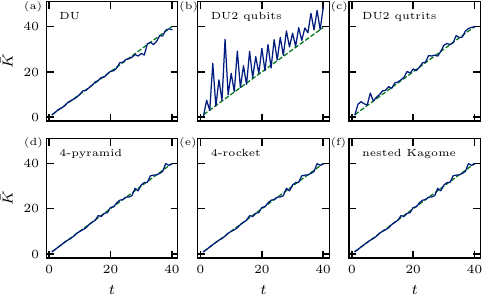}
    \caption{Numerically computed spectral form factor for various completely reducible circuits. The dashed lines are the predictions from random matrix theory. Averages are performed over 1000 circuit realizations. (a) Brickwork DU circuit, $q=2$, $N=12$ sites. (b) Kagome lattice CHM circuit $q=2$, $N=12$. This circuit falls into the COE class, therefore $\bar{K}(t)/2$ is plotted for better comparison with the other models. (c) Kagome lattice CHM circuit $q=3$, $N=8$. (d) 4-rocket CHM circuit $q=4$, $N=6$. (e) 4-pyramid CHM circuit $q=4$, $N=6$. (f) Nested Kagome CHM circuit $q=4$, $N=6$.}
    \label{fig:sff_plots}
    \vspace{-\baselineskip}
\end{figure*}

So far we have discussed the robustness to integrability-breaking interactions as a distinguishing property of completely reducible circuits. However, it is crucial to stress that these circuits are generically non-integrable despite this robustness, which is confined only to particular quantities. In this section, we present numerical evidence that completely reducible circuits show signatures of quantum chaos in their quasienergy spectra~\cite{DAlessio2016,Haake2018}. We investigate the spectral form factor (SFF) and show that it is in agreement with the expectation from random matrix theory. Moreover, we show that completely reducible circuits exhibit a form of maximal quantum chaos termed \emph{critical quantum chaos}~\cite{Bertini2018}.

The SFF is the Fourier transform of the two-level correlation function~\cite{Haake2018}. It is given by
\begin{equation}
    K(t) = \abs{\tr\mathcal{U}^t}^2=\sum_{m,n=1}^{\mathcal{D}}e^{i(\phi_m-\phi_n)t},
\end{equation}
where $\mathcal{U}$ denotes the time-evolution operator of the system and $\{e^{i\phi_m}|m=1\dots \mathcal{D}\}$ are its eigenvalues. Because the SFF is not self averaging, it has to be computed by averaging over an appropriate ensemble of circuits. We denote the averaged SFF by $\bar{K}(t)$.

We numerically compute the averaged SFF in various completely reducible circuits (Fig.~\ref{fig:sff_plots}). To reduce the numerical requirements, we use slightly different parametrizations of the spacetime lattice circuits introduced in the main text. These parametrizations live in a compressed local Hilbert space of $d^{N/2}$ as opposed to $d^N$. They are based on biunitary connections, a certain generalization of DU gates. This procedure, as well as the concrete parametrizations, are discussed in App.~\ref{app:biunitary}. 

We observe that the SFF conforms well to the random matrix prediction, consistent with quantum chaotic dynamics.
Furthermore, we note that generic many-body systems are only expected to conform to the random matrix SFF after a time scale $t_{\rm{Th}}$ known as the Thouless time, leading to a characteristic dip-ramp behavior~\cite{Thouless1977,Altshuler1986,Chan2018}. In brickwork DU circuits, the Thouless time vanishes, implying agreement with random matrix theory on large quasienergy scales~\cite{Bertini2018}. We find similar behavior in other completely reducible circuits, where the characteristic dip is absent. Deviations of the maximally chaotic behavior are present for the Kagome lattice circuit on qubits [see Fig.~\ref{fig:sff_plots}(b)], where we observe a difference between odd and even timesteps. While $\bar{K}(t)$ for odd steps matches well with the random matrix theory prediction, early-time deviation is present for even steps. We attribute this effect to a particularity of the qubit model, as the dip is absent for the same lattice geometry when using qutrits [Fig.~\ref{fig:sff_plots}(c)]. It would be interesting to further investigate this effect, where the availability of analytical results for dual-unitary circuits and their similar spectral statistics suggests that analytical results on the SFF of completely reducible circuits could be within reach.

\section{Relation to knot theory}
\label{sec:knots}

In this section, we point out a relation between completely reducible dynamics and knot theory. 
Specifically, we show that certain link diagrams associated to completely reducible circuits must be unlinked. We also discuss the computational complexity of reducing tensor networks of dual-unitary gates and give an efficient algorithm to check complete reducibility. 
Accessible introductions to knot theory can be found in Refs.~\cite{Adams2004,Kauffman2001}. For any spacetime lattice circuit of DU gates we associate a particular link (a knot consisting of multiple distinct components) to the tensor network diagram $Z_2(m,n)$. We show that if $Z_2(m,n)$ is completely reducible, then the associated link is equivalent to the unlink. In other words, the link is not linked. In this case, $Z_2(m,n)$ is given by the Kauffman bracket of the associated link. This conversely implies that any diagram for which the link is non-trivial cannot be completely reducible.

We motivate the mapping to the link by recalling that for $Z_2(m,n)$ completely reducible, we can choose the SWAP gate to use in place of every DU gate in the diagram without altering the result. 
We now associate each SWAP gate with a crossing of two strands [c.f. Eq.~\eqref{eq:SWAP_crossing}]. Here we additionally have to fix an orientation and we distinguish between gates and their Hermitian conjugate as
\begin{align}\label{eq:gate_to_crossing}
    U=\,\vcenter{\hbox{\includegraphics[width=0.053\textwidth]{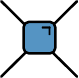}}}\,\, \to \,\, \vcenter{\hbox{\includegraphics[width=0.053\textwidth]{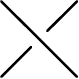}}}\,,\,\,\,\,
    U^\dagger=\,\vcenter{\hbox{\includegraphics[width=0.053\textwidth]{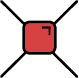}}} \,\,\to\,\, \vcenter{\hbox{\includegraphics[width=0.053\textwidth]{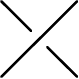}}}\,.
\end{align}
If the diagram is completely reducible, this means that we can reduce it to a product of disconnected overlaps of permutations, with each closed cycle yielding a factor of $d$. The deformations we use in the tensor network diagram language are unitarity and dual unitarity. In the language of knot theory, these operations correspond to Reidemeister II moves. Specifically, unitarity corresponds to the move
\begin{align}
    &\vcenter{\hbox{\includegraphics[width=0.053\textwidth]{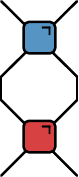}}} \quad \to \quad  \vcenter{\hbox{\includegraphics[width=0.053\textwidth]{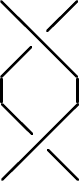}}} \quad = \quad \vcenter{\hbox{\includegraphics[width=0.034\textwidth]{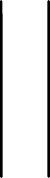}}}\,\,,
\end{align}
and dual-unitarity corresponds to
\begin{align}
    &\vcenter{\hbox{\includegraphics[width=0.067\textwidth]{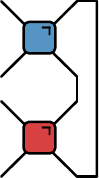}}} \quad \to \quad  \vcenter{\hbox{\includegraphics[width=0.067\textwidth]{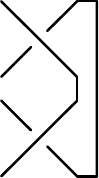}}} \quad = \quad \vcenter{\hbox{\includegraphics[width=0.0215\textwidth]{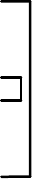}}}\,\,.
\end{align}
Complete reducibility implies that, using only unitarity and dual-unitarity, the full tensor network can be mapped to a product of overlaps, and each overlap $\vcenter{\hbox{\includegraphics[width=0.038\textwidth]{figs/circle_square_rot.pdf}}}$ in turn corresponds to an unknot.
Taken together, complete reducibility implies that the link obtained by expressing the unfolded Eq.~\eqref{eq:z_alpha} in terms of the SWAP gates~\eqref{eq:gate_to_crossing} can be deformed into the unlink using only Reidemeister II moves. Note that after unfolding, diagrams of this type can always be drawn in a plane without any crossing of lines. This is important for the link we associate with the diagram to be unique.

We illustrate this connection using the brickwork circuit of Eq.~\eqref{eq:z_alpha_reduced}. Performing the substitution to SWAP gates~\eqref{eq:gate_to_crossing} in the unfolded diagram with $\alpha=2$, we directly observe that $Z_2(m,n)$ is simply determined by the number of closed loops $N_\ell$ in the diagram:
\begin{widetext}
\begin{align}
    Z_2(m,n) = \left(\frac{1}{d}\right)^{2(m+n)}\,\,\vcenter{\hbox{\includegraphics[width=0.19\textwidth]{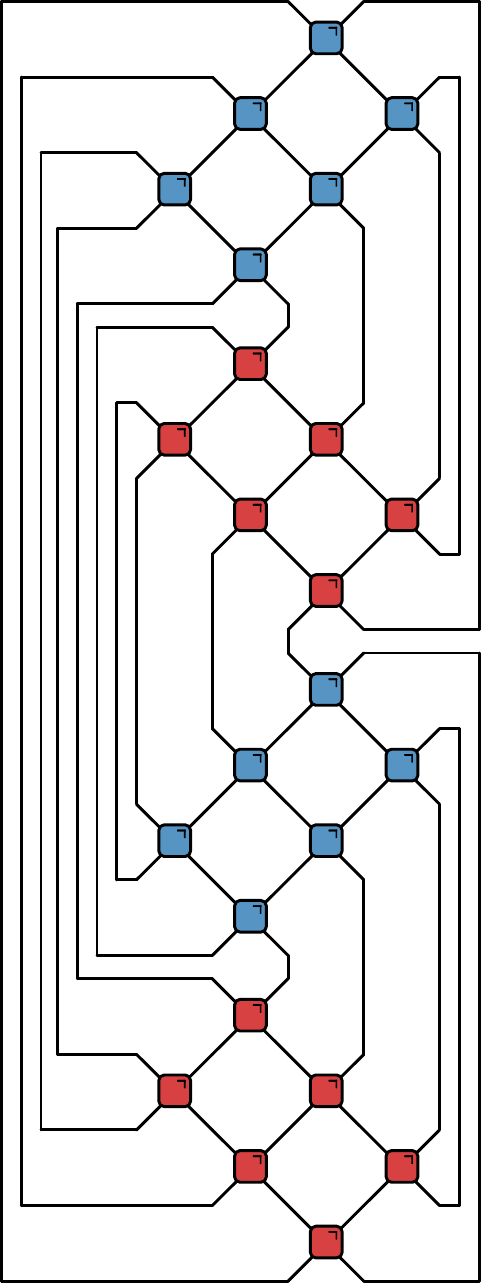}}}\,\, = \left(\frac{1}{d}\right)^{2(m+n)}\,\,\vcenter{\hbox{\includegraphics[width=0.19\textwidth]{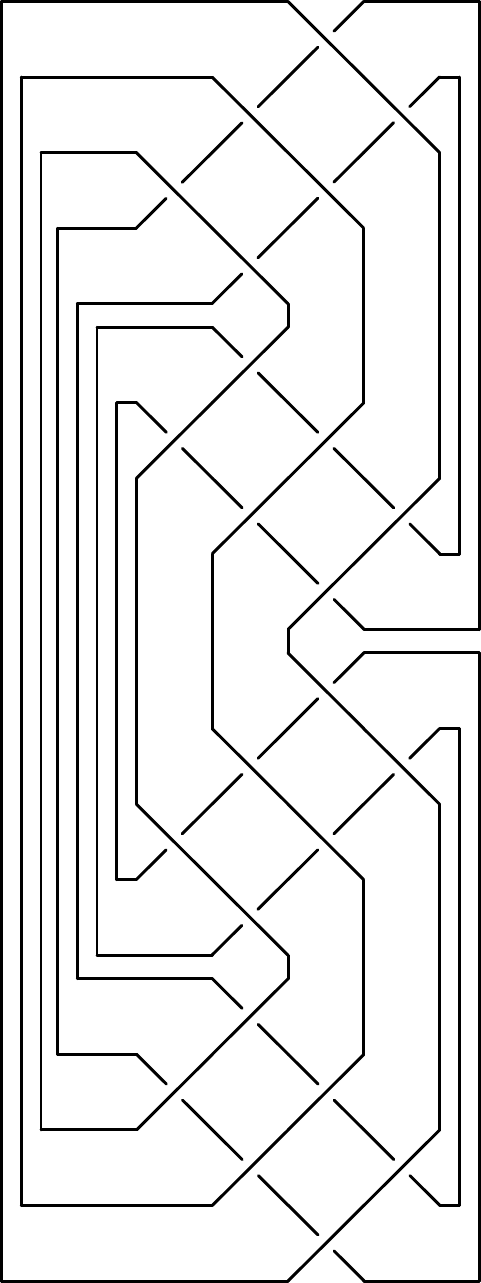}}}\,\, = \left(\frac{1}{d}\right)^{2(m+n)-N_\ell}.
\end{align}
\end{widetext}
Each of these loops corresponds to a single unknotted link in spacetime, and we denote the full link as $L_{m,n}$.
If the diagram is completely reducible, this means that we can reduce this link to a product of disconnected overlaps of permutations, with each closed cycle yielding a factor of $d$.
Complete reducibility implies that this link can be deformed into the unlink using only Reidemeister II moves.
This shows that completely reducible diagrams correspond to trivial links in spacetime, as illustrated in the example above, for which it can be directly seen that no knots occur (as also required for the ELT to be independent of the R\'enyi index $\alpha$). 

The links we have defined in this manner live in a replicated spacetime, which distinguishes them from previous ways of encoding links in quantum information that are based on constructing a unitary evolution operator from gates forming a representation of the braid group~\cite{Kauffman2001a,Aharonov2006,Shor2007}. While for these protocols the link invariant is encoded in the trace of the unitary operator or a particular amplitude, the links we define here are directly related to entanglement. DU gates generically do not form a representation of the braid group, which is why only trivial links yield invariant quantities. However, we can obtain a DU representation of the braid group by choosing DU gates that satisfy the set-theoretic Yang-Baxter equation~\cite{Gombor2022,Zhang2024,MorralYepes2025}. These gates lead to superintegrable non-ergodic dynamics, where any tensor network $Z_2$ constitutes a link invariant.

We briefly highlight that diagrams that cannot be reduced are non-trivially linked. Consider for example the following diagram, with its map to a link and its equivalent under ambient isotopy:
\begin{align*}
    \,\,\vcenter{\hbox{\includegraphics[width=0.064\textwidth]{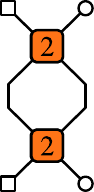}}} \to \vcenter{\hbox{\includegraphics[width=0.107\textwidth]{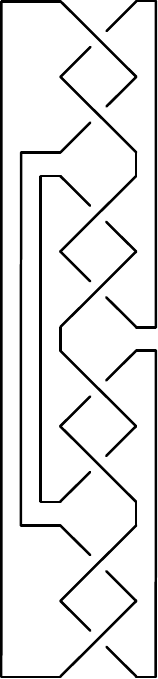}}} \to \vcenter{\hbox{\includegraphics[width=0.212\textwidth]{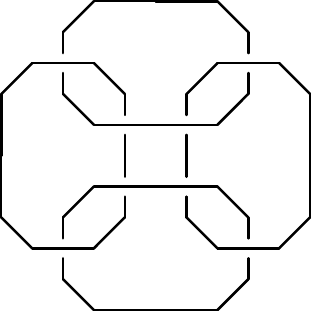}}} \label{eq:one_defect_extended_link}
\end{align*}
This constitutes a non-trivially linked link, as can be seen by examining the linking number of the component drawn of the top with one of the middle components.

The Kauffman bracket of a link $L$ is a polynomial $[L](A)$ that can be determined from a projection of the link~\cite{Kauffman1987}. It is invariant under Reidemeister II and III moves, but not under Reidemeister I moves. We define the variable $A$ from the local Hilbert space dimension by fixing $A^2+A^{-2}=-d$. For completely reducible diagrams, the Kauffman bracket of the link $[L_{m,n}](A)$ yields
\begin{align}
    Z_2(m,n) = \left(\frac{1}{d}\right)^{2(m+n)}[L_{m,n}](A), \quad A^2+A^{-2}=-d.
\end{align}
After applying Reidemeister II moves to unlink the diagram, each unknot is assigned a factor of $-(A^2+A^{-2})$. Setting $-(A^2+A^{-2})=d$ and taking into account the normalization, we recover the expected result.

It is tempting to draw comparisons to the theory of integrable systems, which is also closely related to knot theory~\cite{Witten1989,Reshetikhin1990,Bazhanov1996,Wadati1993,Costello2013,Costello2018}. A large class of integrable models is based on the Yang-Baxter equation. While in the circuit manipulations above, unitarity and dual unitarity are the equivalents of the Reidemeister II move, the Yang-Baxter equation is the equivalent of the Reidemeister III move. 
Furthermore, in topological field theories expectation values are link invariants, e.g., for $2+1$ dimensional Chern-Simons theory the Jones polynomial is realized through expectation values of Wilson lines~\cite{Witten1989}. 
Similarly, for completely reducible circuits, the ELT is determined by the Kauffman polynomial. Nevertheless -- and despite the robustness of $Z_2$ to integrability-breaking interactions -- many aspects of the dynamics of completely reducible circuits are fundamentally different to integrable dynamics. In particular, this was already highlighted by showing that the spectral form factor, an indicator of random-matrix like level statistics, reproduces the expectations of quantum chaos.

To conclude this section, we briefly discuss the complexity of determining whether a circuit is completely reducible.
In practice, given a diagram of finite size, determining if it is completely reducible or not is possible in polynomial time.
The algorithmic complexity of the general task of determining if a link is equivalent to the unlink (the unknotting problem) is however an open problem~\cite{Lackenby2016,Hass1999}. Known algorithms for unknotting require at least exponential time in the number of crossings. The knots and links defined by the $Z_2$ diagrams possess a special structure, such that the problem of determining if such a diagram is completely reducible is only polynomially hard. The reason is that in the folded representation, simplifications can only appear at the boundaries of the diagram.
Given a diagram $Z_2(m,n)$ of a general spacetime lattice, which need not be periodic in space or time, we move along the boundary and check if a reduction using (dual) unitarity is possible. If so, we apply the reduction. This is performed until no reductions are possible, while keeping track of contracted permutations. At this point, either the diagram has been reduced completely, or this is not possible and the irreducible core of the diagram remains.

\section{Conclusion and outlook}
\label{sec:conc}

Various exactly solvable models of chaotic quantum many-body dynamics have appeared in the literature in the past years, each with a characteristic shape of information spreading and varying degrees of solvability.
This research was kickstarted by dual-unitary circuits, in which space-time duality constrains all quantum information to spread with maximal velocity, and followed by various generalizations of dual-unitarity that can broadly be described as multi-unitary dynamics.
These generalizations aim to preserve the solvability of dual-unitary circuits while avoiding some of the pathologies of dual-unitarity, in order to more closely approximate `generic' quantum chaotic models.

In this work we first introduced various examples of previously unknown solvable models, obtained by arranging dual-unitary gates on particular spacetime lattices.
The resulting models were shown to exhibit anywhere from two to five distinct rays of information spreading in spacetime, going beyond the two and three rays expected from known solvability conditions (dual-unitary and DU2 dynamics respectively), with an entanglement line tension that is piecewise linear with kinks at the discrete directions of information spreading. 

The main aim of this work was to present a common framework through which these models could be understood and extended, and to present a complete picture of how their spreading of quantum information can be understood through entanglement membrane theory. 
To do so, we introduced the concept of \emph{completely reducible circuits} as a class of exactly solvable models of non-integrable dynamics generalizing usual dual-unitary brickwork circuits. 
We showed that completely reducible circuits share a common phenomenology that stems from their similarity to non-interacting dynamics.
While these circuits are generically quantum chaotic, scrambling, and do not possess any local conservation laws, certain quantities such as the operator entanglement of the time evolution operator nevertheless remain unchanged with respect to the non-interacting evolution. 
These fully quantify operator spreading and entanglement growth, and our results suggest that exactly solvable models may possess inherent features distinguishing them from generic systems. In particular, in all solvable non-integrable models identified so far the information flow is along a finite number of directions in spacetime, while generic systems distribute information continuously. 
This connection in turn allowed us to relate the curvature of the entanglement line tension to the density of information flow.
Making precise statements which phenomenology is out of reach of solvable models would be a major step forward. 
While we were able to systematically identify criteria for complete reducibility, it remains an open challenge to give a complete enumeration of spacetime lattices giving rise to completely reducible dynamics. Since complete reducibility is a geometric property of the lattice, it might be possible to find necessary conditions in terms of graph theoretic quantities.

We additionally uncovered a surprising connection between the theory of such exactly solvable non-integrable dynamics and the mathematical theory of knots. Completely reducible circuits correspond to knots, or more precisely links,  which are not knotted, and for which the entanglement line tension can be related to the Kauffman polynomial. This constitutes a first step towards an algebraic point of view on solvable non-integrable systems, and knot theory could in turn be used to identify additional completely reducible models. It might also lead to further unexpected connections to the theory of Yang-Baxter integrable systems, which is closely tied to knot theory. 

Possible extensions are plenty. 
While we focused on spacetime lattices of dual-unitary gates, it is also possible to consider spacetime lattices of perfect tensors~\cite{pastawski2015holographic}. These satisfy additional unitarity conditions, allowing for further simplifications in the evaluation of e.g. the entanglement line tension, since various irreducible diagrams can be further reduced for such a choice of gates. The corresponding lattices have natural connection with HaPPY codes~\cite{pastawski2015holographic} and hyperinvariant tensor networks~\cite{evenbly_hyperinvariant_2017}, using hyperbolic lattices of perfect tensors and dual-unitary gates respectively as models for holography. These models can in turn be interpreted as completely reducible and share much of the phenomenology of the models discussed here, as apparent in e.g. their flat entanglement spectrum.
Complete reducibility can more generally be considered in cases where the underlying spatial lattice is a general graph or of a higher dimension. It is, however, not a priori clear which elementary quantum gates to apply in this case -- DU gates acting on individual bonds or gates acting on plaquettes of the lattice (such as ternary unitary gates~\cite{Milbradt2023} or multi-site tree-unitary gates~\cite{Breach2025}).
Another direction would be to develop the theory of completely reducible circuits in the absence of any spatial or temporal periodicity. Kasim and Prosen~\cite{Kasim2023} have investigated DU circuits in random spacetime geometries by considering DU gates situated at the intersections of randomly placed lines. This way of generating a random lattice automatically preserves complete reducibility because worldlines never change direction. It would be interesting to find random geometries leading to more complex phenomena beyond the ones observed in periodic lattices.

Another possible direction is the inclusion of measurements into the spacetime lattice formalism. Studying the interplay of unitary evolution and measurements has been an active research topic in recent years, leading to the discovery of various novel phenomena, such as entanglement transitions and charge-sharpening transitions~\cite{Li2018,Skinner2019,Chan2019,Agrawal2022}. However, these phenomena are typically out of reach of exact treatments. A relevant problem is hence to find instances of exactly solvable non-unitary dynamics. A natural way of including measurements into the spacetime lattice formalism would be to add projections on Bell states as the spacetime dual of the identity gate. We mention here that Refs.~\cite{ippoliti2021postselectionfree} and~\cite{Claeys2022} studied the interplay of forced measurements of this form with DU dynamics. Both return completely reducible dynamics, where the feedback mechanism used in the latter work to maintain the solvability of the circuits can be understood through the lens of our formalism as a way of removing crossings that spoil complete reducibility.

We focused on the entanglement line tension to obtain insight into the macroscopic entanglement dynamics of the models under study. Complete reducibility of the ELT additionally implies that dynamical two-point correlations can be solved exactly and allows the analysis of scrambling {via} the tripartite mutual information~\cite{Schnaack2019,Bertini2020}. It would be interesting to study other quantities in completely reducible circuits, e.g. the dynamics of observables and entanglement after quenches~\cite{Piroli2020}. So-called solvable initial states play a special role in quench dynamics since they lead to a flat entanglement spectrum, inviting comparisons with stabilizer (Clifford) dynamics and motivating studies of nonstabilizerness and magic in the presented circuits, for which it is expected that solvable states can similarly be found.
Of particular interest would also be the further investigation of spectral signatures of quantum chaos~\cite{Bertini2018}. While our numerical computations show agreement of the spectral form factor with the predictions of random matrix theory and the absence of a Thouless time, an analytical proof is lacking beyond brickwork DU circuits. Initial explorations indicate that completely reducible circuits in general only enable the analytical calculation of local quantities, while global quantities such as the spectral form factor are out of reach, leaving the vanishing Thouless time in these models as on open question. 
A closely related question is the generation of randomness in our models, as quantified by e.g. anticoncentration~\cite{Dalzell_2022,Claeys_2025_Fock_space} or the convergence to unitary designs~\cite{Brandao_2016,Haferkamp_2022_randomquantum}.
The lattice models introduced in this work can be viewed as brickwork dual-unitary circuits in which certain dual-unitary gates are replaced by the identity gate in a periodic manner, and it is not a priori clear how this modifies randomness generation.
To conclude, we also note that these models can be directly implemented in current digital quantum computation devices, such that the predicted phenomenology and possible deviations when moving away from complete reducibility are experimentally accessible.

\begin{acknowledgments}
We wish to acknowledge useful discussions with Cecilia De Fazio, J. Alexander Jacoby, Grace Sommers, and Jiangtian Yao. S. A. R. is supported by the European Research Council (ERC, AdG DebuQC). All authors acknowledge support from the Max Planck Society.
\end{acknowledgments}

\begin{widetext}
\appendix

\section{Dynamical correlation functions}
\label{app:correlations}

In this appendix, we highlight how dynamical correlation functions can be exactly obtained in completely reducible circuits. 
On the level of tensor network diagrams, correlation functions have boundary conditions which depend on the specific operators being probed. Because the boundary conditions are not purely composed of permutation states, the tensor network diagrams representing correlation functions can in general not be completely reduced. However, we argue that correlation functions can still be computed exactly in completely reducible circuits, because the problem can be reduced to a one-dimensional contraction. We further argue that correlations vanish everywhere but along the rays of information flow, where the derivative of the ELT is discontinuous, and present several examples.

The dynamical two-point correlation function of two traceless operators $\sigma,\rho$ at distance $x$ is defined as
\begin{align}
    C_{\sigma\rho}(x,t)=\frac{1}{q^L}\tr\left[\mathcal{U}(t)^\dagger\sigma(0)\mathcal{U}(t)\rho(x)\right].
\end{align}
Considering a brickwork unitary circuit of (not necessarily dual-unitary) gates, this quantity is represented by a tensor-network diagram of the form~\cite{Bertini2025}
\begin{align}
    C_{\sigma\rho}(x,t)=\,\,\vcenter{\hbox{\includegraphics[height = .27\textwidth]{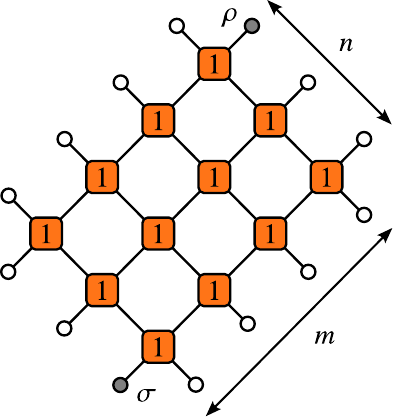}}}\,\,, \label{eq:correlation}
\end{align}
where $m=(t+x)/2,\,n=(t-x+2)/2$ and we have introduced vectorized operators as
\begin{align}
    \vcenter{\hbox{\includegraphics[height=0.031\textheight]{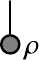}}} \,\equiv\, \frac{1}{\sqrt{q}}\,\, \vcenter{\hbox{\includegraphics[height=0.031\textheight]{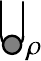}}}\,.
\end{align}
This form of the diagram is obtained by using unitarity to reduce its area to the intersection of the light cones emanating from the initial and final operators, similar to the manipulations used to arrive at Eq.~\eqref{eq:z_alpha_reduced}.
Evaluating the correlator for $x=t$, corresponding to the edge of the causal light cone, each diagram in the sequence is one-dimensional:
\begin{align}
    C_{\sigma\rho}(x=t,t)=\,\,\vcenter{\hbox{\includegraphics[height = .19\textwidth]{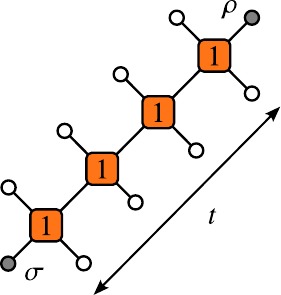}}}\,\,. \label{eq:correlation_lc}
\end{align}
A one-dimensional diagram can be contracted by matrix multiplication with a complexity of $\mathcal{O}(q^2 t)$. Hence, the two-point correlator along $x=t$ is exactly solvable for any choice of gate. If we however choose to evaluate $C_{\sigma\rho}$ along $x=vt$ for any $v$ with $\abs{v}<1$, then the side lengths of the diagrams grow as $n\sim(1-v)t/2$ and $m\sim(1+v)t/2$. For generic (non-DU) gates, the complexity of contracting two-dimensional tensor networks then grows exponentially in $t$~\cite{Schuch2007}. 

\textbf{Brickwork DU circuits.}
When the gates satisfy additional relations, the contraction of Eq.~\eqref{eq:correlation} can be easier. The prototypical example is given by DU gates. Away from the light-cone edge, for any $n>1$, Eq.~\eqref{eq:correlation} can be contracted using Eqs.~\eqref{eq:dual_unitarity_folded} to yield exactly zero
\begin{align}
    \vcenter{\hbox{\includegraphics[height=.25\textwidth]{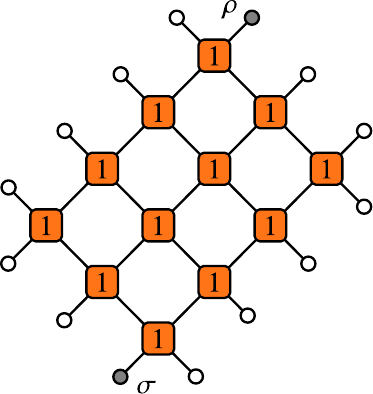}}}\,\,=\,\,\vcenter{\hbox{\includegraphics[height=.25\textwidth]{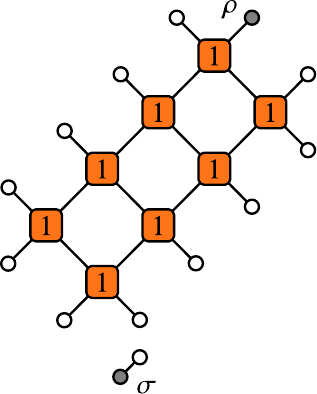}}}\,\,=\frac{\tr[\sigma]}{q}\frac{\tr[\rho]}{q} = 0.
\end{align}
This is consistent with the picture that no information flows inside the light cone. Along the light-cone edge, where information flow is expected because of the kinks in the ELT being at $v=\pm1$, the correlator is given by Eq.~\eqref{eq:correlation_lc} and is generically non-vanishing. 
The solvability of the correlator can in fact be traced back to the complete reducibility of the dynamics. To see this, we decompose the tensor network as
\begin{align}
    C_{\sigma\rho} = \bra{L}T_{\sigma\rho}\ket{R} = \,\,\vcenter{\hbox{\includegraphics[height=0.28\textwidth]{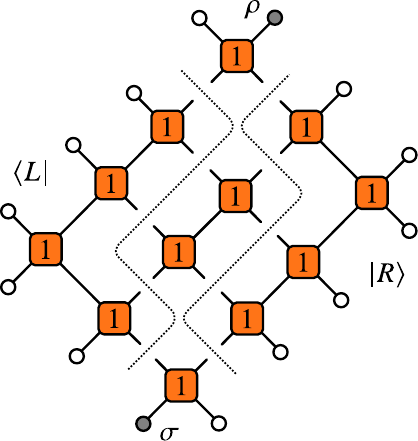}}}\,\,.
\end{align}
The objects $\bra{L}$ and $\ket{R}$ contain the influence of the many-body system on itself when seen as a bath. They are known as \emph{influence matrices}~\cite{Lerose2021,Sonner2021,Foligno2023}. Importantly, the same influence matrices govern the behavior of the ELT~\cite{Rampp2024}. Because the ELT is completely reducible, the influence matrix must have a simple form enabling an efficient evaluation of the correlator. For example, in the brickwork DU case the influence matrices reduce to those of completely depolarizing baths:
\begin{align}
    \vcenter{\hbox{\includegraphics[height=0.17\textwidth]{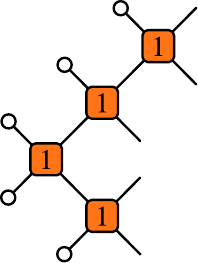}}}\,\,=\,\,\vcenter{\hbox{\includegraphics[height=0.17\textwidth]{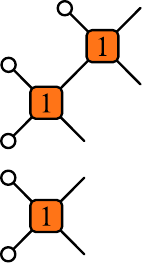}}}\,\,=\,\,\vcenter{\hbox{\includegraphics[height=0.17\textwidth]{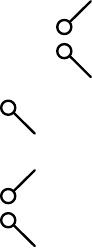}}}\,\,.
\end{align}
In general, we expect influence matrices in completely reducible circuits to have area-law temporal entanglement. In the following, we illustrate the resulting correlation functions for various classes of completely reducible circuits.

\textbf{4-pyramid lattice.} 
We already observed that the ELT in the 4-pyramid lattice [Eq.~\eqref{eq:4pyr_elt}] has kinks at $\abs{v}=1/3$ and $\abs{v}=1$. We now consider correlation functions of operators that occupy a single site in the chain of $d$-qudits (as also generally the case in dual-unitary circuits). Along $v=1/3$ the correlation function can be simplified through the repeated application of unitarity~\eqref{eq:unitarity_folded} and dual-unitarity~\eqref{eq:dual_unitarity_folded} to
\begin{align}
    \vcenter{\hbox{\includegraphics[height=0.5\textwidth]{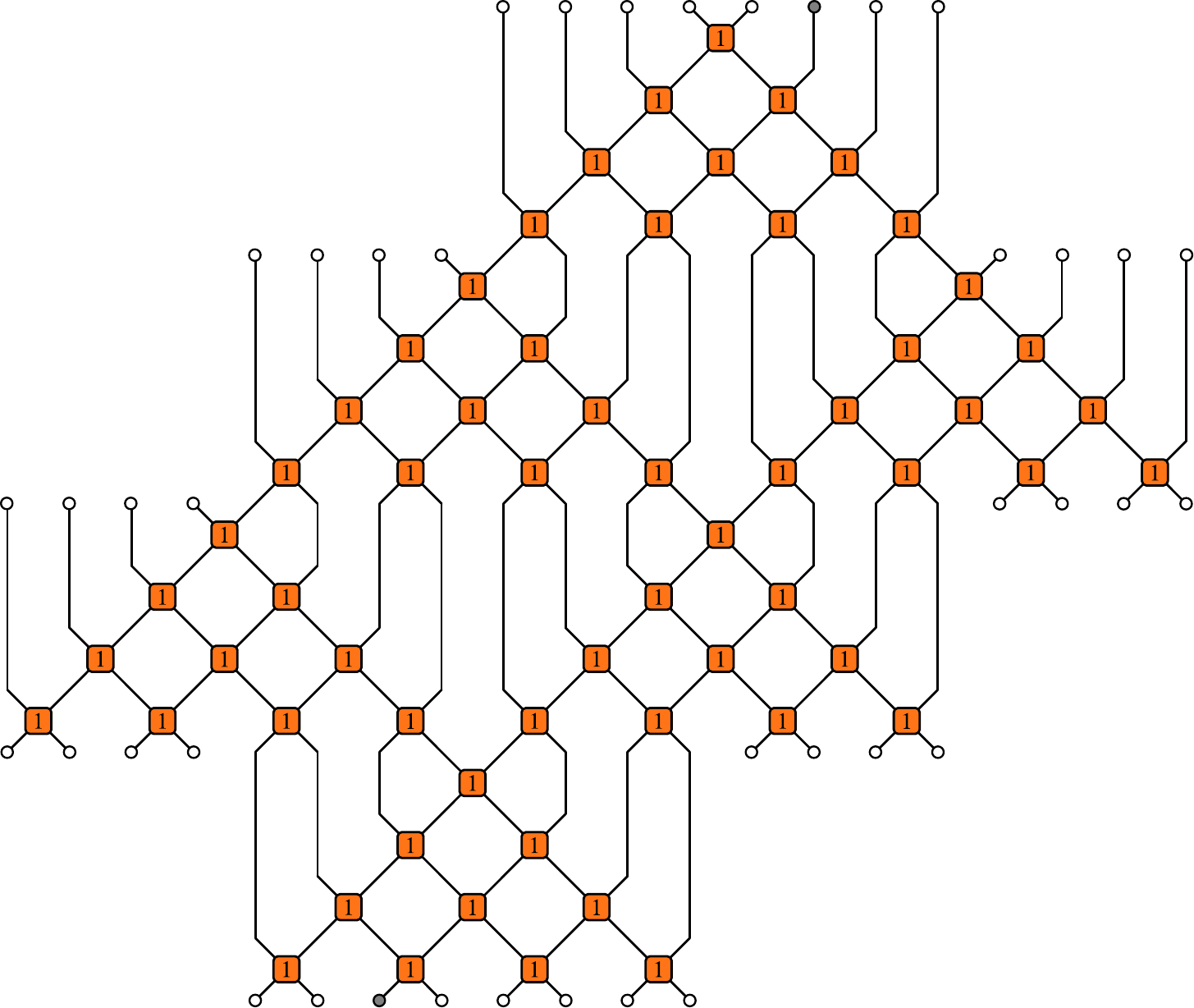}}}\,\,\,\,\,=\!\!\!\!\!\!\!\!\!\!\!\!\!\!\vcenter{\hbox{\includegraphics[height=0.5\textwidth]{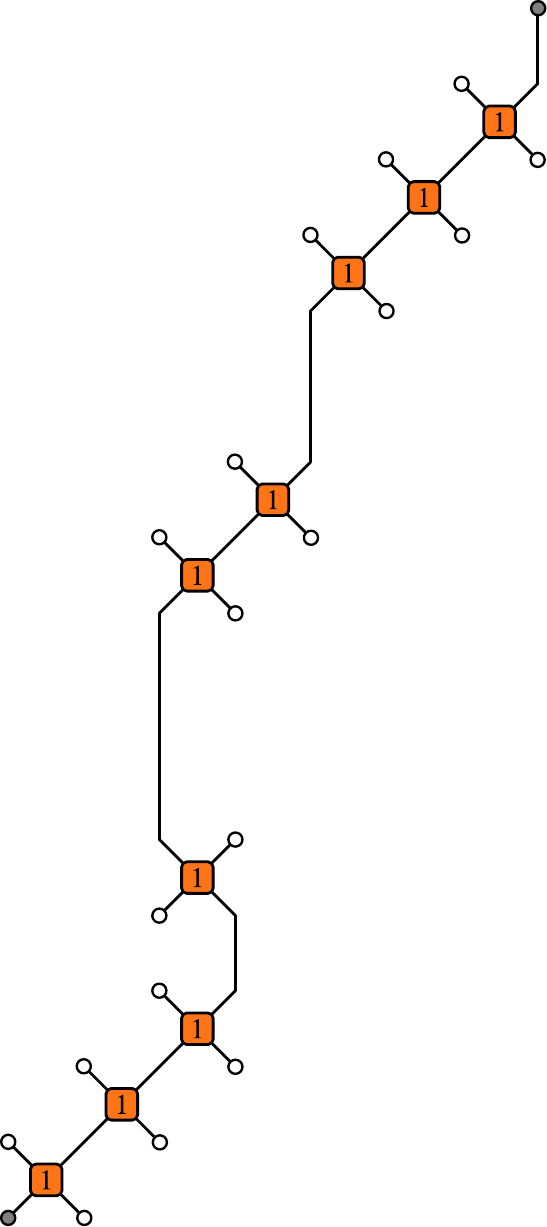}}}
\end{align}
The correlation function can again be reduced to a one-dimensional contraction as a multiplication of finite-dimensional matrices. More generally, we observe that non-vanishing correlations occur when the initial and final operator are connected by a worldline of the associated SWAP circuit. The worldline then defines a decomposition of the tensor network into influence matrices corresponding to perfectly dephasing baths. In completely reducible circuits, the one-site part of an operator is hence restricted to move along worldlines and feels the influence of the remaining many-body system only as a perfectly Markovian bath.

Along $v=1$, where the dynamics is generically solvable, we observe the same phenomenology of the diagram being reduced to the worldline by using the same approach:
\begin{align}
    \vcenter{\hbox{\includegraphics[height=0.38\textwidth]{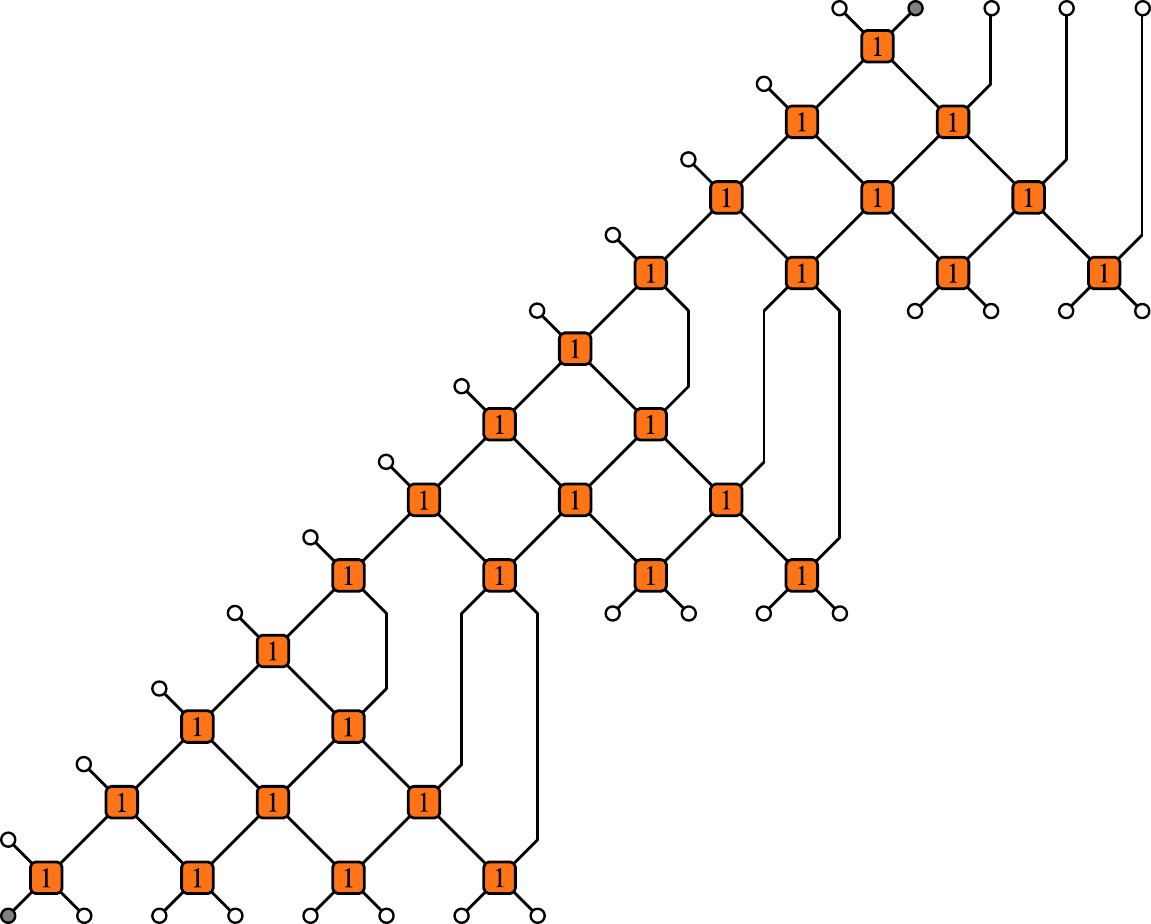}}}\,\,\,=\!\!\!\!\!\!\!\!\!\!\!\!\!\!\!\!\!\!\!\!\!\!\vcenter{\hbox{\includegraphics[height=0.38\textwidth]{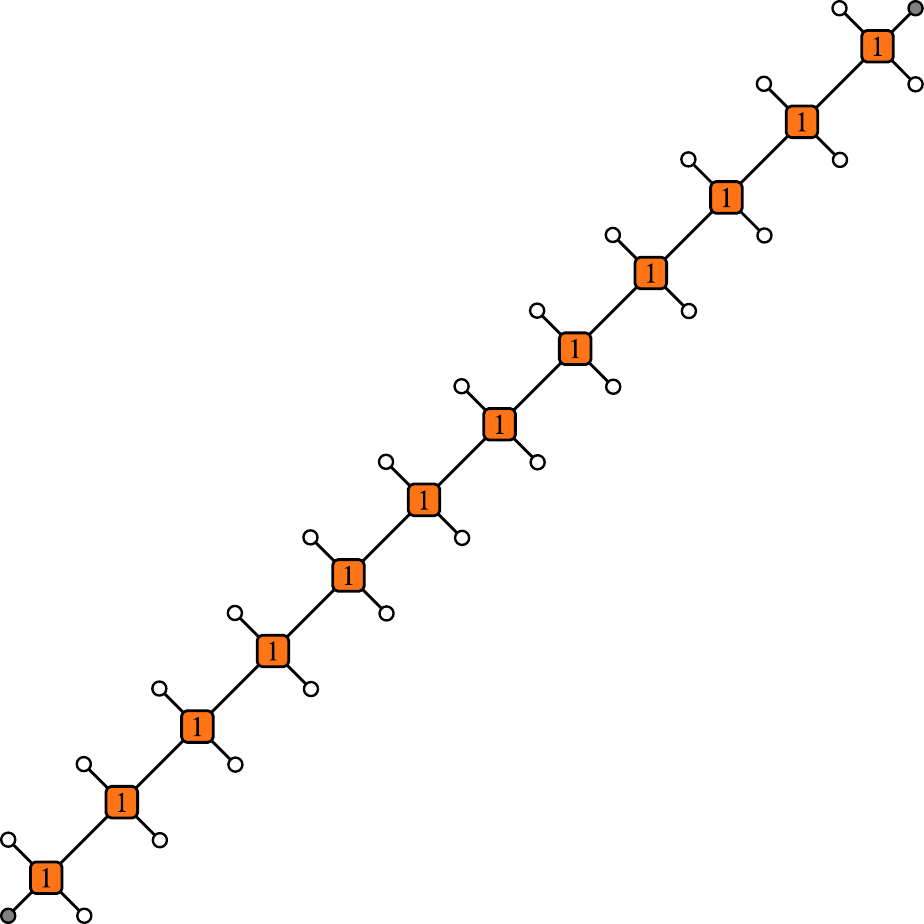}}}
\end{align}

\textbf{4-rocket lattice.} 
In the 4-rocket lattice the information flows along $\abs{v}=1/3$ and $\abs{v}=1$. Along $v=1/3$, we observe that the correlation function again simplifies to a product of finite-dimensional matrices (a property that is preserved for arbitrarily late times) through the repeated use of Eqs.~\eqref{eq:unitarity_folded} and~\eqref{eq:dual_unitarity_folded}:
\begin{align}
    \vcenter{\hbox{\includegraphics[height=0.6\textwidth]{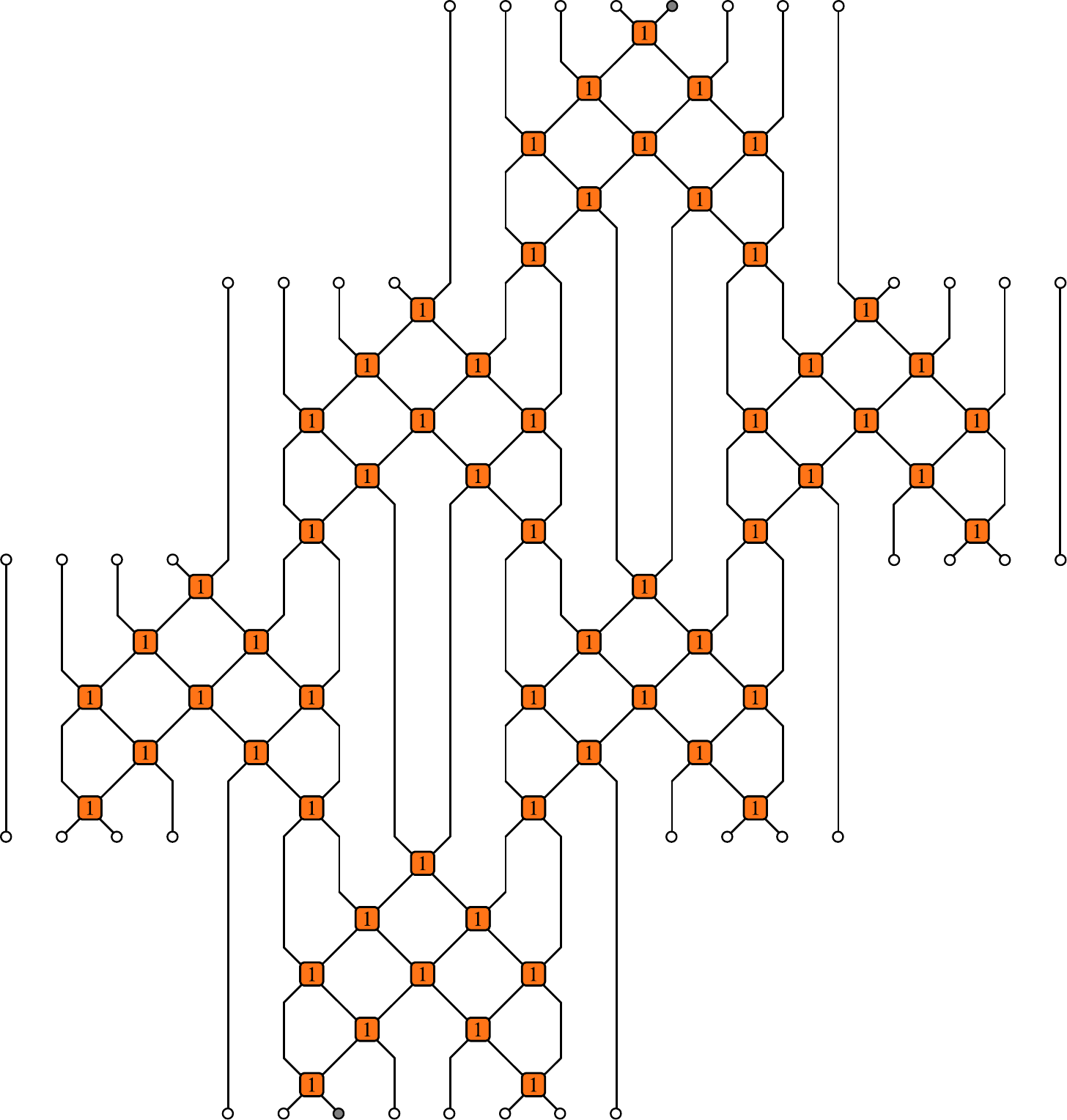}}}\,\,\,\,=\!\!\!\!\!\!\!\!\!\!\!\!\!\!\!\!\!\!\!\!\vcenter{\hbox{\includegraphics[height=0.6\textwidth]{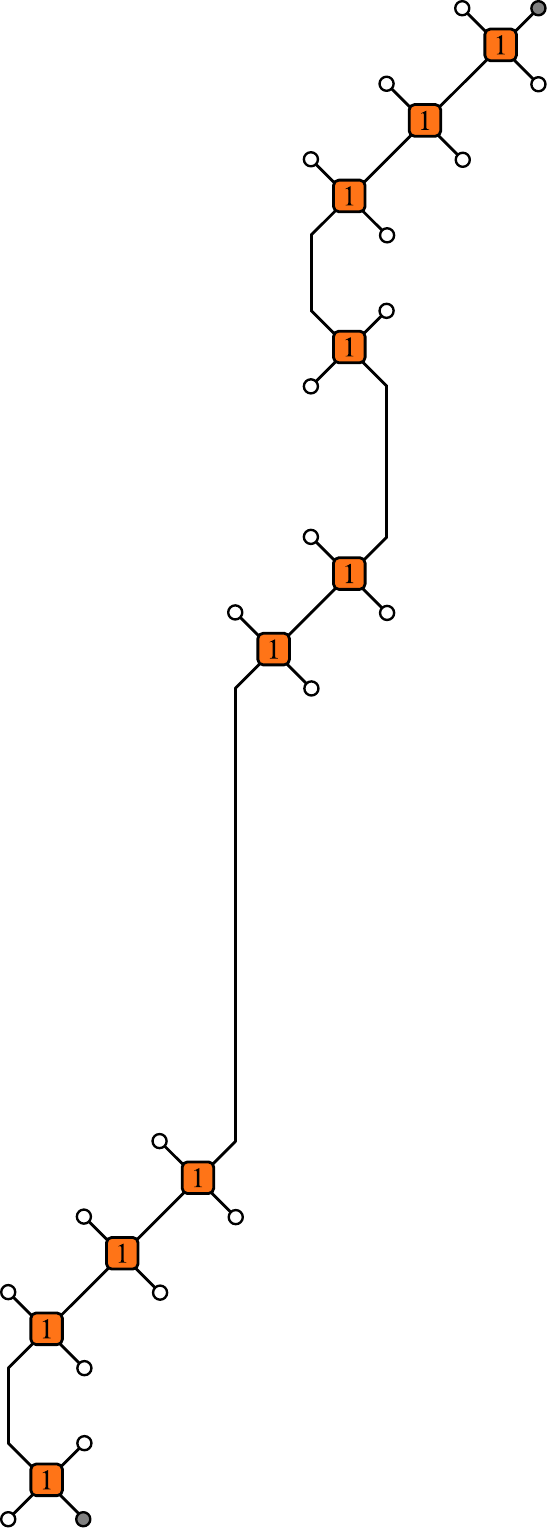}}}
\end{align}

\textbf{5-ray lattice.} In the 5-ray lattice, information flows along $\abs{v}=0,1/3,1$ [c.f. Eq.~\eqref{eq:5ray_elt}]. Along $v=0$ the correlator can be simplified to
\begin{align}
    \vcenter{\hbox{\includegraphics[height=0.47\textwidth]{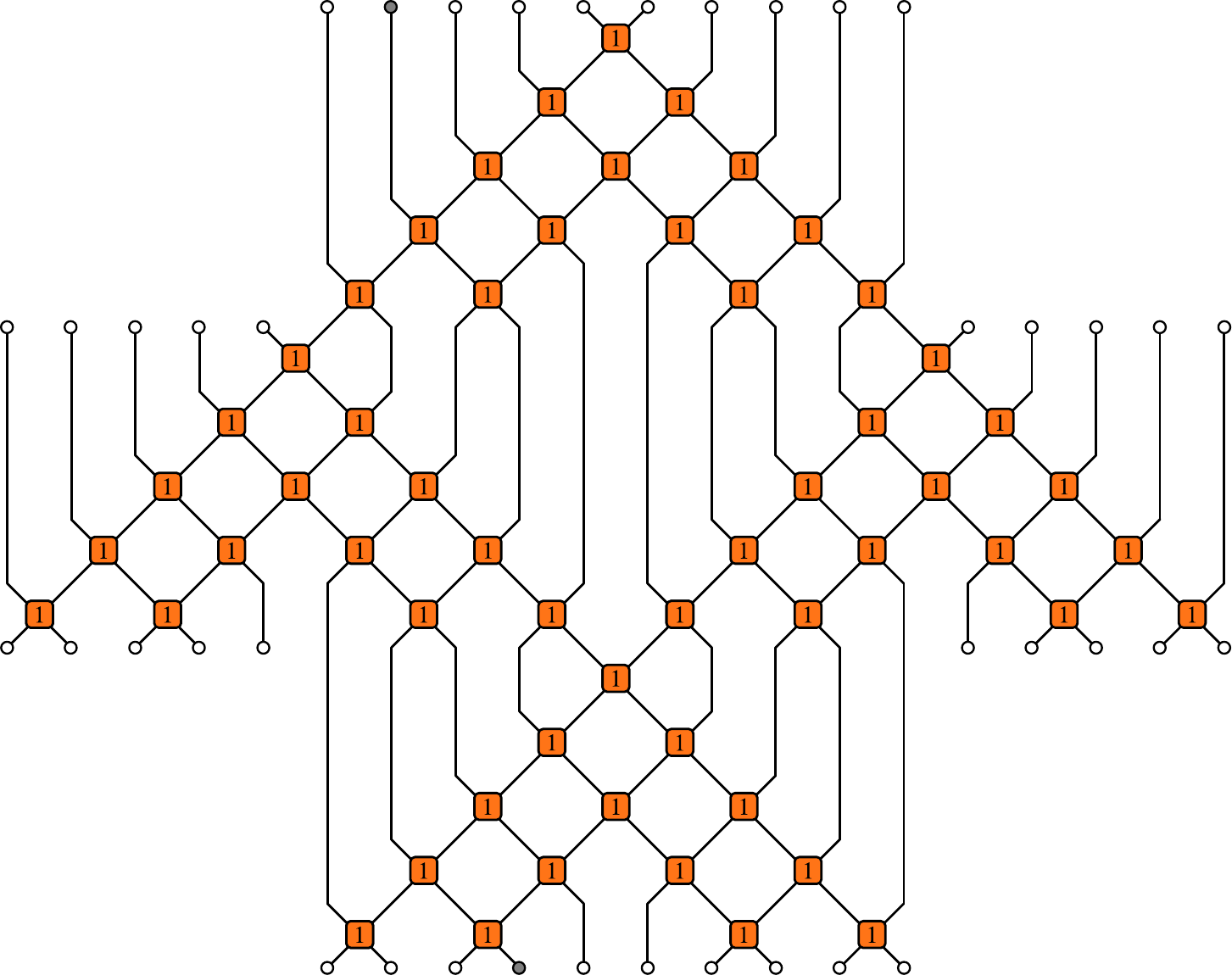}}}\,\,\,=\,\,\vcenter{\hbox{\includegraphics[height=0.47\textwidth]{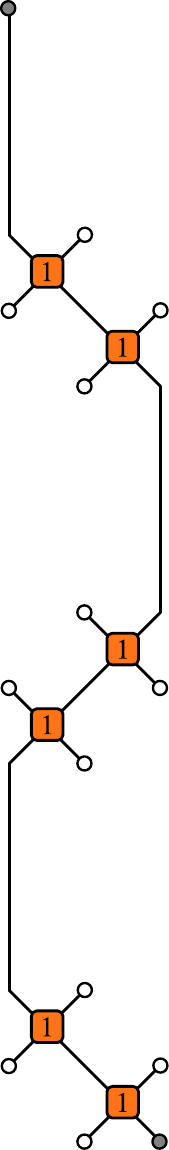}}}
\end{align}
and along $v=1/3$ a similar simplification results in
\begin{align}
    \vcenter{\hbox{\includegraphics[height=0.6\textwidth]{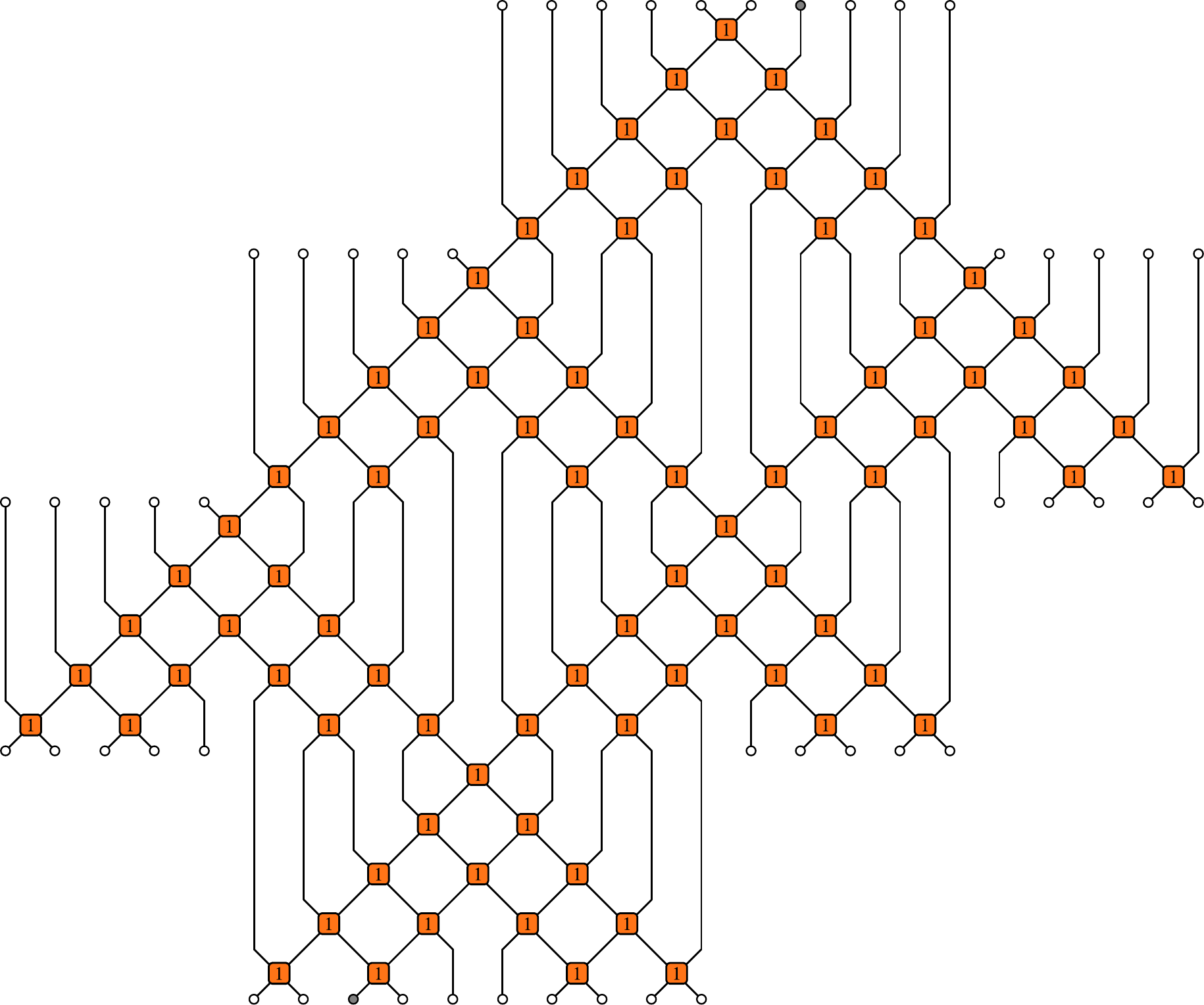}}}\,\,\,\,=\!\!\!\!\!\!\!\!\!\!\!\!\!\!\!\!\!\!\!\!\!\!\vcenter{\hbox{\includegraphics[height=0.6\textwidth]{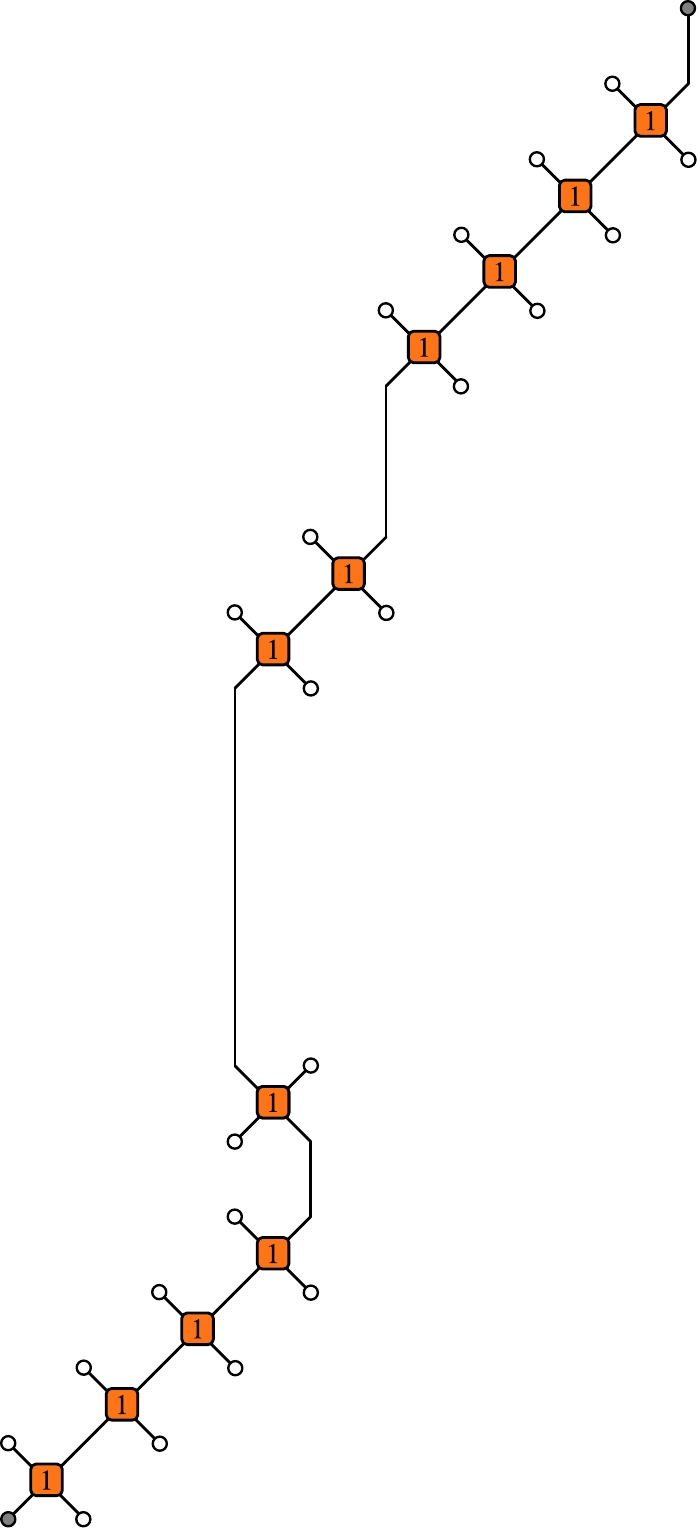}}}
\end{align}
These return the expected result: the effect of all gates that are not on the worldline connecting both operators can be captured by a perfectly Markovian influence matrix.

\section{Further examples}
\label{app:more_examples}

\textbf{Two infinite families.} We here present two families of spacetime lattice circuits generalizing the 4-pyramid and the 4-rocket gates, respectively. These families are formed from a base gate of size $N$ for any $N\geq4$. We show that the circuits have completely reducible dynamics with information flowing along four directions in spacetime. We also show that the entanglement dynamics exhibited by these circuits is distinct for each $N$, in the sense that the ELTs cannot be transformed into each other by coordinate transformations. 

To construct the first family, for any $N\geq4$ we define the base gate
\begin{align}
    U_N = \,\,\vcenter{\hbox{\includegraphics[width=0.34\textwidth]{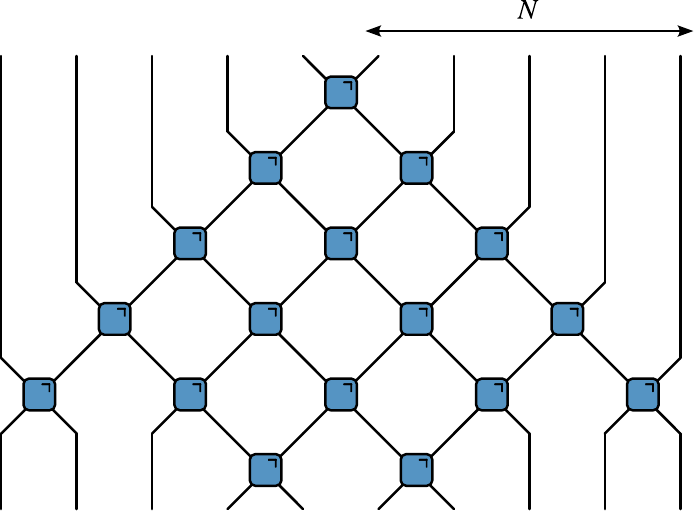}}}\,\,. \label{eq:family_1}
\end{align}
We construct $U_N$ by forming a $(N-1)$ by $(N-1)$ square of DU gates and delete the bottommost gate. We add a qudit to the left and right of the gate and apply a DU gates respectively from the bottom acting on these outermost bonds, connecting the square to the additional qudit. For $N=4$, this yields the base gate of the 4-pyramid lattice introduced in Eq.~\eqref{eq:4pyr}.

These gates lead to completely reducible dynamics, as can be seen by systematic contraction of diagrams.
Analyzing the worldlines of the circuits shows that the information flow is along the four directions $v=\pm\frac{N-3}{N-1},\pm1$ in spacetime. In accordance with this, the ELT has four kinks and reads [see also Fig.~\ref{fig:elt_sketches}(b,c)]
\begin{align}
    \mathcal{E}(v) = \begin{cases}
        ({N-2})/{N}, & \abs{v}\leq ({N-3})/({N-1}),\\
        ({1+(N-1)\abs{v}})/{N}, & ({N-3})/({N-1}) <\abs{v}\leq 1,\\
        \abs{v}, & \abs{v}>1.
    \end{cases} \label{eq:family_elt}
\end{align}

This construction provides an infinite number of solvable circuits with inequivalent entanglement dynamics. Although the information flows along four directions in spacetime for all $N\geq4$, the ELTs for different $N$ cannot be transformed into each other by a spacetime coordinate transformation. This is the case because any such transformation that does not involve mixing of the space and time coordinates preserves the ratio of velocities. Such a transformation is of the form $(x,t)=(ax',bt')$ and thus $v=\frac{a}{b}v'$, but the ratio of velocities of the information flow directions stays invariant. We excluded transformations mixing space and time which break the mirror symmetry of the ELT, and therefore cannot map the ELT of $U_N$ to the ELT of $U_{N'}$.

To obtain the second family, we define the base gate
\begin{align}
    V_N = \,\,\vcenter{\hbox{\includegraphics[width=0.34\textwidth]{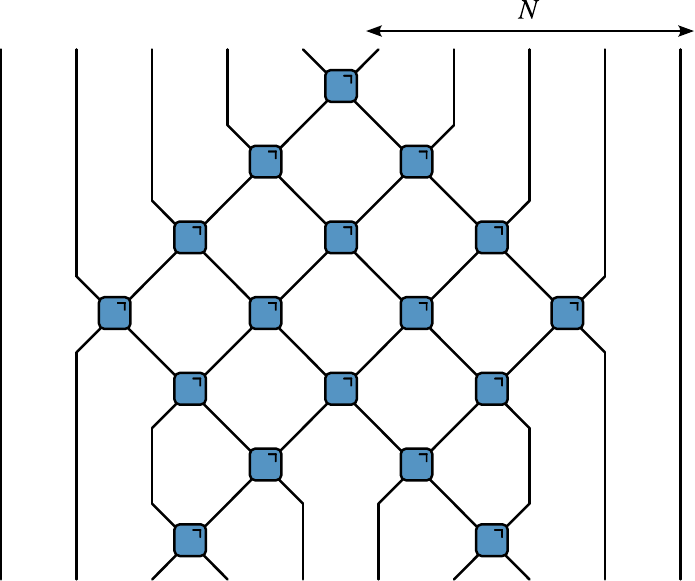}}}\,\,. \label{eq:family_2}
\end{align}
Similar to the construction of $U_N$, we first form a $(N-1)$ by $(N-1)$ square of DU gates and delete the bottommost gate. Then, we add two ancilla qudits and apply a single DU gate respectively from the bottom on the bonds next nearest to the center bond.
These gates lead to completely reducible dynamics. 
As for $U_N$, the information flow goes along four directions in spacetime, $\abs{v}=(N-3)/(N-1)$ and $\abs{v}=1$. The ELT of $V_N$ is the same as for $U_N$, given in Eq.~\eqref{eq:family_elt}.

\textbf{Tables of base gates.} In the following we tabulate some base gates with $2\leq N\leq3$ that have not been discussed in detail in the main text. We exclude cases where two vertices of the lattice are connected by more than one edge. We list their level of solvability and give their ELT if it is known. We identify two new classes of completely reducible DU2 gates inequivalent to the Kagome lattice base gate. 
\begin{figure*}[t]
    \centering
    \includegraphics[width = 0.75\textwidth]{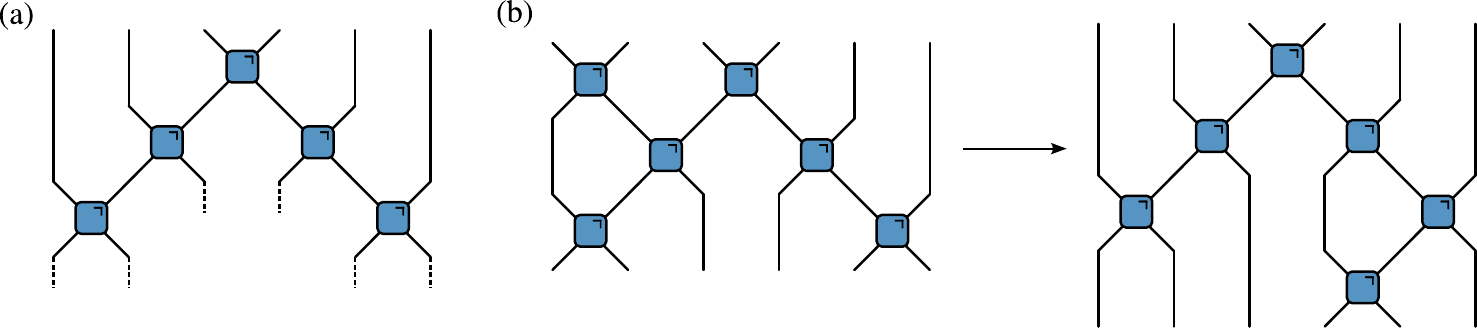}
    \caption{Illustration of standard form of base gates. (a) For a base gate to be in standard form, there cannot be a DU gate above the diagonals. (b) Example of transformation of a gate to standard form.}
    \label{fig:standardform}
\end{figure*}

We are interested in finding different completely reducible circuits that show distinct physical properties. However, as mentioned earlier, distinct base gates can lead to the same spacetime lattice in the bulk. This is a consequence of the unit cell of a lattice not being unique. We therefore briefly discuss different ways in which distinct base gates can lead to equivalent lattices. We consider three notions of equivalence: (i) gauge equivalence, (ii) equivalence under shearing and scaling, (iii) equivalence under blocking.
Any brickwork unitary circuit possesses a gauge invariance. For any two-site unitary gate $U\in U(q^2)$ and local transformations $u,v\in U(q)$, the gate 
\begin{equation}
    U'=(u\otimes v)U(v^\dagger\otimes u^\dagger),
\end{equation}
gives rise to the same dynamics as $U$ upon appropriate redefinition of observables. We exploit this property on the level of the brickwork circuit formed by the base gates. It implies that gauge transformed circuits have the same lattice structure in the bulk. We thus restrict ourselves to classify only gauge-inequivalent base gates. This means that we can take the base gates to be in a standard form depicted in Fig.~\ref{fig:standardform}.
Distinct base gates may also give rise to lattices which can be transformed into each other by a coordinate transformation of the form $(x',t')=(ax+bt,cx+dt)$. Nevertheless, having distinct representations in terms of base gates can be useful for constructing circuits with different biunitary connection, such that we still list such base gates.
Another way in which lattices can be equivalent is by blocking DU gates. DU gates acting on Hilbert spaces of possibly unequal dimensions $(q_1,q_2)$ can be composed diagonally to yield new DU gates~\cite{Borsi2022}. For example, diagonally composed DU gates acting on a $(q,q)$-dimensional dimensional Hilbert space can be combined to return a single DU gate acting on a $(q,q^2)$-dimensional Hilbert space:
\begin{align}
    \vcenter{\hbox{\includegraphics[width=0.22\textwidth]{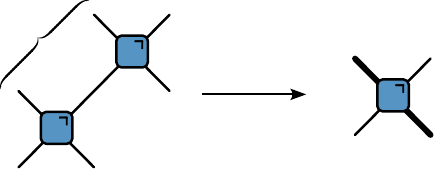}}}\,\,.
\end{align}
By identifying such diagonal compositions, some lattices can be cast in a simpler form. An example is given in Fig.~\ref{fig:blocking}.

Given these equivalences, we now list various $N=2$ base gates which are not discussed in the main text.
The following gate is completely reducible and equivalent to a sheared square lattice, as also apparent in the ELT:
\begin{align}
    \vcenter{\hbox{\includegraphics[width = .12\textwidth]{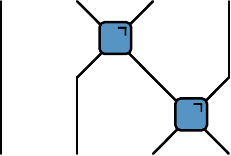}}}\,\,, \qquad \mathcal{E}(v) = \begin{cases}({1+\abs{v}})/{2},&v<\frac{1}{3},\\v,&v\geq\frac{1}{3}.\end{cases}
\end{align}
The following gate is completely reducible and corresponds to DU2 dynamics on the Kagome lattice:
\begin{align}
    \vcenter{\hbox{\includegraphics[width = .12\textwidth]{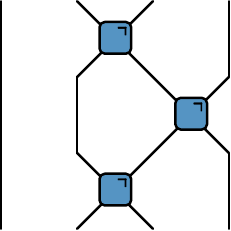}}}\,\,, \qquad \mathcal{E}(v) = \frac{1+\abs{v}}{2}.
\end{align}
It is also possible for dynamics to be completely reducible only in specific regions in spacetime. The following gate leads to dynamics that is completely reducible for $v>0$, as also apparent from the observation that it satisfies the DU2 condition in one direction:
\begin{align}
    \vcenter{\hbox{\includegraphics[width = .12\textwidth]{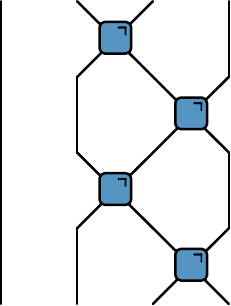}}}\,\,, \qquad \mathcal{E}(v) = \begin{cases}({1+v})/{2},,&v\geq0\\?&v<0.\end{cases}
\end{align}
The following gate satisfies the DU2 condition in the opposite direction and leads to completely reducible dynamics for $v<0$:
\begin{align}
    \vcenter{\hbox{\includegraphics[width = .12\textwidth]{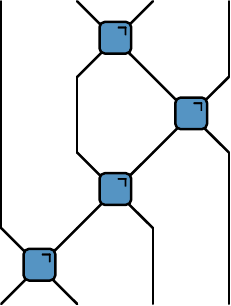}}}\,\,, \qquad \mathcal{E}(v) = \begin{cases}({1-v})/{2},&v\leq0,\\?&v>0.\end{cases}
\end{align}
Additional $N=3$ base gates can also be found. The following gate is completely reducible and equivalent to a coordinate-transformed dual-unitary circuit:
\begin{align}
    \vcenter{\hbox{\includegraphics[width = .2\textwidth]{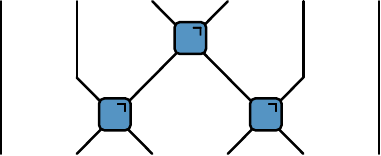}}}\,\,, \qquad \mathcal{E}(v) = \begin{cases}\frac{1}{3},&\abs{v}<\frac{1}{3},\\\abs{v},&v\geq\frac{1}{3}.\end{cases}
\end{align}
The following completely reducible gate satisfies the DU2 condition and is inequivalent to the Kagome lattice construction:
\begin{align}
    \vcenter{\hbox{\includegraphics[width = .2\textwidth]{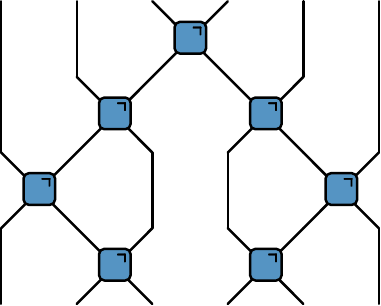}}}\,\,, \qquad \mathcal{E}(v) = \frac{1+2\abs{v}}{3}.
\end{align}
Another inequivalent but completely reducible gate that satisfied the DU2 condition is given by:
\begin{align}
    \vcenter{\hbox{\includegraphics[width = .2\textwidth]{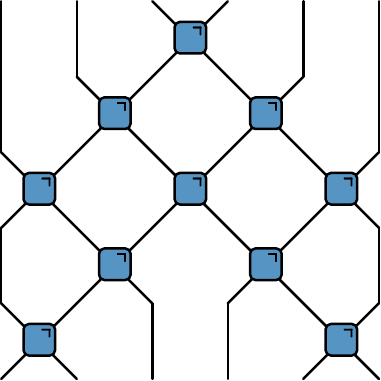}}}\,\,, \qquad \mathcal{E}(v) = \frac{2+\abs{v}}{3}.
\end{align}
To conclude, we note that the following gate is solvable for $|v| \geq 1/3$ and not solvable for $v < 1/3$, since it satisfies the DU3 condition:
\begin{align}
    \vcenter{\hbox{\includegraphics[width = .2\textwidth]{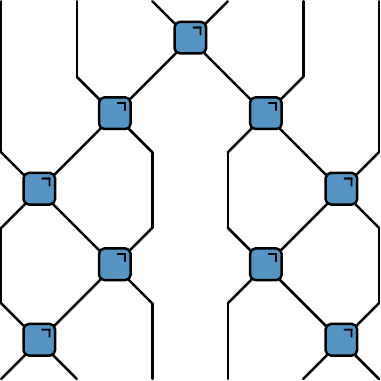}}}\,\,.
\end{align}

\begin{figure*}[t]
    \centering
    \includegraphics[width = 0.7\textwidth]{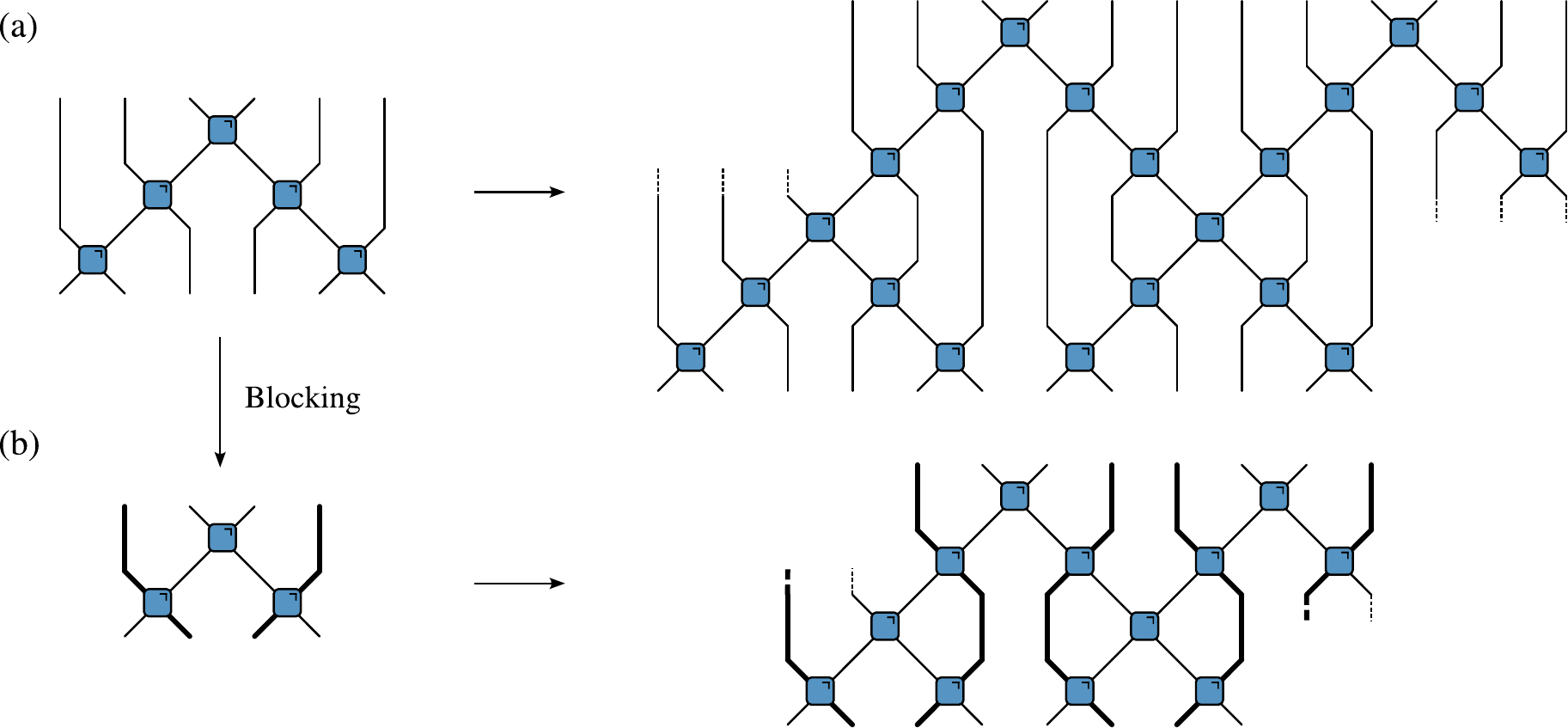}
    \caption{Illustration of equivalence of lattices under blocking of DU gates. The base gate (a) can be brought into a form equivalent to the Kagome lattice (b) by diagonal composition of a subset of DU gates.}
    \label{fig:blocking}
\end{figure*}

\section{Generalization to biunitary connections}
\label{app:biunitary}

In this section we discuss the generalization of spacetime lattice circuits to biunitary connections. Biunitary connections are objects first introduced in pure mathematics that can be regarded as generalizations of dual-unitary gates~\cite{Reutter2019}. We define spacetime lattice circuits of biunitary connections by placing arbitrary biunitary connections (rather than DU gates) on the vertices of a spacetime lattice in a consistent manner, following Refs.~\cite{Claeys2024,Rampp2025}. These circuits can then be contracted in the same manner as spacetime lattices of DU gates, leading to analogous results. The advantage of using the formalism of biunitary connections is that it enables us to provide alternative parametrizations of solvable models, some of which are easier to practically implement either in experiment or numerics. In particular, it enables us to compress the local Hilbert space and find base gates acting on a smaller number of qudits, but with analogous dynamics to the base gates composed of DU gates. For concreteness, we present base gates analogous to the $N=4$ base gates, which here act on a local Hilbert space of size $d^{2}$ as opposed to $d^4$, enabling their implementation in brickwork circuits of ququads.

For a detailed introduction to biunitarity and the shaded calculus we refer the reader to Refs.~\cite{Reutter2019,Claeys2024}. Before introducing biunitary connections, we give a brief overview of the diagrammatic formalism known as the shaded calculus. The shaded calculus is similar to familiar tensor network notation, with the addition of shaded areas. In addition to wires, shaded areas also carry an index that has to be equal for areas shared between distinct tensors and is summed over when the area is closed. This formalism provides a convenient notation to represent the equality of indices that are not summed over, as well as contractions where more than two indices are equal. An exemplary diagram together with its tensor network counterpart is given by
\begin{align}
    \vcenter{\hbox{\includegraphics[width = .2\columnwidth]{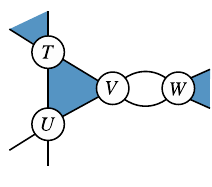}}} \qquad \Rightarrow \qquad 
     \sum_{bef}\vcenter{\hbox{\includegraphics[width = .2\columnwidth]{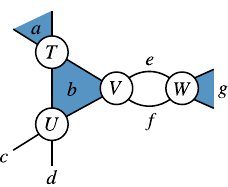}}}\, = \sum_{bef} T_{ab} U_{bcd} V_{bef} W_{efg}\,.
\end{align}

In the shaded calculus, biunitary connections are represented by four-valent vertices, with a shading pattern corresponding to either of the following diagrams
\begin{align}
    \vcenter{\hbox{\includegraphics[width = .5\columnwidth]{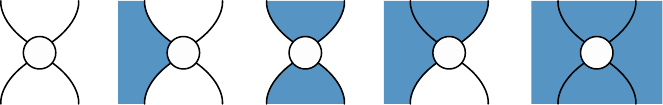}}}
\end{align}

For our purposes, the relevant biunitary connections are dual-unitary gates and complex Hadamard matrices.
Dual-unitary gates are recovered from the shaded calculus in the absence of shading
\begin{align}
    \vcenter{\hbox{\includegraphics[width = .07\textwidth]{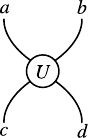}}}\,\, =\,\,  \vcenter{\hbox{\includegraphics[width = .073\textwidth]{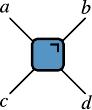}}} \,\,= U_{ab,cd}\,.
\end{align}

Shading two opposing regions of a biunitary results in complex Hadamard matrices (CHM),
\begin{align}
    \vcenter{\hbox{\includegraphics[width = .063\textwidth]{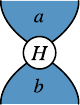}}}\,\, = \,\,\vcenter{\hbox{\includegraphics[width = .0251\textwidth]{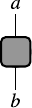}}}\,\, = H_{ab}.
\end{align}
A complex Hadamard matrix (CHM) is a  $q \times q$ matrix $H$ that is proportional to a unitary matrix and with all matrix elements having unit modulus~\cite{Tadej2006}, i.e. $|H_{ab}|=1, \forall a,b$, which fixes $H^{\dagger} H = H H^{\dagger} = q \mathbbm{1}$.
Depending on the orientation of these connections in a quantum circuit, complex Hadamard matrices define either single-site unitary gates or two-site controlled phase gates.
A similar decomposition appears in the kicked Ising model at the self-dual point, a paradigmatic dual-unitary circuit~[X].

Using the freedom to shade circuit diagrams, we can find instances of base gates that implement physically analogous dynamics in a smaller local Hilbert space. We illustrate this here for the 4-pyramid lattice. For this lattice, the circuit diagram admits a bipartite shading:
\begin{align}
    \vcenter{\hbox{\includegraphics[width=0.42\textwidth]{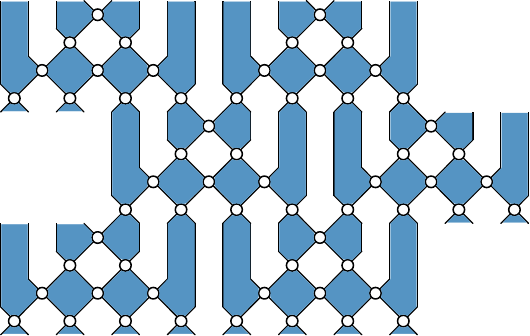}}}
\end{align}
The corresponding circuit can be expressed purely in terms of CHMs as
\begin{align}
    \vcenter{\hbox{\includegraphics[width=0.6\textwidth]{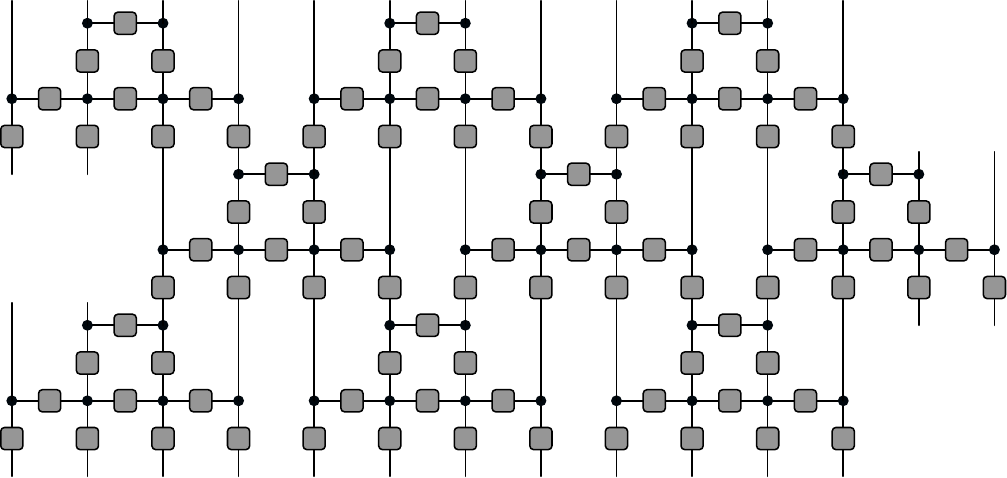}}}
\end{align}
and has a corresponding base gate
\begin{align}
    U_{abcd,efgh} = \,\,\vcenter{\hbox{\includegraphics[width=0.2\textwidth]{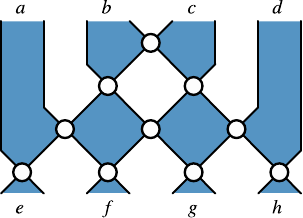}}}\,\,=\,\,\vcenter{\hbox{\includegraphics[width=0.166\textwidth]{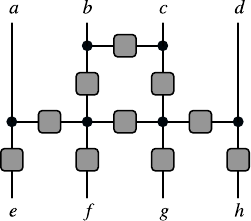}}}\,\,, \label{eq:4pyr_chm}
\end{align}
which acts on a $d^{2}$-dimensional local Hilbert space as opposed to a $d^4$-dimensional one. In particular, the lowest attainable Hilbert space dimension is four. 
For the Kagome lattice, this procedure was used in Ref.~\cite{Rampp2025} to derive a family of DU2 gates given by the base gate
\begin{align}
    U_{ab,cd} = \,\,\vcenter{\hbox{\includegraphics[width=0.087\textwidth]{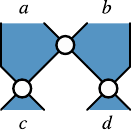}}}\,\,=\,\,\vcenter{\hbox{\includegraphics[width=0.066\textwidth]{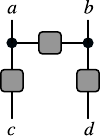}}}\,\,, \label{eq:du2_chm}
\end{align}
The shaded version of the gate Eq.~\eqref{eq:gate_n4_rocket} reads
\begin{align}
    U_{abcd,efgh} = \,\,\vcenter{\hbox{\includegraphics[width=0.2\textwidth]{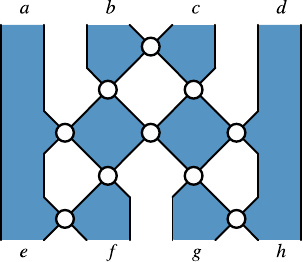}}}\,\,=\,\,\vcenter{\hbox{\includegraphics[width=0.158\textwidth]{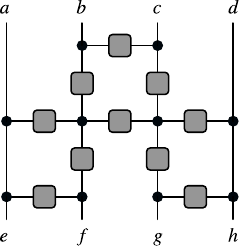}}}\,\,, \label{eq:4rock_chm}
\end{align}
and the shaded version of the nested Kagome gate reads 
\begin{align}
    U_{abcd,efgh} = \,\,\vcenter{\hbox{\includegraphics[width=0.2\textwidth]{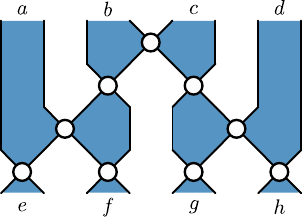}}}\,\,=\,\,\vcenter{\hbox{\includegraphics[width=0.166\textwidth]{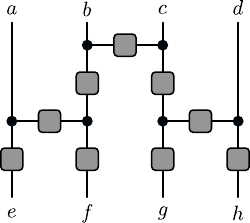}}}\,\,. \label{eq:4nested_chm}
\end{align}

\end{widetext}

\bibliography{lattices}

\end{document}